\documentclass[twocolumn]{aastex6}
\pdfoutput=1
\usepackage{amsmath,color,textcomp,url,graphicx,subfigure}
\newcommand{\kms}{\hbox{km\,s$^{-1}$}}
\newcommand{\Mjup}{$M_{\mathrm{Jup}}$}

\newcommand{\Msol}{$M_{\odot}$}
\newcommand{\masyr}{$\mathrm{mas}\,\mathrm{yr}^{-1}$}
\newcommand{\teff}{$T_{\rm eff}$}

\DeclareMathOperator\erfc{erfc}

\slugcomment{To appear in ApJS.}

\shorttitle{THE INITIAL MASS FUNCTION OF THE TW~HYA ASSOCIATION}
\shortauthors{Gagn\'e et al.}

\begin{document}

\title{BANYAN. IX. THE INITIAL MASS FUNCTION AND PLANETARY-MASS OBJECT SPACE DENSITY OF THE TW~HYA ASSOCIATION}

\author{Jonathan Gagn\'e\altaffilmark{1,2}, Jacqueline K. Faherty\altaffilmark{1,3}, Eric E. Mamajek\altaffilmark{4,5}, Lison Malo\altaffilmark{6,7}, Ren\'e Doyon\altaffilmark{7}, Joseph C. Filippazzo\altaffilmark{8}, Alycia J. Weinberger\altaffilmark{1}, Jessica K. Donaldson\altaffilmark{1}, S\'ebastien L\'epine\altaffilmark{9}, David Lafreni\` ere\altaffilmark{7}, \'Etienne Artigau\altaffilmark{7}, Adam J. Burgasser\altaffilmark{10}, Dagny Looper\altaffilmark{11}, Anne Boucher\altaffilmark{7}, Yuri Beletsky\altaffilmark{12}, Sara Camnasio\altaffilmark{13}, Charles Brunette\altaffilmark{7}, Genevi\` eve Arboit\altaffilmark{7}}
\affil{\altaffilmark{1} Carnegie Institution of Washington DTM, 5241 Broad Branch Road NW, Washington, DC~20015, USA; jgagne@carnegiescience.edu\\
\altaffilmark{2} NASA Sagan Fellow\\
\altaffilmark{3} NASA Hubble Fellow\\
\altaffilmark{4} Department of Physics \& Astronomy, University of Rochester, Rochester, NY 14627, USA\\
\altaffilmark{5}Jet Propulsion Laboratory, California Institute of Technology, 4800 Oak Grove Drive, Pasadena, CA 91109, USA\\ 
\altaffilmark{6} Canada-France-Hawaii Telescope, 65-1238 Mamalahoa Hwy, Kamuela, HI~96743, USA\\
\altaffilmark{7} Institute for Research on Exoplanets, Universit\'e de Montr\'eal, D\'epartement de Physique, C.P.~6128 Succ. Centre-ville, Montr\'eal, QC H3C~3J7, Canada\\
\altaffilmark{8} Space Telescope Science Institute, 3700 San Martin Dr, Baltimore, MD 21218, USA\\
\altaffilmark{9} Department of Physics and Astronomy, Georgia State University, 25 Park Place, Atlanta, GA 30302, USA\\
\altaffilmark{10} Center for Astrophysics and Space Sciences, University of California, San Diego, 9500 Gilman Dr., Mail Code 0424, La Jolla, CA~92093, USA\\
\altaffilmark{11} New York University Tisch School of the Arts, 721 Broadway, 10th floor, New York, NY 10003, USA\\
\altaffilmark{12} Las Campanas Observatory, Carnegie Institution of Washington, Colina el Pino, Casilla 601 La Serena, Chile\\
\altaffilmark{13} Department of Physics and Astronomy, Hunter College, City University of New York, NY 10065, USA}

\begin{abstract}

A determination of the initial mass function (IMF) of the current, incomplete census of the 10\,Myr-old TW~Hya association (TWA) is presented. This census is built from a literature compilation supplemented with new spectra and 17 new radial velocities from on-going membership surveys, as well as a re-analysis of Hipparcos data that \replaced{yielded one new early-type bona fide member (HR~4334, A2\,Vn)}{confirmed HR~4334 (A2\,Vn) as a member}. Though the dominant uncertainty in the IMF remains census incompleteness, a detailed statistical treatment is carried out to make the IMF determination independent of binning, while accounting for small number statistics. The \added{currently known }high-likelihood members\deleted{ of TWA }are fitted by a log-normal distribution with a central mass of $0.21^{+0.11}_{-0.06}$\,\Msol\ and a characteristic width of $0.8^{+0.2}_{-0.1}$\,dex in the 12\,\Mjup--2\,\Msol\ range, whereas a Salpeter power law with $\alpha = 2.2^{+1.1}_{-0.5}$ best describes the IMF slope in the 0.1--2\,\Msol\ range. This characteristic width is higher than \replaced{previous IMF determinations for}{other} young associations, \replaced{and consistent with previous observations that TWA displays a flatter than usual IMF.}{which may be due to incompleteness in the current census of low-mass TWA stars. }A tentative overpopulation of isolated planetary-mass members similar to 2MASS~J11472421--2040204 and 2MASS~J11193254--1137466 is identified\replaced{ in comparison to the stellar members of TWA. These results}{: this} indicates that there might be as many as $10_{-5}^{+13}$ similar\deleted{ low-mass }members of TWA with hot-start model-dependent masses estimated at $\sim$\,5--7\,\Mjup\replaced{. Most of them}{, most of which} would be too faint to be detected in 2MASS\deleted{: TWA~41 and TWA~42 are exceptions, as they lie on the nearer side ($\sim$\,29--33\,pc) of the association}. Our new radial velocity measurements corroborate the membership of 2MASS~J11472421--2040204, and secure TWA~28 (M8.5\,$\gamma$), TWA~29 (M9.5\,$\gamma$) and TWA~33 (M4.5\,e) as members. \replaced{A search for new planetary-mass TWA members in the BASS-Ultracool survey yielded the discovery of 2MASS~J09553336--0208403, a young L7-type interloper unrelated to TWA}{The discovery of 2MASS~J09553336--0208403, a young L7-type interloper unrelated to TWA, is also presented}. 
\end{abstract}


\section{INTRODUCTION}

The study of young moving groups and associations has received much attention in recent years, in particular for their potential in hosting the brightest very low-mass substellar objects in the solar neighborhood. The youngest of these associations within 100\,pc is TW~Hya (TWA hereafter; e.g., see \citealp{1989ApJ...343L..61D,1997Sci...277...67K,2003ApJ...599..342S,2004AA...425L..29C,2005ApJ...634.1385M,2004ARAA..42..685Z,2013ApJ...762..118W,2014AA...563A.121D,2016arXiv161001667D}), at an age of $10 \pm 3$\,Myr \citep{2015MNRAS.454..593B}. Although this association has been well studied over more than a decade, many of its low-mass members are still missing since they are too faint to have been detected by the Hipparcos mission \citep{1997AA...323L..49P}, and only a first estimate of its initial mass function (IMF) has been presented \citep{2011PhDT.......245L}.

Recent discoveries have identified the first few isolated planetary-mass objects in the solar neighborhood (e.g., \citealp{2013ApJ...777L..20L,2013ApJS..205....6M,2014ApJ...785L..14G,2014ApJ...783..121G,2015ApJ...808L..20G,2015ApJS..219...33G}) that are members of young associations. Their young age means that they still retain more heat from their formation process, and thus they are intrinsically brighter than similar objects at the age of the field. Field-age planetary mass objects are expected to have temperatures in the range $\sim$\,250--500\,K, which correspond to the spectral class Y \citep{2008AA...482..961D,2011ApJ...743...50C,2011ApJS..197...19K,2012ApJ...753..156K,2012ApJ...759...60T}. Such objects are extremely faint, even at near-infrared (NIR) wavelengths (e.g., $M_K \geq 20$; \citealt{2015ApJ...799...37L}). Younger planetary-mass objects are typically much brighter ($M_K \sim$\,12--16; e.g., \citealp{2013ApJ...777L..20L,2015ApJS..219...33G,2015ApJ...808L..20G}).

Two recent discoveries in particular demonstrate the interest of TWA as a laboratory for understanding this isolated planetary-mass population. 2MASS~J11193254--1137466 \citep{2015AJ....150..182K,2016ApJ...821L..15K} and 2MASS~J11472421--2040204 \citep{2016ApJ...822L...1S} are both candidate members of TWA with spectral types L7 that display signs of youth, and \replaced{estimated with}{with estimated} masses as low as 5--7\,\Mjup. Their close distances to the Sun (29--33\,pc) place them at the nearer side of the TWA spatial distribution. This result, as well as some recent indications that planetary-mass objects could be more abundant than expected in the young Tucana-Horologium Association \citep{2015ApJS..219...33G}, calls for an update of the IMF of TWA and an estimate of the space density of its planetary-mass members.

\replaced{In this Section, the IMF of TWA is constructed and compared to two functional forms that are widely used in the literature.}{Various functional forms have been used in the literature to characterize the IMF of a stellar population.} The first\deleted{ one }was introduced by \cite{1955ApJ...121..161S}:
\begin{align}
	\phi\left(\log _{10}m\right) = \frac{\mathrm{d}N}{\mathrm{d}\log _{10}m} = \phi_0\,m^{1-\alpha},\label{eqn:salpeterimf}
\end{align}
\noindent where $\phi_0$ is the density of stars per logarithm mass per pc$^{3}$ at $m = 1$\,\Msol.

The Salpeter IMF has been used to represent the IMF of field stars with masses in the range 0.4--10\,\Msol\ with a slope of $\alpha = 2.35$ \citep{1955ApJ...121..161S}. However, steeper slopes have been measured for the more massive stars (e.g., $\alpha = 2.7 \pm 0.2$ in the 1.1--1.6\,\Msol\ range or $\alpha = 3.1 \pm 0.2$ in the 1.6--4\,\Msol\ range; \citealt{2003MNRAS.343.1231S}), whereas low-mass stars yield shallower slopes (e.g., $\alpha = 1.05$ in the 0.1--1.0\,\Msol\ range; \citealt{1997AJ....113.2246R}, or $\alpha = 1.2 \pm 0.3$ in the 0.1--0.7\,\Msol\ range; \citealt{2010ARA&A..48..339B}).

\replaced{In order to represent this slope variation as a function of the mass regime with a single functional form, log-normal distributions have been introduced \citep{1979ApJS...41..513M,2005ASSL..327...41C}}{Log-normal distributions have been introduced \citep{1979ApJS...41..513M,2005ASSL..327...41C} to represent this slope variation as a function of the mass regime with a single functional form}:
\begin{align}
	\phi\left(\log _{10}m\right) = \frac{\phi_t}{\sigma\sqrt{2\pi}}\exp{\left(-\frac{\left(\log _{10}m-\log _{10}m_c\right)^2}{2\sigma^2}\right)},\label{eqn:lognormalimf}
\end{align}
\noindent where $m_c$ represents a characteristic mass where the IMF peaks, $\sigma$ represents the characteristic width of the IMF in log space, and $\phi_t$ is the space density of objects per unit logarithm mass at the peak of the IMF. Additional functional forms have also been introduced, such as multi-segmented power laws \citep{1993MNRAS.262..545K,2001MNRAS.322..231K}, or a combination of a log-normal with power laws \citep{2010ARA&A..48..339B}.

Fitting such an IMF \replaced{on}{to} field stars with masses in the range 0.1--50\,\Msol\ has yielded typical values of $m_c = 0.1$\,\Msol\ and $\sigma = 0.7$\,dex \citep{1979ApJS...41..513M}. A subsequent determination of the field IMF from the Sloan Digital Sky Survey \citep{2011AJ....141...98B} yielded $m_c = 0.18$\,\Msol\ and $\sigma = 0.34$\,dex when counting all components of multiple systems, whereas an IMF treating systems as single objects yielded $m_c = 0.25$\,\Msol\ and $\sigma = 0.28$\,dex \citep{2010AJ....139.2679B}. Most field IMFs yield characteristic masses in the range 0.15--0.25\,\Msol\ \citep{2005ASSL..327...41C}.

IMF determinations in the brown dwarf regime ($m \leq 0.075$\,\Msol) have yielded slopes of $\alpha \leq 0$ \citep{2008ApJ...676.1281M,2010AA...522A.112R,2010MNRAS.406.1885B}, indicating that the space density of brown dwarfs decreases with decreasing masses.

Current evidence suggests that the IMFs of young associations are similar to that of the field. \cite{2005ASSL..327...41C} demonstrated that several young stellar associations are well described by a log-normal IMF similar to that of field stars, with $m_c = 0.25$\,\Msol\ and $\sigma = 0.55$\,dex (valid for $m \leq 1$\,\Msol; see also \citealp{2003ApJ...586L.133C,2003A&A...400..891M}). \cite{2012EAS....57...45J} further demonstrated this by obtaining a log-normal IMF with $m_c = 0.25$\,\Msol\ and $\sigma = 0.52$\,dex that adequately represents the IMF of several young stellar associations (e.g., \citealp{2002AA...395..813B,2003AA...400..891M,2004MNRAS.351.1401J,2005RMxAC..24..217B,2006AA...448..189D,2007AA...471..499M,2007ApJS..173..104L,2009MNRAS.392.1034O,2009AIPC.1094..912C}), with the exception of Upper~Scorpius, which might have an excess of brown dwarfs \citep{2007MNRAS.374..372L}. It must be noted, however, that this last IMF was constructed with candidate members of Upper~Scorpius that were not confirmed with spectroscopy, and could thus be subject to a high level of contamination from reddened background stars or extragalactic sources.

This paper presents a set of new spectroscopic and kinematic observations that furthers the census of TWA members. An updated compilation of its members across spectral types A0--L7 is presented, which is then used to determine its IMF and to estimate the space density of its isolated planetary-mass members. In Section~\ref{sec:newmembers}, new candidate members from the BASS-Ultracool and SUPERBLINK-south surveys and a re-analysis of Hipparcos are presented. \replaced{New observations that consist of low- and high-resolution, optical and near-infrared spectroscopy, are detailed in Section~\ref{sec:obs}}{In Section~\ref{sec:obs}, new observations that consist of low- and high-resolution, optical and near-infrared spectroscopy are detailed}. \replaced{These}{In Section~\ref{sec:data_an}, these} new data are used to assign spectral types, assess signs of low-gravity, and measure the radial velocity of several TWA candidate members that originate from various surveys\replaced{ (Section~\ref{sec:data_an}).}{.} \replaced{The}{In Section~\ref{sec:candcomp}, the} final list of TWA members and candidates is detailed\replaced{ in Section~\ref{sec:candcomp},}{,} and their physical properties are estimated in Section~\ref{sec:physpar}.\added{ In Section~\ref{sec:twa_completeness}, the completeness of the current census of TWA members is discussed. }\replaced{The}{In Section~\ref{sec:imf}, the} IMF of TWA is constructed and discussed\replaced{ in Section~\ref{sec:imf}.}{.}\deleted{ The 2MASS completeness limit of TWA objects is calculated in Section~\ref{sec:tmass_completeness}}\replaced{, and}{In Section~\ref{sec:density}}, the space density of its isolated planetary-mass members is \replaced{presented}{assessed}\replaced{ in Section~\ref{sec:density}.}{.} This paper is concluded in Section~\ref{sec:conclusion}.

\section{NEW MEMBERS OF TW~HYA}\label{sec:newmembers}

This section presents an update on TWA candidate members from the BANYAN All-Sky Survey (BASS; \citet{2015ApJ...798...73G}; Section~\ref{sec:bass}), and reports new candidate members of TWA that were uncovered by three new surveys. Section~\ref{sec:buc} describes the BASS-Ultracool survey that targets members of young moving groups with spectral types later than $\sim$\,L5; Section~\ref{sec:lspm} describes a search for young low-mass, stellar moving group members from the SUPERBLINK-south proper motion catalog; and Section~\ref{sec:hip} describes a re-analysis of the Hipparcos \citep{1997AA...323L..49P} survey data with the BANYAN~II tool for bright members of young moving groups.

\subsection{BASS}\label{sec:bass}

The BANYAN All-Sky Survey (BASS) was initiated by \cite{2015ApJ...798...73G} to identify new $\sim$\,M5--L5 candidate members of young moving groups in the solar neighborhood, including TWA. The BASS survey is based on a cross-match of the \emph{Two Micron All-Sky Survey} (2MASS; \citealt{2006AJ....131.1163S}) with the AllWISE \citep{2010AJ....140.1868W,2014ApJ...783..122K} survey, which yielded proper motion measurements with a precision of $\sim$\,5--15\,\masyr. An initial set of 98\,970 potential nearby $>$\,M5 dwarfs was constructed from several selection criteria, e.g. a good 2MASS and AllWISE photometric quality, a $J-K_S$ color consistent with spectral types M5--L5, no optical $B2$-band detection in USNO--A2.0, a proper motion larger than 30\,\masyr\ and a sky position located further away than 15\textdegree\ from the Galactic plane. See \citealt{2015ApJ...798...73G} for the full details on selection criteria.

Moving group membership probabilities were assessed for all targets in this sample using the Bayesian Analysis for Nearby Young AssociatioNs~II tool\footnote{Available at \url{www.astro.umontreal.ca/\textasciitilde gagne/banyanII.php}} \citep{2014ApJ...783..121G,2013ApJ...762...88M}. This resulted in a sample of 983 candidate members with Bayesian probabilities larger than 10\% and estimated false-positive probabilities below 50\%. As BANYAN~II provides an estimate of the distance of each target assuming membership to the most probable moving group, further selection cuts based on the sequence of known young M5--L5 dwarfs in two color-magnitude diagrams (absolute $W1$ versus $J-K_S$ and absolute $W1$ versus $H-W2$) rejected 435 candidate members and divided the remaining candidates in two samples. The main BASS sample consists of 273 objects (including 54 TWA candidates) that are at least 1$\sigma$ redder than the field-age sequences in both color-magnitude diagrams, and the low-priority BASS (LP-BASS) sample consists of 275 objects (including 33 TWA candidates) that are redder than the field-age sequence by less than 1$\sigma$. The BASS survey has an expected completeness of $\sim$\,72\% for TWA, based on the selection criteria mentioned above. The estimated false-positive rates are estimated to be below 30\% and 80\% for the main BASS and LP-BASS samples.

While the BASS survey was in construction and the selection criteria were still being refined, a subset of 312 candidate members (including 27 TWA candidates) was collected from partial and/or more permissive selection criteria; these candidates are designated as the PRE-BASS sample.

An initial spectroscopic follow-up of 182 candidate members of moving groups (including 36 TWA candidates) that were selected from BASS (106 targets), LP-BASS (27 targets) and PRE-BASS (49 targets) was presented by \cite{2015ApJS..219...33G}. A fraction of 21\% of these targets were found to be contaminants, consisting of old low-mass stars and brown dwarfs, or reddened background objects. A total of 18 TWA candidates were found to have spectroscopic signatures of youth, and 4 more were found to have a spectral type earlier than M5, for which the NIR spectroscopic follow-up could not constrain their age. Twelve of the TWA candidates with spectroscopic signatures of youth originated from the main BASS catalog, and three more originated from LP-BASS. A detailed spectroscopic follow-up of two TWA candidates from the main BASS sample (2MASS~J12074836--3900043 and 2MASS~J12474428--3816464) was also presented by \cite{2014ApJ...785L..14G}. A total of 70 TWA candidate members from the LP-BASS, BASS and PRE-BASS samples have not yet been investigated for signs of low-gravity using spectroscopy, nor have they benefitted from radial velocity or parallax measurements. New radial velocity measurements, low-resolution near-infrared spectra and low-resolution optical spectra are presented in this paper for 9, 12 and 4 of these targets, respectively.

\subsection{BASS-Ultracool}\label{sec:buc}

The BASS-Ultracool survey was recently initiated to identify the late-type ($>$\,L5) members of young moving groups from a cross-match of 2MASS and AllWISE that were missed by the BASS survey due to the color and photometric quality cuts that were imposed on the 2MASS catalog entries (see Section~\ref{sec:bass} and \citealp{2015ApJ...798...73G,2015ApJS..219...33G}).

This ongoing survey is performed in multiple steps, the first of which consists of a re-analysis of 2MASS and AllWISE astrometry and photometry. This is done by cross-matching all entries of 2MASS and AllWISE, and using BANYAN~II, without using photometry as an input, to assess the possible membership of all sources with $W1-W2 > 0.2$. Including photometry in the BANYAN~II analysis is avoided in this particular case, because the NIR sequences of young $>$\,L5 brown dwarfs are poorly constrained at this time. The main survey results will be described in a future paper (J.~Gagn\'e et al., in preparation). This survey has already identified the first isolated T dwarf bona fide member of a young moving group: the AB~Doradus member SDSS~J111010.01+011613.1, \replaced{that has}{with} an estimated mass of 10--12\,\Mjup\ \citep{2015ApJ...808L..20G}.

A total of 8 candidate members of TWA were identified to date in this first step of the BASS-Ultracool survey, which are listed in Table~\ref{tab:buc}. \added{Only the candidate members with BANYAN~II probabilities above 60\% were considered. }\replaced{Their}{The} TWA Bayesian membership \replaced{probability}{probabilities} obtained from BANYAN~II are listed, along with\deleted{ their }statistical distances (see \citealt{2014ApJ...783..121G} for a detailed description of how the membership probabilities are obtained). Estimated spectral types are also presented, based on a comparison of 2MASS and AllWISE photometry at the most probable TWA statistical distance with the young sequences of \cite{2015ApJS..219...33G}.

BASS-UC~121 (2MASS~J11193254--1137466) has been independently discovered by \cite{2015AJ....150..182K} and confirmed as a likely candidate member of TWA with a radial velocity measurement by \cite{2016ApJ...821L..15K}. They note that it likely is the nearest known member of TWA, located at a kinematic distance of $28.9 \pm 3.6$\,pc. It has a very late spectral type (L7), and displays an unusually red NIR slope and weak \ion{K}{1} absorption doublets at 1.117 and 1.125\,$\mu$m, which are both signs of low gravity and thus youth (e.g., see \citealp{2006ApJ...639.1120K,2009AJ....137.3345C,2013ApJ...772...79A,2013ApJ...777L..20L}). At the age of TWA ($10 \pm 3$\,Myr), this object has an estimated mass of 5--7\,\Mjup.

BASS-UC~56 (2MASS~J11472421--2040204) has\deleted{ also }been independently discovered by \cite{2016ApJ...822L...1S} as a young L7 candidate member of TWA. The similarities between this object and 2MASS~J11193254--1137466 are remarkable, with their similar spectral types, distances (31--33\,pc) and estimated masses (6--13\,\Mjup). However, this object did not yet have a radial velocity confirmation of its TWA membership before the present work.

\added{The remaining six candidate members were not previously known, and will be discussed in Section~\ref{sec:data_an} in light of new data presented in this work.}

\begin{splitdeluxetable*}{lcccccBcccccccc}
\tabletypesize{\scriptsize}
\tablecolumns{14}
\tablecaption{Candidate Members of TWA from the BASS-Ultracool Survey \label{tab:buc}}
\tablehead{\colhead{BASS-UC} & \multicolumn{4}{c}{2MASS} & \colhead{} & \multicolumn{3}{c}{AllWISE} & \colhead{$\mu_\alpha\cos\delta$} & \colhead{$\mu_\delta$} & \colhead{Estimated} & \colhead{Bayesian} & \colhead{Stat.}\\
\cline{2-5}
\cline{7-9}
\colhead{Name} & \colhead{Designation} & \colhead{$J$} & \colhead{$H$} & \colhead{$K_S$} & \colhead{} & \colhead{Designation} & \colhead{$W1$} & \colhead{$W2$} & \colhead{(\masyr)} & \colhead{(\masyr)} & \colhead{SpT\tablenotemark{a}} & \colhead{Prob. (\%)\tablenotemark{b}} & \colhead{Distance (pc)}}
\startdata
BASS-UC~51 & 09553336--0208403 & $17.11 \pm 0.24$ & $> 15.56$ & $14.79 \pm 0.12$ & & 095533.26--020841.6 & $13.93 \pm 0.03$ & $13.39 \pm 0.03$ & $-123.9 \pm 14.8$ & $-105.6 \pm 15.5$ & L8 & 77.1 & $24.1 \pm 2.8$\\
BASS-UC~55 & 11063147--4201251 & $15.28 \pm 0.06$ & $14.44 \pm 0.07$ & $14.09 \pm 0.08$ & & 110631.37--420125.1 & $13.87 \pm 0.03$ & $13.65 \pm 0.03$ & $-101.6 \pm 7.9$ & $-0.3 \pm 7.7$ & L0 & 94.7 & $43.0_{-5.6}^{+6.0}$\\
BASS-UC~121 & 11193254--1137466 & $> 17.29$ & $15.606\pm 0.14$ & $14.62 \pm 0.11$ & & 111932.43--113747.7 & $13.55 \pm 0.03$ & $12.88 \pm 0.03$ & $-148.5 \pm 15.4$ & $-98.1 \pm 14.7$ & L8 & 65.1 & $25.3 \pm 2.8$\\
BASS-UC~56 & 11472421--2040204 & $> 17.51$ & $15.76\pm 0.11$ & $14.87 \pm 0.11$ & & 114724.10--204021.3 & $13.72 \pm 0.03$ & $13.09 \pm 0.03$ & $-121.6 \pm 11.1$ & $-74.2 \pm 11.9$ & L8 & 84.5 & $31.3 \pm 3.6$\\
BASS-UC~57 & 12021801--3110348 & $14.91 \pm 0.04$ & $14.20 \pm 0.03$ & $13.84 \pm 0.05$ & & 120217.92--311035.1 & $13.62 \pm 0.03$ & $13.41 \pm 0.03$ & $-98.6 \pm 7.6$ & $-27 \pm 7$ & M9.5 & 96.7 & $44.6_{-4.8}^{+5.2}$\\
BASS-UC~58 & 12162481--2742007 & $14.84 \pm 0.04$ & $14.25 \pm 0.04$ & $13.85 \pm 0.05$ & & 121624.74--274201.1 & $13.66 \pm 0.03$ & $13.43 \pm 0.03$ & $-77.5 \pm 6.7$ & $-31.2 \pm 6.5$ & M9 & 94.8 & $49.8_{-5.2}^{+5.6}$\\
BASS-UC~59 & 12194846--3232059 & $15.66 \pm 0.06$ & $14.97 \pm 0.06$ & $14.59 \pm 0.08$ & & 121948.39--323206.1 & $14.38 \pm 0.03$ & $14.16 \pm 0.04$ & $-70.2 \pm 7.7$ & $-15.9 \pm 8.3$ & L0 & 90.8 & $56.2 \pm 6.4$\\
BASS-UC~60 & 12454194--3903106 & $15.46 \pm 0.05$ & $14.84 \pm 0.05$ & $14.41 \pm 0.07$ & & 124541.87--390310.9 & $14.22 \pm 0.03$ & $13.95 \pm 0.04$ & $-73.1 \pm 7.3$ & $-22.3 \pm 7.4$ & M9.5 & 95.3 & $58.2_{-6.8}^{+7.2}$\\
\enddata
\tablenotetext{a}{Photometric spectral types estimated from 2MASS and AllWISE photometry and the BANYAN~II statistical distances (see \citealp{2014ApJ...783..121G} and \citealp{2015ApJ...798...73G} for a detailed explanation).}
\tablenotetext{b}{Based on kinematics alone.}
\tablecomments{See Section~\ref{sec:buc} for more details.}
\end{splitdeluxetable*}

\subsection{SUPERBLINK-south}\label{sec:lspm}

A search for new members of nearby, young associations was performed using the SUPERBLINK proper motion survey of the Southern hemisphere (SUPERBLINK-south; S.~L\'epine et al. in preparation). The SUPERBLINK-south catalog includes stars with total proper motions above 40\,\masyr\ in the region $-33 <$\,Decl.\,$< 0$, to an approximate magnitude limit of $V < 20$\added{, and provides proper motions with a typical precision of $\sim$\,11\,\masyr}. All stars are matched against the 2MASS all-sky point source catalog, and include optical blue (IIIaJ) and red (IIIaF) photographic magnitude estimates from the USNO-B1.0 catalog.

The \added{403\,085} SUPERBLINK-south catalog entries were supplemented with $I$-band photometric data from the AAVSO Photometric All-Sky Survey (APASS-DR9; \citealt{2016yCat.2336....0H}). \added{Potential young and nearby low-mass stars were selected by applying the following criteria:~(1) A NUV detection in the Galaxy Evolution Explorer (GALEX) survey \citep{2005ApJ...619L...1M} must be present; (2) The declination must be located above $-40$\textdegree and the right ascension must be located between 10\,h and 14\,h; (3) All 2MASS to WISE colors must be consistent with a spectral type of M0 or later \citep{2013ApJS..208....9P}; (4) The apparent 2MASS $K_S$ band magnitude must be fainter than 11.5; and (5) The AAVSO $I$-band magnitude must have error bars of 0.2\,mag or smaller. These criteria yielded a set of 714 catalog entries.}

The BANYAN~I tool \citep{2013ApJ...762...88M} was used to identify candidate members, using $I_C$ and 2MASS $J$-band magnitudes, proper motions and sky positions as input observables.\added{ This analysis yielded fifteen TWA candidate members with a probability of 90\% or more, and for which the absolute NUV magnitude is consistent with a young low-mass star (see Figure~1 of \citealt{2011ApJ...727...62R}).}

\added{The known member TWA~5~A (M2\,IVe) was recovered, as well as the BASS candidate member 2MASS~J10585054--2346206 and the LP-BASS candidate member 2MASS~J10542303--1507082. One further candidate member (2MASS~J12000160--1731308) is presented in this paper. Three of the remaining 11 candidates were observed and rejected as members, and the remaining eight have not been observed yet; these 11 objects will be the subject of a future publication to identify young moving group members from SUPERBLINK-South.}

\added{New high-resolution optical spectra were obtained for the three SUPERBLINK-South targets described above, which are detailed in Section~\ref{sec:obs_espadons}. Their TWA membership in light of these new data is discussed in Section~\ref{sec:rv2}. The full SUPERBLINK-south catalog will be presented in S.~L\'epine et al. (in preparation).}

\subsection{A Re-Analysis of Hipparcos}\label{sec:hip}

\begin{figure}
	\centering
	\includegraphics[width=0.465\textwidth]{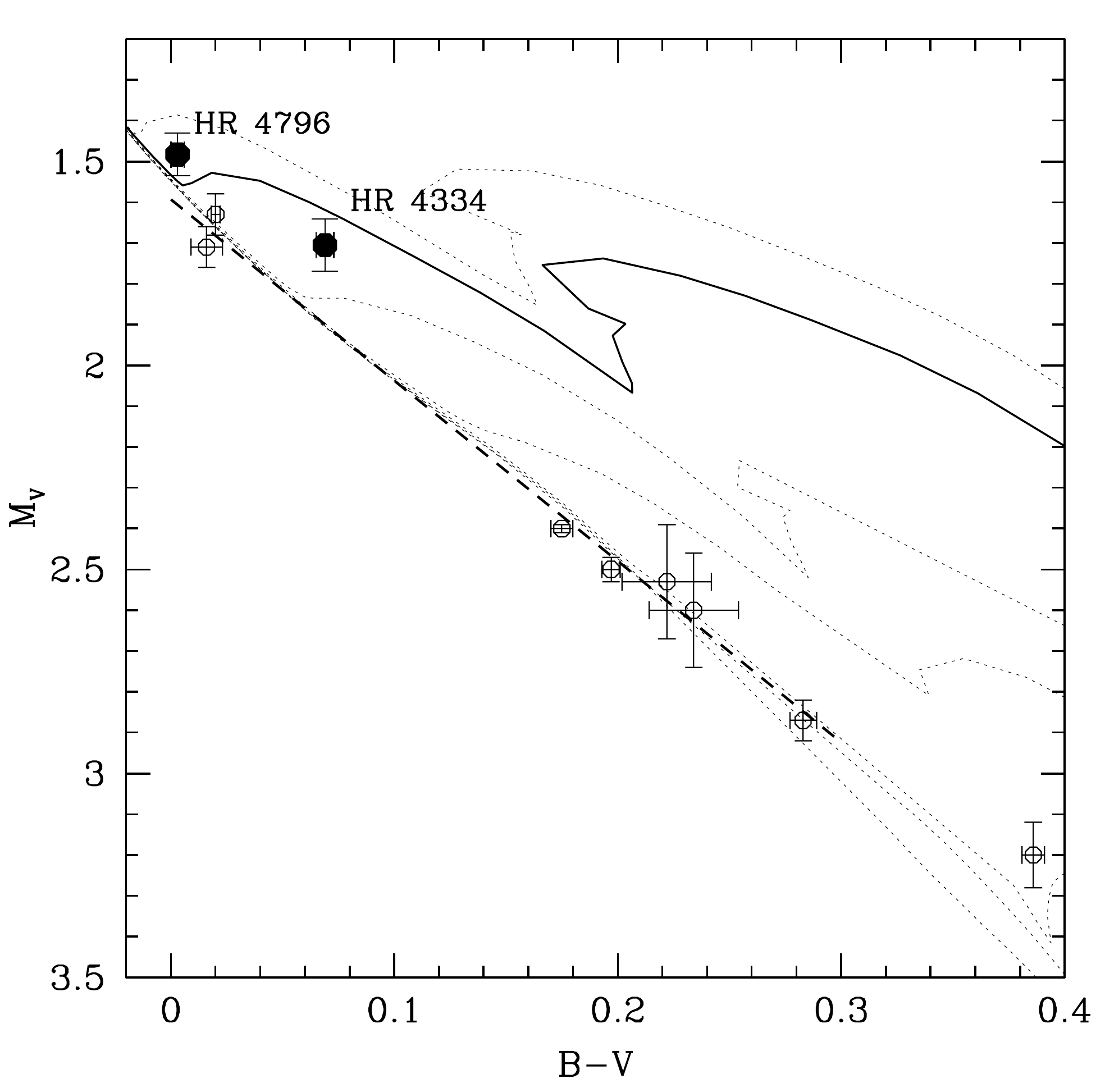}
	\caption{Color-magnitude diagram for young A-type stars. The filled circles are HIP~54477 (HR~4334) and TWA~11 (HR~4796). The open circles are six A-type members of the $\beta$~Pictoris Moving Group from \citeauthor{2014MNRAS.445.2169M} (\citeyear{2014MNRAS.445.2169M}; age 23\,$\pm$\,3\,Myr). Isochrones from the MIST/MESA tracks of \cite{2016ApJ...823..102C} are shown, ranging in age from 7--25\,Myr ($\log\left({\rm age}/{\rm yr}\right) = 6.85, 6.9, 7.0, 7.1, 7.2, 7.3, 7.4$). The $\log\left({\rm age}/{\rm yr}\right) = 6.9$ isochrone (8\,Myr) is drawn as a solid thick line. The thick dashed line is an empirical fit to the $\beta$~Pictoris moving group stars ($M_V = 1.594 + 4.431\left(B-V\right)$; between $0.0 < B-V < 0.3$), and demonstrates that the \cite{2016ApJ...823..102C} solar metallicity isochrones do an excellent job fitting the $\beta$~Pictoris zero age main sequence. Both HR~4796 and HR~4334 appear to be slightly pre-main sequence, as expected for $\sim$\,7--10\,Myr-old stars, \replaced{and we conclude that}{thus} HR~4334 has a color-magnitude position that is consistent with that expected for an A2\,V TWA member. See Section~\ref{sec:hip} for more information.}
	\label{fig:hip}
\end{figure}

A re-analysis of the Hipparcos catalog was performed \deleted{in order }to identify any missing bright members of TWA. In a first step, the proper motions and distances of all entries were used to identify candidate members with a TWA BANYAN~II probability above 90\%. The resulting 17 objects were then parsed to gather existing radial velocity measurements \deleted{in order }to refine the BANYAN~II membership probability. The candidate members that were identified in this way are listed in Table~\ref{tab:hip}. Four of these are known members of TWA (TWA~1, TWA~4, TWA~9 and TWA~11). Nine remaining objects have kinematics that are possibly consistent with TWA, although not all of them have \replaced{RV}{radial velocity} measurements. In the remainder of this section, we discuss each of these\deleted{ seven }targets individually, \deleted{in order }to determine whether their global characteristics are consistent with membership to TWA.

\textbf{HIP~50032} is a moderately active \citep[$\log R^\prime_{HK} = -4.38$][]{2010ApJ...725..875I} star with a low lithium abundance (EW($\lambda$6707) = 30\,m\AA) compared to pre-main sequence K-type stars \citep{2003AN....324..543T}. Its absolute magnitude ($M_V = 5.93$) calculated using its Hipparcos parallax \citep{2007A&A...474..653V} is consistent with a main sequence star. Its chromospheric activity, coronal X-ray emission ($\log L_X/L_{\rm bol} \simeq -4.5$), lithium abundance, and HR diagram position are all consistent with it being a Hyades-age interloper unrelated to TWA.

\textbf{HIP~52776} is a K4.5\,V(k) star of modest chromospheric activity \citep[$\log R^\prime_{HK} = -4.45$;][]{2006AJ....132..161G}. Its \replaced{lacking}{lack of} detectable lithium \citep[$A({\rm Li}) < 0.31$;][]{2015A&A...576A..69D} suggests an age older than \added{that of }the Pleiades ($\sim$\,120\,Myr; \citealt{1998ApJ...499L.199S}). It is a wide (234$''$) companion \replaced{with}{to} HIP~52787, a K1\,V(k) star of similarly modest chromospheric activity \citep[$\log R^\prime_{HK} = -4.42$;][]{2006AJ....132..161G} and detectable Li (EW($\lambda$6707) = 110\,m\AA; \citealt{2006A&A...460..695T}), similar to that of $\sim$\,0.2--0.3\,Gyr-old clusters \replaced{like}{such as} M7, M34, and M35. \replaced{HIP~52787's rotation period of 6.579\,days \citep{2012AcA....62...67K}}{The rotation period of HIP~52787 (6.579\,days; \citealt{2012AcA....62...67K})} is consistent with a \replaced{gyrochronology}{gyrochronological} age of $\sim$\,0.3\,Gyr using the calibration of \cite{2008ApJ...687.1264M}, and its $\log R^\prime_{HK}$ is consistent with an age of $\sim$\,0.4\,Gyr \citep{2008ApJ...687.1264M}. The absolute $V$ magnitudes \replaced{for}{of} HIP~52776 ($M_V = 7.34$) and HIP~52787 ($M_V = 5.70$)\added{,} calculated using Hipparcos photometry and astrometry \citep{1997ESASP1200.....E,2007A&A...474..653V}\added{,} place both stars squarely on the main sequence. We conclude that the HIP~52776 + HIP~52787 pair are young ($\sim$\,0.3\,Gyr) dwarf interlopers unrelated to TWA.

\textbf{HIP~54095} is a F2\,V star according to \cite{1982mcts.book.....H}, which translates to F3.5\,V on the modern MK system \citep{2016MNRAS.461..794P}. \cite{2015ApJ...804..146D} estimate an isochronal age in the 1$\sigma$ range 0.8--2.7\,Gyr. It is detected in the {\it ROSAT} All-Sky Survey \citep{2016A&A...588A.103B}, however its X-ray luminosity is \replaced{unimpressive}{not particularly high} for its bolometric luminosity ($\log \left(L_X/L_{bol}\right) = -5.4$, $\log L_X = 28.94$), and its X-ray emission is rather soft (HR1 $= -1.0$). There are multiple photometric metallicity estimates consistent with the star being slightly metal poor, with [Fe/H] in the range $-0.28$ \citep{2003ApJ...595.1206S} to $-0.15$ \citep{1995BICDS..47...13M}. Most recently, \cite{2011A&A...530A.138C} estimate its metallicity \replaced{to be}{at} [Fe/H] $= -0.19$ and [M/H] = $-0.13$, which would be at odds with that of young ($< 10^8$\,yr-old) stellar associations in the solar neighborhood. Its combination of effective temperature \citep[$T_{\rm eff} = 6802$\,K;][]{2011A&A...530A.138C} and luminosity ($\log L_{\odot} = 0.76$) would be consistent with a somewhat older pre-main sequence star ($\sim$\,14\,Myr; see Fig.\,6 of \citealt{2012ApJ...746..154P} for comparison with F-type stars in the Sco-Cen subgroups), however the star seems to be somewhat under-luminous ($\sim$\,0.2\,dex) if it were\deleted{ really }10\,Myr-old.\deleted{ At this point, }There is \added{currently }no corroborating evidence to suggest that this star is coeval with TWA. It appears to be a slightly metal-poor main sequence early-type F dwarf.

\textbf{HIP~54477} (HR~4334) is a rapidly rotating A2\,Vn star \replaced{within}{with} $v \sin i \simeq 230$\,\kms\ \citep{1995ApJS...99..135A}. The star was previously proposed as a candidate TWA member by \citet[][\S2.5]{2005ApJ...634.1385M}, but \added{it was }not assigned a TWA number\footnote{HIP~54477 (as HD~96819) is accidentally listed with the age of the $\beta$~Pictoris moving group ($\sim$\,23\,Myr) in Table~1 of \cite{2015MNRAS.453.2378M} rather than the TWA group ($\sim$\,10\,Myr).}. In Figure~\ref{fig:hip},\deleted{we plot }the color-magnitude position of HR~4334\added{ is} compared to HR~4796 (known TWA member; TWA~11) and six A-type members of the $\sim$\,23\,Myr-old $\beta$ Pictoris moving group from \citep[][their Table~3]{2014MNRAS.445.2169M}. HR~4334 is consistent with being a 9$\pm$1\,Myr-old pre-main sequence star just above the zero-age main sequence as defined empirically using the A-type $\beta$~Pictoris members, and theoretically using the solar-composition MIST isochrones of \cite{2016ApJ...823..102C}. \replaced{We find that HR~4334's color-magnitude position is}{The color-magnitude position of HR~4334 is located} where one would predict an A2-type TWA member to lie. Since the kinematic properties of HIP~54477 are also a\deleted{ very }good match to those of TWA (its BANYAN~II\deleted{ Bayesian }membership probability is 98.7\%\added{ when distance and radial velocity are included}), we suggest that it is a new bona fide member\deleted{ member} of TWA.

\textbf{HIP~54690} (CD--28~8704) is classified as a K5\,V star by \cite{1972AJ.....77..486U}. Although its kinematics match those of TWA members, its position on a $V-J$ ($1.96 \pm 0.03$; \citealp{1992A&A...258..217E,2006AJ....131.1163S}) versus $M_V$ ($7.1 \pm 0.2$) color-magnitude diagram is fully consistent with that of a main sequence star (e.g., see \citealt[][their Figure~10]{2015MNRAS.454..593B}). A TWA-aged star with this $V-J$ color would be expected to be at least 1\,mag brighter. We conclude that HIP~54690 is an older interloper unrelated to TWA.

\textbf{HIP~55516} has been classified as a G8\,IV/V \citep{1978mcts.book.....H} or G9\,V \citep{1982PASP...94..304L} star, and its absolute magnitude ($M_V = 5.2 \pm 0.2$) calculated using its revised Hipparcos parallax of $\varpi = 15.69 \pm 1.09$\,mas \citep{2007A&A...474..653V} confirms its dwarf status. For its color \citep[$B-V = 0.76 \pm 0.02$;][]{1997ESASP1200.....E}, a $\sim$\,10\,Myr TWA member should have $M_V \simeq 4.3$, hence the star is a magnitude too faint to be a pre-main sequence TWA member. The star has not been detected in any X-ray surveys, nor has it been flagged as a variable star. Its predicted kinematic distance ($\sim$\,42\,pc) based on its proper motion and the TWA velocity vector does not match its trigonometric parallax distance ($\sim$\,64 pc), despite its BANYAN~II probability being high (99.8\%). This apparent discrepancy between the two methods arises from the fact that HIP~55516 \replaced{is devious}{deviates} from the mean $UVW$ position of TWA members (by $9 \pm 3$\,\kms) in the direction where the BANYAN~II model ellipsoid\added{ of TWA} is most elongated. The isochrone mismatch precludes HIP~55516 from being as young as TWA, and we thus conclude that it is likely an older interloper star unrelated to TWA.

\textbf{HIP~58290} (HD~103840) is a G3\,V star \citep{1978mcts.book.....H}, which on the modern MK system translates to approximately G1.5\,V \citep{2016MNRAS.461..794P}. It has no X-ray detection in any reported survey, \replaced{shows}{displays a} low projected rotation ($v \sin i = 2.7$\,\kms; \citealt{2004A&A...418..989N}) and \replaced{has}{a} dwarf surface gravity ($\log g = 4.58$; \citealt{2011A&A...530A.138C}). The star has a low metallicity ([Fe/H] = $-0.33$), which is corroborated by its intrinsic faintness ($M_V = 5.22 \pm 0.07$), \replaced{which situates}{situating} it $\sim$\,0.6 mag below the main sequence of \cite{2004AJ....128.1273W,2005AJ....129.1776W} for its $B-V$ color ($0.61 \pm 0.01$). We conclude that HIP~58290 is an old, inactive, metal-poor interloper.

\textbf{HIP~59077} is classified \added{as a }G8/K0\,V\added{ star} by \citep{1982mcts.book.....H}, and has a wide, faint common proper motion companion 56$''$ away with colors consistent with\deleted{ being }a DA white dwarf \citep{2013AJ....146...76H}. HIP~59077 is \replaced{slow}{slowly} rotating \citep[$v \sin i = 1.8$\,\kms][]{2005ESASP.560..571G} with\added{ an} absolute magnitude $M_V = 5.8 \pm 0.1$, placing it near the main sequence. These indicators suggest that HIP~59077 is an older interloper unrelated to the $\sim$\,10 Myr-old TWA.

\textbf{HIP~59257} is a F6\,V star \citep{1978mcts.book.....H} that is listed by \cite{2000MNRAS.313...43H} and \cite{2011MNRAS.416.3108R} as a member of the $\sim$\,16\,Myr-old \citep{2002AJ....124.1670M} Lower Centaurus Crux (LCC) OB association. Its trigonometric distance of $84.9 \pm 9.7$\,pc would make it the furthest member of TWA, and is more consistent with a membership to LCC. It is likely that this star is thus an interloper from LCC that obtains a high Bayesian membership probability to TWA\deleted{ simply }because no model of LCC is included in BANYAN~II.

A detailed consideration of the properties of all new Hipparcos candidate members that were uncovered in this section revealed them to be unrelated interlopers, with the exception of HIP~54477. \replaced{This star had been previously reported as a candidate member by \cite{2005ApJ...634.1385M}, and it is thus concluded that it is a new A2-type bona fide member of TWA.}{A new assessment of its age based on the solar-composition MIST isochrones confirms that it is a bona fide member of TWA.}\added{ After the completion of this survey for new members using Hipparcos data, the \emph{Gaia} Data Release 1 \citep{2016arXiv160904303L} provided more precise trigonometric distances and proper motions for a number of targets detailed in this section. Although these new data did not affect the conclusions presented in this section, they are taken into account for refining membership probabilities in Section~\ref{sec:candcomp} where the list of TWA candidates and members is compiled.}

\begin{splitdeluxetable*}{llccccccBcccccccc}
\tablecolumns{16}
\tabletypesize{\scriptsize}
\tablecaption{Candidate Members of TWA from Hipparcos \label{tab:hip}}
\tablehead{\colhead{HIP} & \colhead{Other} & \colhead{Spectral} & \colhead{RA} & \colhead{DEC} & \colhead{$\mu_\alpha\cos\delta$} & \colhead{$\mu_\delta$} & \colhead{Trig.} & \colhead{HIP} & \colhead{RV} & \colhead{} & \colhead{Bayesian} & \colhead{$B-V$} & \colhead{$M_V$} & \colhead{Consistent} & \colhead{Member-}\\
\colhead{Number} & \colhead{Names} & \colhead{Type} & \colhead{(hh:mm:ss.sss)} & \colhead{(dd:mm:ss.ss)} & \colhead{(\masyr)} & \colhead{(\masyr)} & \colhead{Dist. (pc)} & \colhead{Number} & \colhead{(\kms)} & \colhead{Ref.} & \colhead{Prob. (\%)\tablenotemark{a}} & \colhead{(mag)} & \colhead{(mag)} & \colhead{Age?\tablenotemark{b}} & \colhead{ship\tablenotemark{c}}}
\startdata
50032 & HD~88656 & K2\,V & 10:12:52.77748 & -28:30:48.1785 & $-49.84  \pm 0.75$ & $-24.65 \pm 0.86$ & $42.7 \pm 1.7$ & 50032 & $8.2 \pm 0.2$ & (1) & 96.7 & $0.88 \pm 0.02$ & $5.93 \pm 0.09$ & N & R\\
52776 & BD--21~3153 & K4.5\,V(k) & 10:47:25.38730 & -22:17:12.1792 & $-126.1 \pm 1.4$ & $-30.3 \pm 1.2$ & $32.6 \pm 1.5$ & 52776 & $\cdots$ & $\cdots$ & 99.9 & $1.16 \pm 0.02$ & $7.3 \pm 0.1$ & N & R\\
52787 & HD~93528 & K0\,V & 10:47:31.15457 & -22:20:52.9160 & $-124.0 \pm 0.9$ & $-28.2 \pm 0.8$ & $34.5 \pm 1.0$ & 52787 & $23.4 \pm 1.7$ & (2) & 0.0 & $0.83 \pm 0.02$ & $5.70 \pm 0.06$ & $\cdots$ & R\\
53911 & TWA~1 & K6\,Ve & 11:01:51.90671 & -34:42:17.0323 & $-66.2 \pm 1.9$ & $-13.9 \pm 1.5$ & $53.7 \pm 6.2$ & 53911 & $13.4 \pm 0.8$ & (3) & $>$\,99.9 & $0.7 \pm 0.1$ & $7.3 \pm 0.3$ & $\cdots$ & K\\
54095 & HD~96033 & F3.5\,V & 11:04:07.38096 & -40:18:30.9042 & $-51.2 \pm 0.6$ & $-9.1 \pm 0.4$ & $81.8 \pm 4.2$ & 54095 & $\cdots$ & $\cdots$ & 99.8 & $0.369 \pm 0.009$ & $2.9 \pm 0.1$ & N & R\\
54477 & HR~4334 & A2\,Vn & 11:08:43.99954 & -28:04:50.4127 & $-72.8 \pm 0.4$ & $-22.2 \pm 0.5$ & $55.7 \pm 1.6$ & 54477 & $16 \pm 5$ & (4) & 98.7 & $0.069 \pm 0.004$ & $1.70 \pm 0.06$ & Y & BF\\
54690 & CD--28~8704 & K5\,V & 11:11:47.10817 & -29:27:04.1717 & $-101.2 \pm 1.7$ & $-37.8 \pm 1.5$ & $47.8 \pm 4.7$ & 54690 & $\cdots$ & $\cdots$ & 99.9 & $1.10 \pm 0.02$ & $7.1 \pm 0.2$ & N & R\\
55505 & TWA~4; HD~98800 & K4\,V & 11:22:05.28975 & -24:46:39.7571 & $-85.4 \pm 1.7$ & $-33.1 \pm 2.1$ & $44.9 \pm 4.7$ & 55505 & $9 \pm 1$ & (3) & $>$\,99.9 & $1.15 \pm 0.04$ & $5.6 \pm 0.2$ & $\cdots$ & K\\
55516 & HD~98870 & G8\,IV/V & 11:22:14.75215 & -48:56:43.3480 & $-88.8 \pm 0.8$ & $-15.5 \pm 0.8$ & $63.7 \pm 4.4$ & 55516 & $14.1 \pm 0.3$ & (5) & 99.8 & $0.76 \pm 0.02$ & $5.3 \pm 0.2$ & N & R\\
57589 & TWA~9 & K7\,IVe + M1 & 11:48:24.22320 & -37:28:49.1537 & $-52.4 \pm 2.4$ & $-22.9 \pm 1.7$ & $46.8 \pm 5.4$ & 57589 & $9.5 \pm 0.4$ & (3) & 99.9 & $1.6 \pm 0.4$ & $7.8 \pm 0.3$ & $\cdots$ & K\\
58290 & HD~103840 & G1.5\,V & 11:57:15.65407 & -48:44:36.6582 & $-106.7 \pm 0.6$ & $-18.5 \pm 0.6$ & $38.6 \pm 1.3$ & 58290 & $7.9 \pm 0.3$ & (5) & 97.1 & $0.61 \pm 0.01$ & $5.22 \pm 0.07$ & N & R\\
58363 & HD~103933 & F5\,V & 11:58:03.66897 & -31:39:03.3795 & $-92.6 \pm 0.6$ & $-37.1 \pm 0.3$ & $57.2 \pm 1.6$ & 58363 & $15.2 \pm 0.3$ & (4) & 8.2 & $0.478 \pm 0.008$ & $3.30 \pm 0.06$ & $\cdots$ & R\\
59077 & HD~105227 & G8/K0\,V & 12:06:54.69914 & -38:06:23.2107 & $-82.8 \pm 1.1$ & $-45.5 \pm 0.9$ & $51.4 \pm 3.3$ & 59077 & $10.1 \pm 0.4$ & (5) & $>$\,99.9 & $0.77 \pm 0.03$ & $5.8 \pm 0.1$ & N & R\\
59257 & HD~105577 & F6\,V & 12:09:20.45154 & -42:50:18.4649 & $-47.0 \pm 0.5$ & $-10.0 \pm 0.4$ & $87.3 \pm 4.6$ & 59257 & $3.8 \pm 0.3$ & (5) & 99.2 & $0.532 \pm 0.003$ & $3.2 \pm 0.1$ & N & R\\
60239 & HD~107434 & F6\,V & 12:21:09.57045 & -38:18:09.8071 & $-55.6 \pm 0.7$ & $-20.1 \pm 0.5$ & $67.2 \pm 3.2$ & 60239 & $-10.2 \pm 0.3$ & (5) & 0.0 & $0.54 \pm 0.02$ & $4.0 \pm 0.1$ & $\cdots$ & R\\
61327 & HD~109296 & F8 & 12:33:55.44130 & -48:36:05.2101 & $-63.3 \pm 1.0$ & $-24.1 \pm 1.2$ & $84.9 \pm 9.7$ & 61327 & $-8.0 \pm 1.1$ & (6) & 0.0 & $0.61 \pm 0.03$ & $4.8 \pm 0.3$ & $\cdots$ & R\\
61498 & TWA~11; HR~4796 & A0\,V & 12:36:01.03100 & -39:52:10.2270 & $-56.7 \pm 0.3$ & $-25.0 \pm 0.2$ & $72.8 \pm 1.8$ & 61498 & $7 \pm 1$ & (5) & $>$\,99.9 & $0.003 \pm 0.003$ & $1.47 \pm 0.05$ & $\cdots$ & K\\
\enddata
\tablenotemark{a}{Bayesian probability including all available measurements in the literature (position, proper motion, trigonometric distance and radial velocity when available).}
\tablenotetext{b}{This Yes/No flag indicates whether the general properties of this object are consistent with the age of TWA ($\sim$\,10\,Myr). Only the 9 stars with a high Bayesian membership probability that were not previously known as TWA members were investigated.}
\tablenotetext{c}{K: Known member, R: Rejected, BF: New bona fide member.}
\tablecomments{See Section~\ref{sec:hip} for more details.}
\tablerefs{(1)~\citealt{2011AAS...21743412C}, (2)~\citealt{2011AA...531A...8J}, (3)~\citealt{2006AA...460..695T}, (4)~\citealt{2007AN....328..889K}, (5)~\citealt{2006AstL...32..759G}, (6)~\citealt{2005MNRAS.357..497B}.}
\end{splitdeluxetable*}

\section{OBSERVATIONS}\label{sec:obs}

Near-infrared and optical spectra of various resolutions were obtained at 5 facilities\deleted{, in order} to measure the spectral types, \replaced{spectral}{spectroscopic} indications of low-gravity (youth) and/or the radial velocity of TWA candidate members. The targets were selected for follow-up through a variety of \replaced{homogeneous}{heterogeneous} \replaced{programs}{surveys presented in Section~\ref{sec:newmembers}} and were thus not selected in an optimal way from the final list of TWA candidate members compiled in this work.\added{ Most targets were selected from the BASS, LP-BASS and PRE-BASS samples presented in Section~\ref{sec:bass}, and the remaining targets were selected from the BASS-Ultracool and SUPERBLINK-south surveys detailed in Sections~\ref{sec:buc} and \ref{sec:lspm}.} The observations are described in this section, and a detailed observing log is displayed in Table~\ref{tab:obslog}. The new spectra detailed in this section are presented in Appendix~\ref{an:spt}.

\subsection{FIRE at Magellan/Baade}\label{sec:obs_fire}

Three low- and ten mid-resolution near-infrared spectra were obtained for 9 TWA candidate members from 2013 December to 2016 February with the Folded-port InfraRed Echellette (FIRE; \citealp{2008SPIE.7014E..0US,2013PASP..125..270S}) at the Magellan/Baade telescope. The 0\farcs6 slit was used in all cases, either in the high-throughput prism mode (resolving power of $R \sim 450$) or in the high-resolution echelle mode ($R \sim 6\,000$), both providing a wavelength coverage of 0.8--2.45\,$\mu$m. The data were obtained in an ABBA nodding pattern along the slit with two to six exposures of 500 to 900\,s (echelle), or 8 exposures of 40 to 80\,s (prism). This yielded signal-to-noise ratios per pixel of 40 or more, except in one observation at high airmass and with \replaced{quickly}{rapidly} degrading seeing (2MASS~11472421--2040204 on 2016 January 23). A0-type spectral standards were obtained immediately after each science target at a similar airmass to ensure a proper telluric absorption correction.

Several high- and low-voltage internal flat fields were obtained at the beginning of every night, as well as external dome flats, which are used to build the slit illumination function. Several NeAr wavelength calibration lamps were obtained at the beginning of the night for the prism mode, whereas a single ThAr calibration lamp exposure was obtained after every target and telluric standard in\deleted{ the }
echelle mode. The data were reduced with the Interactive Data Language (IDL) Firehose~v2.0 package (\citealp{2009PASP..121.1409B,zenodofirehose}\footnote{Available at \url{https://github.com/jgagneastro/FireHose\_v2/tree/v2.0}}; see \citealt{2015ApJS..219...33G} for more details on this reduction package).

\subsection{GMOS at GEMINI-S and GEMINI-N}\label{sec:gmos}

Optical spectra were obtained with GMOS \citep{2004PASP..116..425H} in queue mode at the Gemini-North and Gemini-South telescopes for 21 TWA candidate members from 2012 December to 2014 April. The R400 grating with a central wavelength setting of 800\,nm were used with the OG515 filter and the 1\farcs0 or 0\farcs75 slits to obtain resolving powers of $R \sim 950$ or $R \sim 1\,250$ covering $\sim$\,5\,900--10\,100\,\AA. Four exposures of 45--1\,500\,s were obtained to achieve signal-to-noise ratios of $\sim$\,30--250 per pixel on the science targets. The two first science exposures were obtained with a central wavelength of 800\,nm, followed by a single flat-field exposure under the same \replaced{setting, and an}{settings. An} additional flat exposure and the remaining two science exposures were obtained with a central wavelength of 805\,nm. This \replaced{ensured}{ensures} that the gaps between individual detectors do not result in gaps in the wavelength coverage of the final science spectra. Two CuAr lamp exposures were typically obtained at the end of each night, one for each central wavelength setting. The standard white dwarfs or bright stars G~191--2~B, LTT~2415, LTT~4816, CD--32~9927 were observed once per semester with each setting, as part of the regular Gemini calibrations and\deleted{ in order} to correct for instrumental response. \replaced{Only}{In most cases, only} the central spectrum region \replaced{were typically}{was} read on the detector to reduce readout time.

A single 10\,s $i$- or $r$-band acquisition image was\deleted{ typically }obtained\added{ before each target observation}, followed by a single 20\,s acquisition exposure with the slit on\replaced{, in order}{that was used} to verify whether the target \replaced{was}{is} a visual binary. In such cases, the position angle was adjusted\deleted{ such as }to place both components within the slit.

The data were reduced with a custom IDL pipeline, which applies bias and flat field corrections, straightens the spectral traces, flags and ignores bad pixels, extracts the spectra on each detector separately using a 1D Moffat profile \citep{1969AA.....3..455M}, performs wavelength calibration using the CuAr lamps and \replaced{stitches}{combines} the spectra from individual detectors\deleted{ together}. Individual exposures are then combined\deleted{ together }and corrected for instrumental response using the Gemini spectral standards. Six visual binaries were flagged from a visual inspection of the acquisition images, and extracted with a special algorithm that fits two Moffat profiles at every spectral position, \replaced{which yields}{yielding} two individual spectra for the respective binary components.\deleted{ The resulting GMOS spectra are displayed in Figure~\ref{fig:gmos_seq}.}

The spectra that were reduced with the standards CD--32~9927 and G~191--2~B initially suffered from slope systematics at $\gtrsim$\,850\,nm. A correction to these systematics was developed for each standard star: this was done by calculating the ratio of each observed spectrum \replaced{with}{to} that of a template M-type dwarf of the same spectral type and surface gravity that was re-sampled at the same resolution as the data. A linear polynomial was fit to the median of all slope corrections\added{ as a function of wavelength} for the science targets that were observed with a given standard star, which was subsequently divided to the \replaced{data}{science spectra}.

\subsection{SPEX at IRTF}\label{sec:obs_spex}

Low-resolution NIR spectra with a wavelength coverage of 0.8--2.45\,$\mu$m\, were obtained for 2 TWA candidate members with SpeX \citep{2003PASP..115..362R} at the IRTF telescope on 2015 December 6. The 0\farcs5 slit was used with the prism mode, yielding a resolving power of $R \sim 120$. Four to eight exposures of 180\,s were obtained in an ABBA nodding pattern along the slit which yielded signal-to-noise ratios of $\sim$\,20 per pixel. This was followed by a standard SpeX calibration sequence consisting of 5 flat field exposures and 2 arc lamp exposures. The A0-type standard stars HD~79752 and HD~91398 were observed immediately after the science targets and at a similar airmass to correct for telluric absorption and instrument response. The data were reduced using the IDL SpeXTool~v4.0~beta package\footnote{Available at \url{http://irtfweb.ifa.hawaii.edu/\textasciitilde spex/}} \citep{2003PASP..115..389V,2004PASP..116..362C}.

\subsection{Flamingos-2 at GEMINI-S}\label{sec:obs_f2}

Low-resolution NIR spectra were obtained with Flamingos-2 \citep{2004SPIE.5492.1196E} at Gemini-South in queue mode for 2 TWA candidate members, in 2015 April. The JH grism was used with the 0\farcs72 slit, providing a resolving power of $R \sim 500$ across 0.9--1.73\,$\mu$m. Sixteen to 36 exposures of 120\,s were obtained, which yielded signal-to-noise ratios above 100 per pixel -- these large numbers of exposures are required\deleted{ in order} to correct for Flamingos-2 systematics such as fringing, which can otherwise artificially affect the spectral morphology. The A0-type spectral standards HD~92699 and HD~105764 were observed immediately after science target exposures and at a similar airmass to provide a correction for telluric absorption. Standard Flamingos-2 calibrations (darks, flat fields and Ar wavelength calibration lamps) were obtained at the end of every night.

\added{The data were reduced using the Red Flamingos IDL pipeline\footnote{Available at \url{https://github.com/jgagneastro/red\_flamingos}} (see \citealt{2015ApJS..219...33G} for details). This pipeline performs spectral extraction, applies standard calibrations and corrects instrumental fringing.}

\begin{deluxetable}{lccccccc}
\tabletypesize{\scriptsize}
\tablecolumns{8}
\setlength{\tabcolsep}{2pt}
\tablecaption{Log of Observations\label{tab:obslog}}
\tablehead{\colhead{2MASS} & \colhead{UT} & \colhead{Slit} & \colhead{$T_{\mathrm{Exp}}$} & \colhead{$N_{\mathrm{Exp}}$} & \colhead{S/N} & \colhead{Standard} & \colhead{}\\
\colhead{Designation} & \colhead{Date} & \colhead{($''$)} & \colhead{(s)} & \colhead{} & \colhead{} & \colhead{Star} & \colhead{Input\tablenotemark{a}}}
\startdata
\sidehead{\textbf{Magellan/Baade FIRE, Echelle}}
08254335--0029110 & 131213 & 0.6 & 1310 & 2 & 65 & HIP~35837 & 1\\
09553336--0208403 & 160123 & 0.6 & 5400 & 6 & 55 & HD~79359 & 2\\
11020983--3430355 & 140512 & 0.6 & 1200 & 2 & 225 & HIP~54890 & 1\\
11472421--2040204 & 160223 & 0.6 & 5454 & 6 & 40 & HIP~61830 & 2\\
11472421--2040204 & 160123 & 0.6 & 5400 & 6 & 15\tablenotemark{b} & HD~79359 & 2\\
11480096--2836488 & 150531 & 0.6 & 3600 & 4 & 50 & HIP~59351 & 1\\
12074836--3900043 & 151222 & 0.6 & 2400 & 4 & 95 & HD~104647 & 1\\
12451416--4429077 & 160223 & 0.6 & 1816 & 4 & 120 & HIP~70402 & 3\\
12563961--2718455 & 140512 & 0.6 & 1800 & 2 & 10 & HD~116699 & 1\\
14112131--2119503 & 140512 & 0.6 & 1000 & 2 & 230 & HIP~69639 & 1\\
\sidehead{\textbf{Magellan/Baade FIRE, Prism}}
11063147--4201251 & 160122 & 0.6 & 320 & 8 & 140 & HD~102338 & 2\\
11472421--2040204 & 160122 & 0.6 & 640 & 8 & 70 & HD~105992 & 2\\
12194846--3232059 & 160122 & 0.6 & 320 & 8 & 80 & HD~102338 & 2\\
\sidehead{\textbf{IRTF SpeX, Prism}}
09553336--0208403 & 151206 & 0.5 & 1440 & 8 & 20 & HD~79752 & 2\\
10212570--2830427 & 151206 & 0.5 & 720 & 4 & 20 & HD~91398 & 1\\
\sidehead{\textbf{Gemini-South Flamingos-2, JH grism}}
11034950--3409445 & 150426 & 0.72 & 1920 & 16 & 500 & HD~92699 & 1\\
12451035--1443029 & 150427 & 0.72 & 4320 & 36 & 400 & HD~105764 & 1\\
\sidehead{\textbf{CFHT, ESPaDOnS, Echelle}}
10190109--2646336 & 160421 & $\cdots$ & 1800 & 2 & 35 & $\cdots$ & 1\\
10284580--2830374 & 160418 & $\cdots$ & 1800 & 2 & 50 & $\cdots$ & 1\\
10585054--2346206 & 160124 & $\cdots$ & 1600 & 2 & 65 & $\cdots$ & 4\\
10585054--2346206 & 160611 & $\cdots$ & 1800 & 1 & 75 & $\cdots$ & 4\\
11023986--2507113 & 160421 & $\cdots$ & 1500 & 1 & 50 & $\cdots$ & 1\\
11152992--2954436 & 160421 & $\cdots$ & 1800 & 2 & 30 & $\cdots$ & 1\\
11382693--3843138 & 160612 & $\cdots$ & 1800 & 2 & 30 & $\cdots$ & 1\\
11393382--3040002 & 160421 & $\cdots$ & 1460 & 1 & 85 & $\cdots$ & 1\\
11423628--3859108 & 160516 & $\cdots$ & 1800 & 2 & 15 & $\cdots$ & 1\\
12000160--1731308 & 160115 & $\cdots$ & 500 & 1 & 65 & $\cdots$ & 4\\
12073145--3310222 & 160612 & $\cdots$ & 1790 & 1 & 50 & $\cdots$ & 5\\
12175920--3734433 & 160516 & $\cdots$ & 1800 & 2 & 20 & $\cdots$ & 1\\
\sidehead{\textbf{Gemini-South and North GMOS, OG515/R400}}
08141769+0253199 & 130212 & 0.75 & 1120 & 4 & 115 & G~191--2~B & 1\\
08144321+2336045 & 140216 & 1.0 & 256 & 4 & 100 & G~191--2~B & 1\\
09512673--2220196 & 121205 & 0.75 & 1520 & 4 & 60 & LTT~2415 & 1\\
10144705--3728151 & 140219 & 1.0 & 180 & 4 & 30 & LTT~4816 & 1\\
10284580--2830374 & 130207 & 0.75 & 400 & 4 & 60 & CD--32~9927 & 1\\
10455263--2819303 & 130208 & 0.75 & 6000 & 4 & 45 & CD--32~9927 & 1\\
10542303--1507082 & 121208 & 0.75 & 360 & 4 & 20 & LTT~2415 & 1\\
10585054--2346206 & 121210 & 0.75 & 400 & 4 & 70 & LTT~2415 & 1\\
11112820--2655027 & 130206 & 0.75 & 400 & 4 & 95 & CD--32~9927 & 1\\
11112984--2713320 & 130204 & 0.75 & 400 & 4 & 65 & CD--32~9927 & 1\\
11195251--3917150 & 140312 & 1.0 & 188 & 4 & 35 & LTT~4816 & 1\\
11504110--2356075 & 140412 & 1.0 & 1152 & 4 & 85 & G~191--2~B & 1\\
11532691--3015414 & 140412 & 1.0 & 440 & 4 & 115 & G~191--2~B & 1\\
12000160--1731308 & 140216 & 1.0 & 256 & 4 & 90 & G~191--2~B & 4\\
12041256+0514128 & 130303 & 0.75 & 600 & 4 & 250 & G~191--2~B & 1\\
12113180--3416537 & 140319 & 1.0 & 256 & 4 & 55 & G~191--2~B & 1\\
12175920--3734433 & 140221 & 1.0 & 180 & 4 & 35 & LTT~4816 & 1\\
12214852--3652349 & 140412 & 1.0 & 256 & 4 & 20 & G~191--2~B & 1\\
12282569--3955014 & 130210 & 0.75 & 400 & 4 & 70 & CD--32~9927 & 1\\
12421948--3805064 & 130207 & 0.75 & 400 & 4 & 70 & CD--32~9927 & 1\\
12532702--3504151 & 140412 & 1.0 & 256 & 4 & 100 & G~191--2~B & 1\\
\enddata
\tablenotetext{a}{1:~BASS survey \citep{2015ApJ...798...73G}, including the LP-BASS and\\ PRE-BASS surveys \citep{2015ApJS..219...33G}; 2:~BASS-Ultracool survey\\ (see Section~\ref{sec:buc}); 3:~\cite{2011PhDT.......245L}; 4:~SUPERBLINK-south (S.~L\'epine\\ et al., in preparation; see Section~\ref{sec:lspm}); 5:~\citep{Elliott:2016dd}}
\tablenotetext{b}{Poor weather conditions.}
\tablecomments{See Section~\ref{sec:obs} for more details.}
\end{deluxetable}
\setlength{\tabcolsep}{5pt}

\subsection{ESPaDOnS at CFHT}\label{sec:obs_espadons}

Two optical high-resolution spectra were obtained for 11 TWA candidates between 2016 January and June. The data were obtained in the Queue Service Observations (QSO) mode, using the ESPaDOnS optical high resolution spectropolarimeter \citep{2006ASPC..358..362D} at the Canada-France-Hawaii Telescope (CFHT). The normal readout speed mode was used along with the \emph{Star+Sky} mode, where a 1\farcs6 optical fiber is placed on the science target and a 1\farcs8 optical fiber is placed on the sky, which yielded a resolving power of $R \sim 67\,000$ over 3\,670--10\,500\,\AA\ across 40 spectral orders. One or two exposures of 500--1800\,s were used, resulting in signal-to-noise ratios of $\sim$\,35--85 per pixel. The data were reduced by the QSO team using the Upena/Libre-Esprit pipeline \citep{1997MNRAS.291..658D}.

\section{DATA ANALYSIS}\label{sec:data_an}

\subsection{Spectral Typing}\label{sec:spt}

Spectral typing was performed using the method described by \cite{2015ApJS..219...33G} and K.~Cruz et al. (submitted to AJ)\footnote{See also \url{https://dx.doi.org/10.6084/m9.figshare.923587.v1} and \url{https://github.com/kelle/NIRTemplates_Manuscript/releases/tag/v1}}, i.e., the spectra were compared to templates\deleted{ that were }built from a median combination of sets of \replaced{known}{previously established} spectral standards. This was done within each of three separate bands ($zJ$, $H$ and $K$) in the case of NIR spectra, or in the full 5\,900--10\,100\,\AA\ range in the case of optical spectra. Optical spectra that displayed H$\alpha$ emission were flagged with the ``e'' suffix. The resulting spectral types are listed in Table~\ref{tab:gmos}.

\replaced{In order}{The slope of all NIR spectra presented here were corrected using available photometry from 2MASS or the Vista Hemisphere Survey (VHS; PI McMahon, Cambridge, UK)} to ensure that no systematic instrumental effects have affected \replaced{the slope of the near-infrared spectra presented here, they were corrected using available photometry from 2MASS or the Vista Hemisphere Survey (VHS; PI McMahon, Cambridge, UK)}{them}. The synthetic magnitudes of the spectra were calculated and compared to the measured values, and a linear relation was fitted to the synthetic-to-measured flux ratio as a function of the logarithm of wavelength. This relation was then used to correct the spectral data, such that deriving synthetic magnitudes from the corrected spectra yielded similar synthetic magnitudes as the 2MASS or VHS measurements.\added{ These corrections have only affected the FIRE data obtained in the prism mode.}

\subsection{H$\alpha$ Emission}\label{sec:halpha}

Young low-mass stars display enhanced H$\alpha$ emission due to strong chromospheric activity, which persists for $\sim$\,400\,Myr \citep{2008AJ....135..785W} and then for an additional $\sim$\,6--7\,Gyr in intermittence \citep{2006AJ....132.2507W,2008AJ....135..785W,2011ApJ...727....6S,2015AJ....149..158S}. The presence of H$\alpha$ emission at a given moment is thus not a strong indication of a very young age, however its absence can be used to constrain the age at $\gtrsim$\,400\,Myr, and thus safely reject candidate members of TWA. There are 7 objects (spectral types M2--M5) in the GMOS data sample presented here that were rejected in this way. The H$\alpha$ equivalent widths are listed in Table~\ref{tab:gmos}. There are no stars in the sample that display H$\alpha$ strong enough to be classified as classical T Tauri stars, according to the criterion of \citeauthor{2003AJ....126.2997B} (\citeyear{2003AJ....126.2997B}; see their Table~1).

\begin{splitdeluxetable*}{lccccccccBlcccccccc}
\tabletypesize{\tiny}
\tablecolumns{18}
\tablecaption{GMOS Optical Spectral Types and Indices\label{tab:gmos}}
\tablehead{\colhead{2MASS} & \colhead{Opt.} & \colhead{Effect on} & \colhead{EW(H$\alpha$)} & \multicolumn{2}{c}{\cite{2009AJ....137.3345C}} & \colhead{} & \multicolumn{2}{c}{\cite{1999ApJ...519..802K}} & \colhead{2MASS} & \multicolumn{8}{c}{\cite{1999ApJ...519..802K}}\\
\cline{5-6}
\cline{8-9}
\cline{11-18}
\colhead{Designation} & \colhead{SpT\tablenotemark{a}} & \colhead{Membership\tablenotemark{b}} & \colhead{(\AA)} & \colhead{K-a} & \colhead{K-b} & \colhead{} & \colhead{Na-a} & \colhead{Na-b} & \colhead{Designation} & \colhead{Rb-a} & \colhead{Rb-b} & \colhead{Cs-a} & \colhead{Cs-b} & \colhead{CrH-a} & \colhead{CrH-b} & \colhead{FeH-a} & \colhead{FeH-b}}
\startdata
08141769+0253199 & M5\,e & LM $\rightarrow$ R & $6.3 \pm 0.4$ & $2.27 \pm 0.02$ & $2.16 \pm 0.02$ &  & $1.21 \pm 0.01$ & $1.35 \pm 0.01$ & 08141769+0253199 & $1.09 \pm 0.01$ & $1.04 \pm 0.01$ & $1.04 \pm 0.01$ & $1.02 \pm 0.01$ & $0.97 \pm 0.01$ & $1.03 \pm 0.01$ & $0.97 \pm 0.01$ & $1.30 \pm 0.01$\\
08144321+2336045 & M4\,e & LM $\rightarrow$ R & $8.0 \pm 0.3$ & $2.00 \pm 0.01$ & $1.80 \pm 0.01$ &  & $1.16 \pm 0.01$ & $1.14 \pm 0.01$ & 08144321+2336045 & $0.97 \pm 0.01$ & $1.04 \pm 0.01$ & $0.99 \pm 0.01$ & $1.01 \pm 0.01$ & $0.99 \pm 0.01$ & $1.02 \pm 0.01$ & $0.98 \pm 0.01$ & $1.20 \pm 0.01$\\
09512673--2220196 A & M5 & CM $\rightarrow$ R & $\cdots$ & $1.98 \pm 0.02$ & $1.83 \pm 0.02$ &  & $1.23 \pm 0.02$ & $1.33 \pm 0.03$ & 09512673--2220196 A & $1.04 \pm 0.01$ & $1.05 \pm 0.01$ & $1.03 \pm 0.01$ & $1.02 \pm 0.02$ & $0.92 \pm 0.01$ & $0.97 \pm 0.03$ & $0.95 \pm 0.01$ & $1.09 \pm 0.04$\\
09512673--2220196 B & M5 & CM $\rightarrow$ R & $\cdots$ & $1.98 \pm 0.02$ & $1.81 \pm 0.01$ &  & $1.23 \pm 0.03$ & $1.33 \pm 0.03$ & 09512673--2220196 B & $1.05 \pm 0.01$ & $1.05 \pm 0.01$ & $1.03 \pm 0.01$ & $1.02 \pm 0.01$ & $0.92 \pm 0.01$ & $0.99 \pm 0.03$ & $0.96 \pm 0.01$ & $1.09 \pm 0.04$\\
10144705--3728151 A & M3\,pec & CM $\rightarrow$ R & $\cdots$ & $1.61 \pm 0.04$ & $1.47 \pm 0.04$ &  & $1.13 \pm 0.02$ & $1.18 \pm 0.03$ & 10144705--3728151 A & $0.98 \pm 0.03$ & $1.04 \pm 0.01$ & $0.96 \pm 0.01$ & $0.99 \pm 0.02$ & $0.97 \pm 0.01$ & $\cdots$ & $0.98 \pm 0.02$ & $1.24 \pm 0.05$\\
10144705--3728151 B & M3\,pec & CM $\rightarrow$ R & $\cdots$ & $1.65 \pm 0.05$ & $1.46 \pm 0.04$ &  & $1.13 \pm 0.02$ & $1.18 \pm 0.03$ & 10144705--3728151 B & $0.98 \pm 0.02$ & $1.04 \pm 0.01$ & $0.97 \pm 0.01$ & $0.98 \pm 0.02$ & $0.96 \pm 0.01$ & $\cdots$ & $0.97 \pm 0.02$ & $1.27 \pm 0.05$\\
10284580--2830374 & M5\,$\gamma$e & CM $\rightarrow$ CM & $7.2 \pm 0.6$ & $2.26 \pm 0.01$ & $2.17 \pm 0.01$ &  & $1.12 \pm 0.01$ & $1.13 \pm 0.01$ & 10284580--2830374 & $1.07 \pm 0.01$ & $1.01 \pm 0.01$ & $1.07 \pm 0.01$ & $1.01 \pm 0.01$ & $0.88 \pm 0.01$ & $0.96 \pm 0.01$ & $0.97 \pm 0.01$ & $1.14 \pm 0.01$\\
10455263--2819303 & M5.5\,$\gamma$e & CM $\rightarrow$ CM & $5.8 \pm 0.7$ & $2.69 \pm 0.01$ & $2.44 \pm 0.01$ &  & $1.09 \pm 0.01$ & $1.21 \pm 0.01$ & 10455263--2819303 & $1.02 \pm 0.01$ & $1.06 \pm 0.01$ & $1.05 \pm 0.01$ & $1.01 \pm 0.01$ & $0.91 \pm 0.01$ & $0.99 \pm 0.02$ & $0.96 \pm 0.01$ & $1.13 \pm 0.03$\\
10542303--1507082 & M5.5\,e & CM $\rightarrow$ R & $5.9 \pm 0.6$ & $2.38 \pm 0.02$ & $2.06 \pm 0.02$ &  & $1.23 \pm 0.03$ & $1.34 \pm 0.03$ & 10542303--1507082 & $1.04 \pm 0.01$ & $1.05 \pm 0.02$ & $1.05 \pm 0.02$ & $1.01 \pm 0.02$ & $0.95 \pm 0.02$ & $1.00 \pm 0.02$ & $0.96 \pm 0.02$ & $1.23 \pm 0.02$\\
10585054--2346206 & M6\,$\gamma$e & CM $\rightarrow$ CM & $8.4 \pm 0.5$ & $2.30 \pm 0.01$ & $2.09 \pm 0.01$ &  & $1.09 \pm 0.01$ & $1.16 \pm 0.01$ & 10585054--2346206 & $1.01 \pm 0.01$ & $1.08 \pm 0.01$ & $1.06 \pm 0.01$ & $1.04 \pm 0.02$ & $0.89 \pm 0.01$ & $1.01 \pm 0.02$ & $0.96 \pm 0.01$ & $1.17 \pm 0.02$\\
11112820--2655027 & M6\,$\gamma$e & CM $\rightarrow$ CM & $16.2 \pm 0.8$ & $2.70 \pm 0.01$ & $2.50 \pm 0.01$ &  & $1.10 \pm 0.01$ & $1.16 \pm 0.01$ & 11112820--2655027 & $1.05 \pm 0.01$ & $1.05 \pm 0.01$ & $1.12 \pm 0.01$ & $1.05 \pm 0.01$ & $0.90 \pm 0.01$ & $0.97 \pm 0.01$ & $0.98 \pm 0.01$ & $1.17 \pm 0.01$\\
11112984--2713320 & M4.5\,e & CM $\rightarrow$ R & $3.7 \pm 0.4$ & $2.33 \pm 0.02$ & $2.09 \pm 0.02$ &  & $1.21 \pm 0.01$ & $1.29 \pm 0.01$ & 11112984--2713320 & $0.98 \pm 0.01$ & $1.06 \pm 0.01$ & $1.00 \pm 0.01$ & $1.00 \pm 0.01$ & $0.95 \pm 0.01$ & $1.03 \pm 0.02$ & $0.99 \pm 0.01$ & $1.18 \pm 0.02$\\
11195251--3917150 A & M2 & CM $\rightarrow$ R & $\cdots$ & $1.49 \pm 0.01$ & $1.38 \pm 0.01$ &  & $1.14 \pm 0.02$ & $1.17 \pm 0.02$ & 11195251--3917150 A & $1.00 \pm 0.01$ & $1.04 \pm 0.01$ & $0.94 \pm 0.01$ & $1.00 \pm 0.02$ & $0.95 \pm 0.01$ & $\cdots$ & $0.97 \pm 0.01$ & $1.13 \pm 0.04$\\
11195251--3917150 B & M2 & CM $\rightarrow$ R & $\cdots$ & $1.48 \pm 0.01$ & $1.34 \pm 0.01$ &  & $1.11 \pm 0.01$ & $1.15 \pm 0.01$ & 11195251--3917150 B & $1.00 \pm 0.01$ & $1.05 \pm 0.01$ & $0.93 \pm 0.01$ & $1.00 \pm 0.01$ & $0.98 \pm 0.01$ & $\cdots$ & $0.97 \pm 0.01$ & $1.21 \pm 0.02$\\
11504110--2356075 & M6\,e & R $\rightarrow$ R & $3.2 \pm 0.3$ & $3.05 \pm 0.03$ & $2.62 \pm 0.02$ &  & $1.34 \pm 0.01$ & $1.47 \pm 0.01$ & 11504110--2356075 & $1.08 \pm 0.01$ & $1.04 \pm 0.01$ & $1.07 \pm 0.01$ & $1.02 \pm 0.01$ & $0.98 \pm 0.01$ & $1.04 \pm 0.01$ & $1.04 \pm 0.01$ & $1.37 \pm 0.01$\\
11532691--3015414 & M4.5\,e & R $\rightarrow$ R & $2.0 \pm 0.3$ & $2.16 \pm 0.01$ & $1.94 \pm 0.01$ &  & $1.24 \pm 0.01$ & $1.33 \pm 0.01$ & 11532691--3015414 & $1.06 \pm 0.01$ & $1.03 \pm 0.01$ & $1.03 \pm 0.01$ & $1.00 \pm 0.01$ & $0.99 \pm 0.01$ & $1.03 \pm 0.01$ & $0.95 \pm 0.01$ & $1.24 \pm 0.01$\\
12000160--1731308 AB & M4\,$\gamma$e & LM $\rightarrow$ LM & $3.9 \pm 0.3$ & $1.88 \pm 0.01$ & $1.74 \pm 0.01$ &  & $1.03 \pm 0.01$ & $1.06 \pm 0.01$ & 12000160--1731308 AB & $0.97 \pm 0.01$ & $1.02 \pm 0.01$ & $0.99 \pm 0.01$ & $1.01 \pm 0.01$ & $0.96 \pm 0.01$ & $1.02 \pm 0.01$ & $0.97 \pm 0.01$ & $1.13 \pm 0.01$\\
12041256+0514128 & M5\,e & R $\rightarrow$ R & $9.6 \pm 0.4$ & $2.42 \pm 0.01$ & $2.28 \pm 0.01$ &  & $1.26 \pm 0.01$ & $1.39 \pm 0.01$ & 12041256+0514128 & $1.07 \pm 0.01$ & $1.05 \pm 0.01$ & $1.05 \pm 0.01$ & $1.03 \pm 0.01$ & $0.98 \pm 0.01$ & $1.02 \pm 0.01$ & $1.00 \pm 0.01$ & $1.32 \pm 0.01$\\
12113180--3416537 & M2 & CM $\rightarrow$ R & $\cdots$ & $1.41 \pm 0.01$ & $1.26 \pm 0.01$ &  & $1.15 \pm 0.01$ & $1.20 \pm 0.01$ & 12113180--3416537 & $1.01 \pm 0.01$ & $1.04 \pm 0.01$ & $0.92 \pm 0.01$ & $1.00 \pm 0.01$ & $0.99 \pm 0.01$ & $1.02 \pm 0.01$ & $0.94 \pm 0.02$ & $1.08 \pm 0.01$\\
12175920--3734433 & M5\,$\gamma$e & CM $\rightarrow$ CM & $7.4 \pm 0.6$ & $2.53 \pm 0.02$ & $2.28 \pm 0.02$ &  & $1.10 \pm 0.02$ & $1.19 \pm 0.02$ & 12175920--3734433 & $0.98 \pm 0.01$ & $1.08 \pm 0.01$ & $1.04 \pm 0.01$ & $1.05 \pm 0.02$ & $0.88 \pm 0.01$ & $\cdots$ & $1.00 \pm 0.01$ & $1.35 \pm 0.04$\\
12214852--3652349 A & M4\,e & CM $\rightarrow$ R & $5.6 \pm 1.9$ & $1.94 \pm 0.09$ & $1.74 \pm 0.08$ &  & $1.22 \pm 0.02$ & $1.30 \pm 0.02$ & 12214852--3652349 A & $1.07 \pm 0.04$ & $1.04 \pm 0.02$ & $1.02 \pm 0.01$ & $1.01 \pm 0.01$ & $0.98 \pm 0.01$ & $1.01 \pm 0.02$ & $0.89 \pm 0.01$ & $1.21 \pm 0.03$\\
12214852--3652349 B & M4\,e & CM $\rightarrow$ R & $5.4 \pm 1.7$ & $1.96 \pm 0.09$ & $1.79 \pm 0.08$ &  & $1.22 \pm 0.03$ & $1.30 \pm 0.03$ & 12214852--3652349 B & $1.06 \pm 0.04$ & $1.03 \pm 0.03$ & $1.01 \pm 0.01$ & $1.01 \pm 0.01$ & $0.98 \pm 0.01$ & $1.00 \pm 0.03$ & $0.95 \pm 0.01$ & $1.22 \pm 0.04$\\
12282569--3955014 A & M4\,e & CM $\rightarrow$ R & $5.2 \pm 0.3$ & $1.91 \pm 0.01$ & $1.68 \pm 0.01$ &  & $1.19 \pm 0.01$ & $1.25 \pm 0.01$ & 12282569--3955014 A & $1.00 \pm 0.01$ & $1.04 \pm 0.01$ & $0.97 \pm 0.01$ & $1.01 \pm 0.01$ & $0.99 \pm 0.01$ & $1.01 \pm 0.02$ & $1.00 \pm 0.01$ & $1.15 \pm 0.01$\\
12282569--3955014 B & M3.5\,e & CM $\rightarrow$ CM & $5.0 \pm 0.3$ & $1.91 \pm 0.01$ & $1.68 \pm 0.01$ &  & $1.20 \pm 0.01$ & $1.24 \pm 0.01$ & 12282569--3955014 B & $1.01 \pm 0.01$ & $1.04 \pm 0.01$ & $0.97 \pm 0.01$ & $1.01 \pm 0.01$ & $1.00 \pm 0.01$ & $0.98 \pm 0.01$ & $1.00 \pm 0.01$ & $1.14 \pm 0.01$\\
12421948--3805064 A & M3\,e & LM $\rightarrow$ LM & $3.2 \pm 0.3$ & $1.85 \pm 0.01$ & $1.65 \pm 0.01$ &  & $1.22 \pm 0.01$ & $1.28 \pm 0.01$ & 12421948--3805064 A & $1.02 \pm 0.01$ & $1.03 \pm 0.01$ & $0.99 \pm 0.01$ & $0.97 \pm 0.01$ & $0.98 \pm 0.01$ & $0.95 \pm 0.01$ & $1.00 \pm 0.01$ & $1.12 \pm 0.01$\\
12421948--3805064 B & M3\,e & LM $\rightarrow$ LM & $3.4 \pm 0.3$ & $1.86 \pm 0.01$ & $1.64 \pm 0.01$ &  & $1.21 \pm 0.01$ & $1.25 \pm 0.01$ & 12421948--3805064 B & $1.02 \pm 0.01$ & $1.02 \pm 0.01$ & $1.01 \pm 0.01$ & $1.00 \pm 0.01$ & $0.98 \pm 0.01$ & $0.95 \pm 0.01$ & $0.99 \pm 0.01$ & $1.12 \pm 0.01$\\
12532702-3504151 & M3\,e & CM $\rightarrow$ CM & $4.5 \pm 0.2$ & $1.56 \pm 0.01$ & $1.43 \pm 0.01$ &  & $1.11 \pm 0.01$ & $1.18 \pm 0.01$ & 12532702-3504151 & $1.03 \pm 0.01$ & $1.03 \pm 0.01$ & $0.99 \pm 0.01$ & $1.00 \pm 0.01$ & $0.98 \pm 0.01$ & $1.01 \pm 0.01$ & $0.96 \pm 0.01$ & $1.13 \pm 0.01$\\
\enddata
\tablenotetext{a}{Optical spectral type. ``e'' indicates H$\alpha$ emission, $\gamma$ indicates a low surface gravity and ``pec'' indicates other peculiar features. See Section~\ref{sec:spt} for more details.}
\tablenotetext{b}{Previous membership status $\rightarrow$ Updated membership status, based on the inclusion of new age constraints based on spectroscopic indices. See Section~\ref{sec:data_an} for more detail. R: Rejected; LM: Low-likelihood candidate member; CM: Candidate member, HM: High-likelihood candidate member; BF: Bona fide member.}
\tablecomments{See Section~\ref{sec:data_an} for more details.}
\end{splitdeluxetable*}

\deleted{ The star that is closest to this criterion is 2MASS~J11112820--2655027 (TWA~37; M6\,$\gamma$e), with EW(H$\alpha$)$ = 16.2 \pm 0.8$\,\AA: at this spectral type, EW(H$\alpha$) $\geq$\,24.1\,\AA\ would be required to categorize it as a classical T Tauri star. The $W3$ and $W4$ WISE magnitudes of TWA~37 are well detected (catalog entry WISE~J111128.13--265502.9) at $W3 = 8.77 \pm 0.02$ mag (46$\sigma$) and $W4 = 8.31 \pm 0.23$ (4.7$\sigma$), however it does not present conclusive signs of an infrared excess. Its $W1-W4$ color ($0.93 \pm 0.23$) is not red enough to respect the $W1-W4 > 1.0$ criterion of \cite{2012ApJ...757..163S} for infrared excess (see also \citealp{2012ApJ...754...39S}), and a comparison with predictions from BT-Settl models places the infrared excess of this source at a significance below 3$\sigma$ in these two photometric bands (see \citealt{2016arXiv160808259B}).}

\begin{figure*}[p]
	\centering
\subfigure{\includegraphics[width=0.488\textwidth]{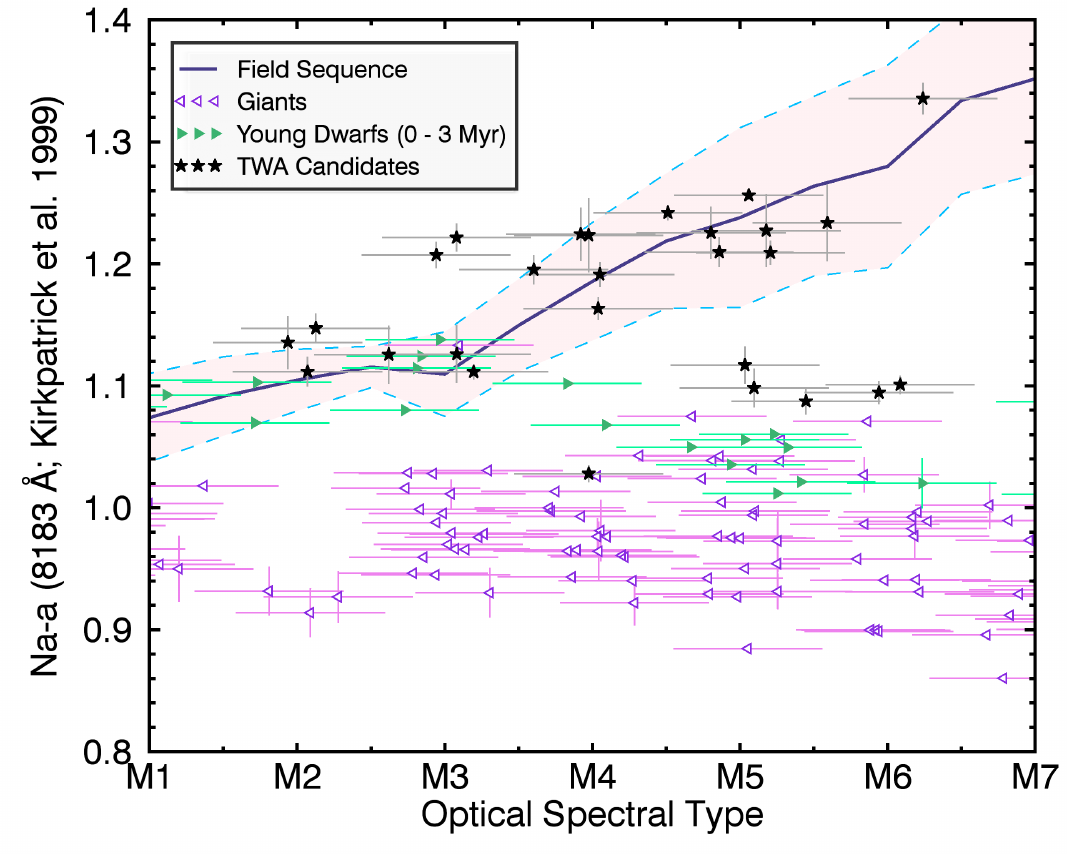}}
\subfigure{\includegraphics[width=0.488\textwidth]{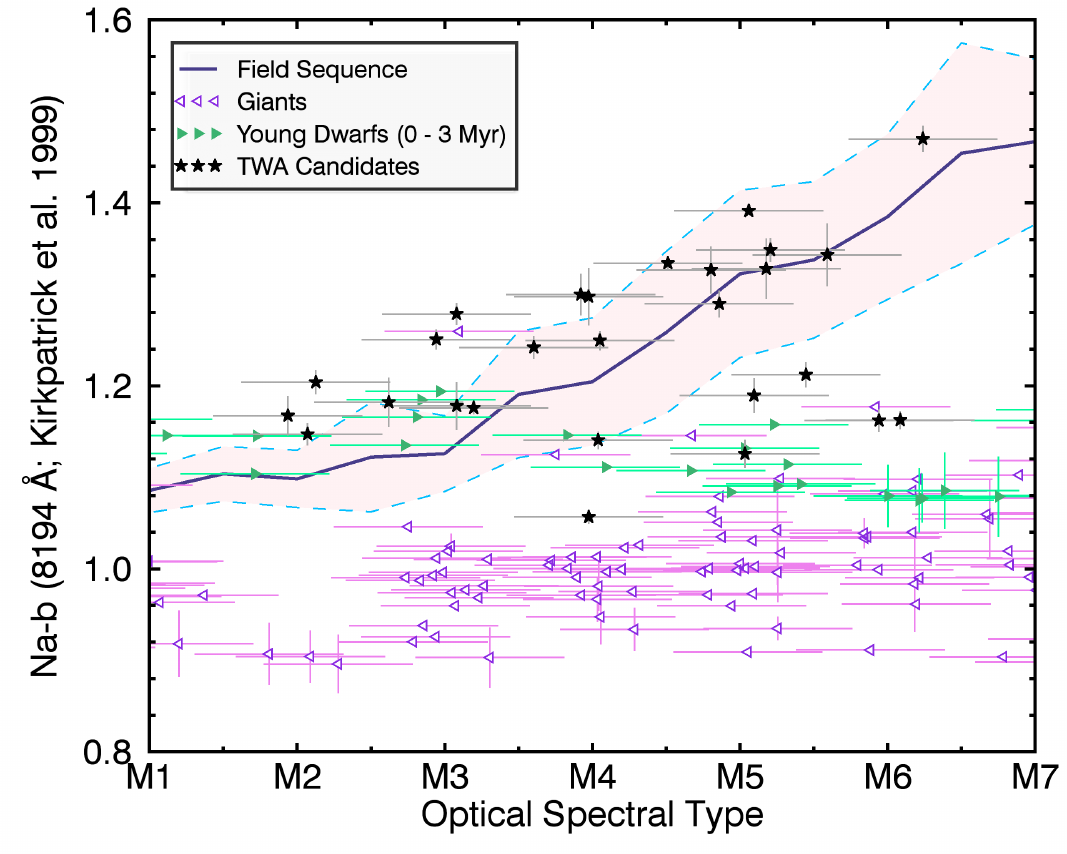}}
\subfigure{\includegraphics[width=0.488\textwidth]{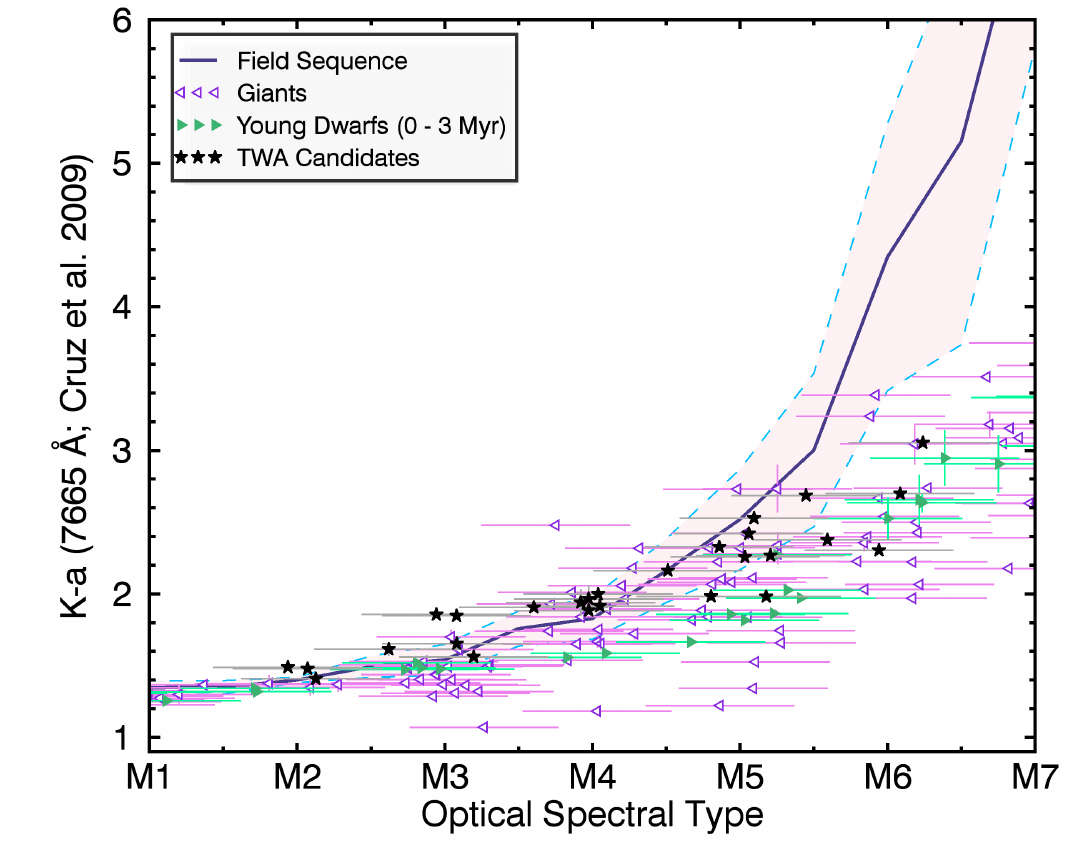}}
\subfigure{\includegraphics[width=0.488\textwidth]{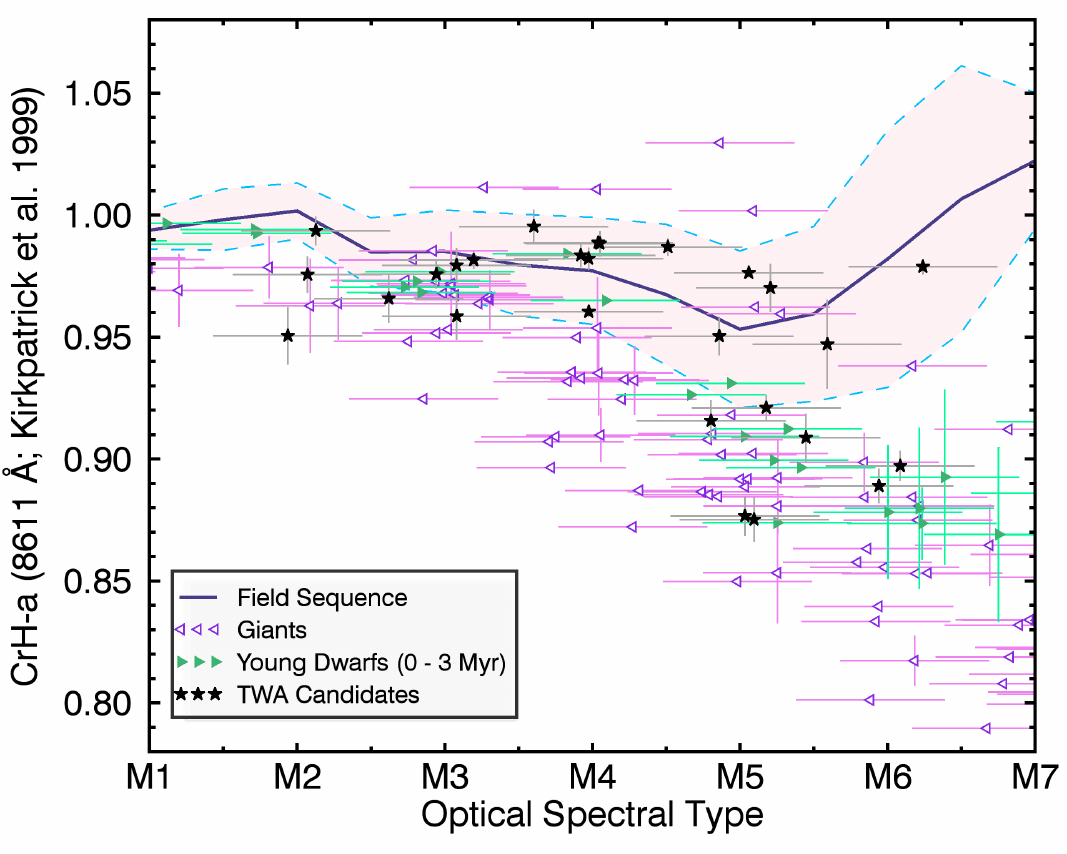}}
	\caption{Gravity-sensitive spectral indices for giant stars (leftwards purple triangles), young ($\leq$\,3\,Myr) dwarfs (rightwards green triangles) and new observations of TWA candidate members (black stars). A sequence of field-age dwarfs is displayed as a thick, dark blue line and its scatter is represented by the light pink region that is encompassed with dashed pale-blue lines. These spectral indices do not allow for a distinction between young and field $\leq$\,M3 dwarfs, however they can identify giants even at these early types. Six of the new $\geq$\,M4 candidates members have spectral indices that are systematically lower than the field sequence, indicating a young age. Twelve others show signs of an older age and can thus be rejected as  candidate members of TWA. See Section~\ref{sec:lowg} for more details.}
	\label{fig:GMOS_inds}
\end{figure*}

\subsection{Signs of Low Surface Gravity}\label{sec:lowg}

Several spectral indices have been developed to identify signs of low \added{surface }gravity in the optical or near-infrared spectra of low-mass stars and brown dwarfs. In the optical, the equivalent width of the \ion{Na}{1} (8183.3 and 8194.8\,\AA), \ion{K}{1} (7665 and 7699\,\AA) and CrH (8611 and 9969\,\AA) absorption lines are known to be weaker in low-gravity atmospheres, due to the weaker effect of pressure broadening (e.g., \citealp{2006ApJ...639.1120K,2009AJ....137.3345C}). In the near-infrared, a similar effect is observed with the \ion{Na}{1} (1.1396$\mu$m), \ion{K}{1} (1.1692, 1.7778, 1.2437 and 1.2529\,$\mu$m) and FeH (0.998\,$\mu$m) absorption lines \citep{2013ApJ...772...79A}. In addition to this, the $H$-band continuum ($\sim$\,1.5--1.7\,$\mu$m) of low-gravity $\geq$ M6 dwarfs displays a typical triangular shape, which is a combined effect of decreased collision-induced absorption of the H$_2$ molecule, and weaker absorption from the FeH molecule at $\sim$\,1.6--1.7\,$\mu$m \citep{2010ApJ...715L.165R,2013ApJ...772...79A}. \cite{2005ARAA..43..195K} and \cite{2006ApJ...639.1120K} have introduced the spectral typing suffixes $\alpha$, $\beta$ and $\gamma$ to identify M and L dwarf with normal gravity, subtle signs of low gravity and strong signs of low gravity, respectively (see also \citealt{2009AJ....137.3345C}).

The gravity-sensitive indices defined by \cite{1999ApJ...519..802K} and \cite{2009AJ....137.3345C} for the new GMOS data are presented in Table~\ref{tab:gmos}. In Figure~\ref{fig:GMOS_inds}, the optical indices derived from the GMOS data are compared with a field sequence, and with spectral indices of known giants and young ($\leq$\,3\,Myr) dwarfs. Six objects display clear signs of low gravity in all of the sequences, and are thus assigned the $\gamma$ suffix. All other 12 targets with spectral types $\geq$\,M4 display spectral indices that are fully consistent with a field age, and are thus rejected as candidate members of TWA. None of the objects display ambiguous signs of low gravity that would justify the assignment of a $\beta$ suffix. It can be seen that the sequences of field and young dwarfs merge at spectral types earlier than M4, and for this reason it is not possible to draw any conclusion regarding the age of the 9 targets that fall in this range using these low-gravity sensitive indices.\deleted{ It can also be noted that the Na-a index of one of the new TWA candidates (2MASS~J12000160--1731308~AB) could be weak enough for it to be a giant star instead of a young brown dwarf, however this possibility is rejected by the presence of H$\alpha$ emission in its spectrum.}

In the near-infrared, the gravity-sensitive spectral indices have been combined by \cite{2013ApJ...772...79A} in a classification scheme that allows to assign a gravity class to $R \sim$\,120--1\,200 spectra. The three possible gravity classes are field-gravity (FLD-G), intermediate-gravity (INT-G) and very low-gravity (VL-G), and have been shown to correspond \replaced{to the same ranges of age as the}{age ranges that are similar to the} $\alpha$, $\beta$ and $\gamma$ suffixes described above. \cite{2015ApJS..219...33G} have also defined distinct NIR spectral templates \replaced{of}{for} the three gravity classes that allow for a gravity classification\replaced{ that is }based on a visual comparison\added{,} as described in Section~\ref{sec:spt}.

There are 3\deleted{ new }late-M targets for which new NIR data is presented in this work. Their names and respective spectral types, gravity classifications, and gravity scores are: 2MASS~J12194846--3232059 (M7 FLD-G; score 0n00); 2MASS~J11063147--4201251 (M8 FLD-G; score 0n00); \added{and }2MASS~J11034950--3409445 (M9 INT-G; score 1n10; see \citealt{2013ApJ...772...79A} for a detailed description of the gravity scores). The first two targets are rejected as TWA candidate members, due to their lack of low-gravity signatures. The visual spectral type classifications are consistent with the \cite{2013ApJ...772...79A} index-based gravity classes for the three targets (M7, M8 and M9\,$\beta$, respectively).

NIR spectral types of new TWA candidates from \cite{2011PhDT.......245L} were revised using the spectral standards of \cite{2015ApJS..219...33G}, since no low-gravity NIR standards or templates were available at the time of their discovery. All subtypes remained unchanged, but all objects were classified with the $\gamma$ gravity class (2MASS~J10455263--2819303, M6\,$\gamma$; 2MASS~J11064461--3715115, M9\,$\gamma$; 2MASS~J11112820--2655027 or TWA~37, M6\,$\gamma$; 2MASS~J12035905--3821402 or TWA~38, M8\,$\gamma$; 2MASS~J12071089--3230537 or TWA~31, M6\,$\gamma$;\added{ and} 2MASS~J12520989--4948280, M8\,$\gamma$).

\begin{deluxetable*}{lccccccc}
\tabletypesize{\scriptsize}
\tablecolumns{8}
\tablecaption{FIRE Radial Velocity Measurements\label{tab:rvs}}
\tablehead{\colhead{2MASS} & \colhead{Other} & \colhead{RV\tablenotemark{a}} & \colhead{$v\sin i$}& \colhead{EW(Li)} & \colhead{S/N} & \colhead{UT Date} & \colhead{Effect on}\\
\colhead{Designation} & \colhead{Name} & \colhead{(\kms)} & \colhead{(\kms)} & \colhead{(m\AA)} & \colhead{per pixel} & \colhead{(ddmmyy)} & \colhead{Membership\tablenotemark{b}}}
\startdata
08254335--0029110 & $\cdots$ & $17 \pm 3$ & $\cdots$ & $\cdots$ & 79 & 131213 & R $\rightarrow$ R\\
09553336--0208403 & BASS-UC~51 & $-20 \pm 4$ & $\cdots$ & $\cdots$ & 60 & 160123 & CM $\rightarrow$ R\\
11020983--3430355 & TWA~28 & $9 \pm 3$ & $\cdots$ & $\cdots$ & 230 & 140512 & HM $\rightarrow$ BF\\
11472421--2040204 & BASS-UC~56 & $9 \pm 5$ & $\cdots$ & $\cdots$ & 15 & 160123 & $\cdots$\\
11472421--2040204 & BASS-UC~56 & $7 \pm 3$ & $\cdots$ & $\cdots$ & 45 & 160223 & $\cdots$\\
11472421--2040204\tablenotemark{c} & BASS-UC~56 & $7 \pm 3$ & $\cdots$ & $\cdots$ & $\cdots$ & $\cdots$ & CM $\rightarrow$ HM\\
11480096--2836488 & $\cdots$ & $6 \pm 3$ & $\cdots$ & $\cdots$ & 55 & 150531 & CM $\rightarrow$ CM\\
12074836--3900043 & $\cdots$ & $6 \pm 3$ & $\cdots$ & $\cdots$ & 100 & 151222 & CM $\rightarrow$ HM\\
12451416--4429077 & TWA~29 & $8 \pm 3$ & $\cdots$ & $\cdots$ & 120 & 160223 & HM $\rightarrow$ BF\\
12563961--2718455 & $\cdots$ & $-19 \pm 4$ & $\cdots$ & $\cdots$ & 15 & 140512 & CM $\rightarrow$ R\\
14112131--2119503 & $\cdots$ & $-13 \pm 3$ & $\cdots$ & $\cdots$ & 230 & 140512 & R $\rightarrow$ R\\
\enddata
\tablenotetext{a}{Radial relocity.}
\tablenotetext{b}{Previous membership status $\rightarrow$ Updated membership status, based on the inclusion of the new radial velocity measurement in BANYAN~II. See Section~\ref{sec:rv} for more detail. R: Rejected; LM: Low-likelihood candidate member; CM: Candidate member, HM:High-likelihood candidate member; BF: Bona fide member. A few candidates have been rejected either by additional information after they were observed with GMOS; this explains the few cases with a mention R\,$\rightarrow$\,R.}
\tablenotetext{c}{Error-weighted combination of the two measurements (weights are determined before the application of the $\pm 3$\,\kms\ systematic error).}
\tablecomments{See Section~\ref{sec:rv} for more details.}
\end{deluxetable*}

\subsection{FIRE/Echelle Radial Velocity Measurements}\label{sec:rv}

Radial velocities were measured for all FIRE-echelle spectra by comparing them with zero-velocity CIFIST 2011 BT-Settl spectra \citep{2015A&A...577A..42B,2012RSPTA.370.2765A}. The IDL implementation of the amoeba  Nelder-Mead downhill simplex algorithm \citep{Nelder:1965in} was used to fit a forward model to the data in several fixed wavelength windows located in the $H$ band, which provides the highest signal-to-noise\deleted{ with current }data. The 4 free parameters that were used in this forward modelling approach are (1) the radial velocity Doppler shift, (2) the characteristic width of the instrumental line spread function, and (3) a two-parameters multiplicative linear correction to the spectral flux density continuum to account for instrumental systematics.

The instrumental LSF \replaced{is}{was} assumed to be Gaussian, and no telluric model \replaced{is}{was} needed as telluric standard observations were used to correct the science spectra\added{ as described in Section~\ref{sec:obs_fire}}. The Firehose pipeline generates spectra that are corrected for barycentric velocity variations and placed in a vacuum wavelength reference, thus removing the need to account for these effects or to determine a wavelength solution as part of the forward model.

The BT-Settl model that minimizes the $\chi^2$ value in \added{the }$H$\replaced{-}{ }band when compared to the science spectrum \replaced{is}{was} selected for the radial velocity determination, and the radial velocity fitting \replaced{is}{was} performed in fifteen 0.02\,$\mu$m-wide segments regularly distributed in the 1.5100--1.5535\,$\mu$m range to account for any systematics and limit the effects of bad pixels. This method is very similar to that used by \cite{2015ApJ...808L..20G} and Burgasser et al. (in preparation) to measure \replaced{RVs}{radial velocities} using FIRE-echelle spectra, and is known to produce RVs that are limited to a $3$\,\kms\ precision due to systematics (this \replaced{number}{assessment of precision} was obtained by performing similar RV measurements on \replaced{RV}{radial velocity} standard stars). The average of \replaced{the}{all} individual 15 measurements \replaced{is}{was} taken as the radial velocity measurement, and their standard deviation\added{,} added in quadrature to the $3$\,\kms\ systematic error\added{,}  \replaced{is}{was} taken as the measurement error.

These new radial velocity measurements are displayed in Table~\ref{tab:rvs}, along with their updating effect on TWA membership, when used as an additional input to BANYAN~II.

\subsection{ESPaDOnS/CFHT Measurements}\label{sec:rv2}

Heliocentric radial velocities and projected rotational velocities for the \replaced{two}{11} ESPaDOnS/CFHT spectra were measured using the method described by \cite{2014ApJ...788...81M}, which consists in \added{performing }a cross-correlation of the data with an observed radial velocity standard star of a similar spectral type. Lithium absorption lines were clearly detected for 8/11 spectra, and \replaced{the}{their} equivalent width\added{s} were measured using the method of \cite{2014ApJ...788...81M}. These detections ensure that the objects in question are younger than $\sim$\,80--200\,Myr, depending on their spectral types.

There are a few objects that were observed only with ESPaDOnS\replaced{, and have}{for which} no spectral types \replaced{yet measured}{were yet determined} in the literature. \replaced{For this reason, the}{The} method of \cite{2006AJ....132..866R} was used to estimate \added{their }spectral types \replaced{from}{based on} the TiO5 spectral index. All resulting measurements are reported in Table~\ref{tab:esp_rvs}.

\subsection{Discussion of Individual Objects}

\added{In this section, we discuss several individual objects for which the new data presented here require special attention.}

\subsubsection{2MASS~J11423628--3859108}\label{sec:J1142}

The ESPaDOnS spectrum of 2MASS~J11423628--3859108 \replaced{is much different}{differs significantly} from that of an M- or later-type dwarf, hence we conclude that it is likely a background star contaminant and reject it from the sample of TWA candidate members.

\subsubsection{2MASS~J11112820--2655027}\label{sec:J1111}

\added{2MASS~J11112820--2655027 (TWA~37; M6\,$\gamma$e) is the star in the GMOS data sample that has an H$\alpha$ equivalent width closest to the \cite{2003AJ....126.2997B} criterion for Classical T~Tauri stars although it does not meet it, with EW(H$\alpha$)$ = 16.2 \pm 0.8$\,\AA: at this spectral type, EW(H$\alpha$) $\geq$\,24.1\,\AA\ would be required to categorize it as a classical T Tauri star.}

\added{The $W3$ and $W4$ WISE magnitudes of TWA~37 are well detected (catalog entry WISE~J111128.13--265502.9) at $W3 = 8.77 \pm 0.02$ mag (46$\sigma$) and $W4 = 8.31 \pm 0.23$ (4.7$\sigma$), however it does not present conclusive signs of an infrared excess. Its $W1-W4$ color ($0.93 \pm 0.23$) is not red enough to respect the $W1-W4 > 1.0$ criterion of \cite{2012ApJ...757..163S} for infrared excess (see also \citealp{2012ApJ...754...39S}), and a comparison with predictions from BT-Settl models places the infrared excess of this source at a significance below 3$\sigma$ in these two photometric bands (see \citealt{2016arXiv160808259B}).}

\subsubsection{2MASS~J12000160--1731308~AB}\label{sec:J1200}

\added{The Na-a index \citep{1999ApJ...519..802K} of 2MASS~J12000160--1731308~AB (M4\,$\gamma$e; see Table~\ref{tab:gmos} and Figure~\ref{fig:GMOS_inds}) could be weak enough for\deleted{ it to be }a giant star instead of a young brown dwarf, however this possibility is rejected by the presence of H$\alpha$ emission in its spectrum (with an H$\alpha$ equivalent width of $3.9 \pm 0.3$\,\AA).}

\subsubsection{2MASS~J09553336--0208403: A New young L7 dwarf}\label{sec:J0955}

Although \replaced{a}{the new} radial velocity measurement \added{presented here} rejects 2MASS~J09553336--0208403 as a \replaced{candidate}{possible} member of TWA (see Section~\ref{sec:rv}), its very red $J-K_S$ color ($2.14 \pm 0.02$), triangular-shaped $H$-band continuum and weak \ion{K}{1} absorption lines (see Table~\ref{tab:J1147_K1} and Figure~\ref{fig:j0955}) are indicative of a low surface gravity. Using the young spectral type--$K_S$-magnitude sequence of \cite{2015ApJS..219...33G}, the spectrophotometric distance of 2MASS~J09553336--0208403 is estimated at $30.5 \pm 9.0$\,pc. A likelihood analysis based on the BT-Settl models (see \citealt{2014ApJ...783..121G}) yields a mass estimate of $18 \pm 6$\,\Mjup\ at this distance \replaced{by}{when} adopting a conservative age range of 1--200\,Myr. \replaced{The}{This} upper age limit \replaced{is}{was} chosen \added{at the approximate boundary } where low-gravity spectral indices become inapparent\added{ in the NIR spectra of brown dwarfs and low-mass stars} (e.g., \citealt{2013ApJ...772...79A}).

\begin{deluxetable*}{lcccccccccc}
\tabletypesize{\scriptsize}
\tablecolumns{11}
\tablecaption{Measurements from ESPaDOnS Optical Spectra\label{tab:esp_rvs}}
\tablehead{\colhead{2MASS} & \colhead{Other} & \colhead{RV} & \colhead{$v\sin i$} & \colhead{EW(H$\alpha$)} & \colhead{EW(Li)} & \colhead{Li Age\tablenotemark{a}} & \colhead{S/N} & \colhead{Spectral} & \colhead{UT Date} & \colhead{Effect on}\\
\colhead{Designation} & \colhead{Name} & \colhead{(\kms)} & \colhead{(\kms)} & \colhead{(\AA)} & \colhead{(m\AA)} & \colhead{(Myr)} & \colhead{/pix} & \colhead{Type} & \colhead{(ddmmyy)} & \colhead{Membership\tablenotemark{b}}}
\startdata
10190109--2646336 & $\cdots$ & $-3.1 \pm 0.3$ & $24 \pm 2$ & $4.0 \pm 0.1$ & $< 27$ & $> 100$ & 35 & M5\,e & 160421 & CM $\rightarrow$ R\\
10284580--2830374 & TWA~34 & $12.4 \pm 0.3$ & $15 \pm 2$ & $8.7 \pm 0.1$ & $630 \pm 20$ & $< 200$ & 50 & M6\,e & 160418 & CM $\rightarrow$ HM\\
10585054--2346206 & $\cdots$ & $8.1 \pm 0.3$ & $22 \pm 2$ & $8.7 \pm 0.1$ & $620 \pm 20$ & $< 200$ & 65 & M6\,e & 160124 & $\cdots$\\
10585054--2346206 & $\cdots$ & $8.2 \pm 0.3$ & $24 \pm 2$ & $7.9 \pm 0.3$ & $680 \pm 60$ & $< 200$ & 75 & M6\,e & 160611 & $\cdots$\\
10585054--2346206\tablenotemark{c} & $\cdots$ & $8.2 \pm 0.2$ & $23 \pm 2$ & $8.6 \pm 0.1$ & $630 \pm 20$ & $< 200$ & $\cdots$ & M6\,e & $\cdots$ & CM $\rightarrow$ CM\\
11023986--2507113 & $\cdots$ & $17.3 \pm 0.3$ & $8.4 \pm 0.9$ & $3.29 \pm 0.06$ & $26 \pm 4$ &  $< 80$ & 50 & M4\,e & 160421 & CM $\rightarrow$ LM\\
11152992--2954436 & $\cdots$ & $13.3 \pm 0.2$ & $5 \pm 1$ & $3.51 \pm 0.08$ & $71 \pm 6$ & $< 80$ & 30 & M4\,e & 160421 & CM $\rightarrow$ CM\\
11382693--3843138 & $\cdots$ & $18.7 \pm 0.4$ & $24 \pm 2$ & $4.9 \pm 0.1$ & $53 \pm 9$ & $< 100$ & 30 & M5\,e & 160612 & CM $\rightarrow$ LM\\
11393382--3040002 & TWA~33 & $5.8 \pm 0.7$ & $15 \pm 2$ & $3.88 \pm 0.07$ & $590 \pm 10$ & $< 150$ & 85 & M5.5\,e & 160421 & HM $\rightarrow$ BF\\
11423628--3859108 & $\cdots$ & $\cdots$ & $\cdots$ & $\cdots$ & $\cdots$ & $\cdots$ & 15 & $<$\,M0 & 160516 & CM $\rightarrow$ R\\
12000160--1731308\tablenotemark{d} & $\cdots$ & $-0.1 \pm 0.8$ & $59 \pm 4$ & $3.9 \pm 0.1$ & $720 \pm 30$ & $< 200$ & 65 & M6\,e & 160115 & R $\rightarrow$ R\\
12073145--3310222 & $\cdots$ & $-9.0 \pm 0.2$ & $3 \pm 1$ & $-0.45 \pm 0.02$ & $< 12$ & $\cdots$ & 50 & $\sim$\,M0 & 160612 & CM $\rightarrow$ R\\
12175920--3734433 & $\cdots$ & $5 \pm 3$ & $32 \pm 5$ & $10.6 \pm 0.9$ & $850 \pm 20$ & $< 200$ & 20 & M6\,e & 160516 & CM $\rightarrow$ HM\\
\enddata
\tablenotetext{a}{Age limit based on the detection of Li, \teff\ estimated from spectral types (see \citealt{2013ApJS..208....9P}) and the lithium depletion boundaries of \cite{1998ASPC..134..394B}.}
\tablenotetext{b}{Previous membership status $\rightarrow$ Updated membership status, based on the inclusion of the new radial velocity measurement in BANYAN~II. See Section~\ref{sec:rv} for more detail. R: Rejected; LM: Low-likelihood candidate member; CM: Candidate member, HM:High-likelihood candidate member; BF: Bona fide member.}
\tablenotetext{c}{Error-weighted combination of the two epochs.}
\tablenotetext{d}{Possible spectral binary.}
\tablecomments{See Section~\ref{sec:rv} for more details.}
\end{deluxetable*}

\subsubsection{2MASS~J10212570--2830427: A young L5\,$\beta$ dwarf}\label{sec:J1021}

\begin{figure}
	\centering
	\subfigure[2MASS~J09553336--0208403]{\includegraphics[width=0.488\textwidth]{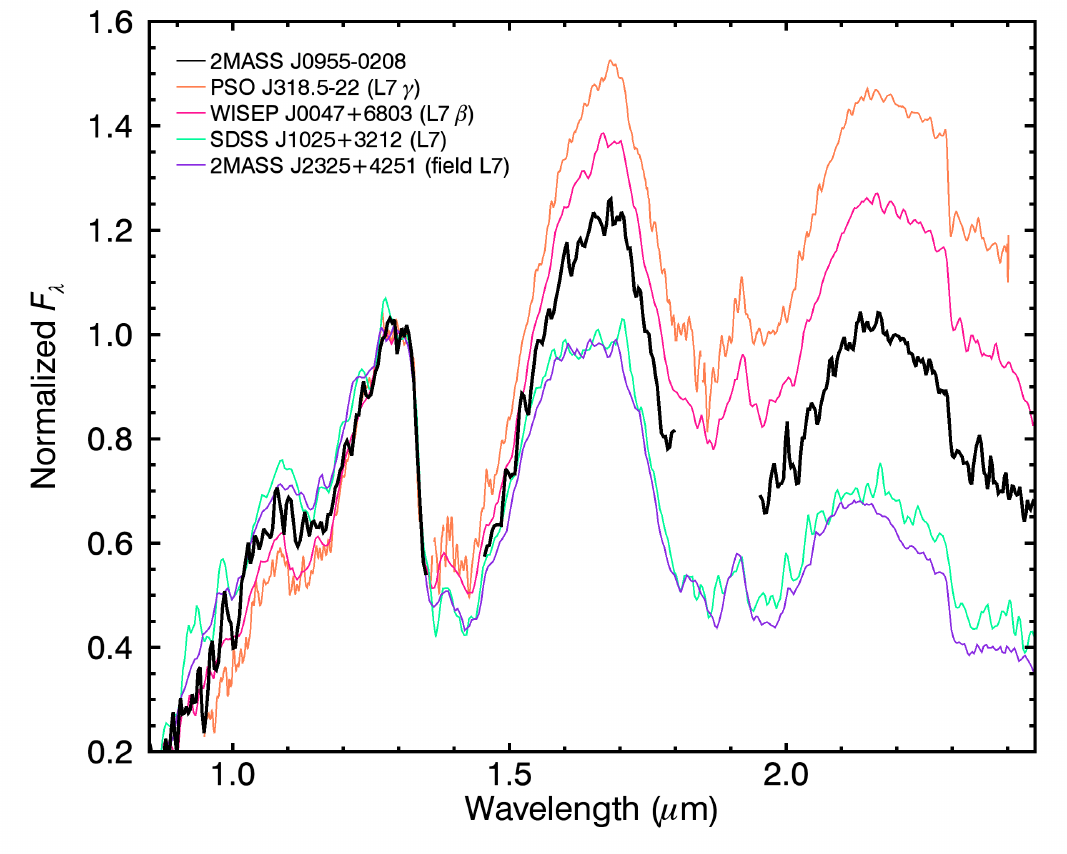}\label{fig:j0955}}
	\subfigure[2MASS~J10212570--2830427]{\includegraphics[width=0.488\textwidth]{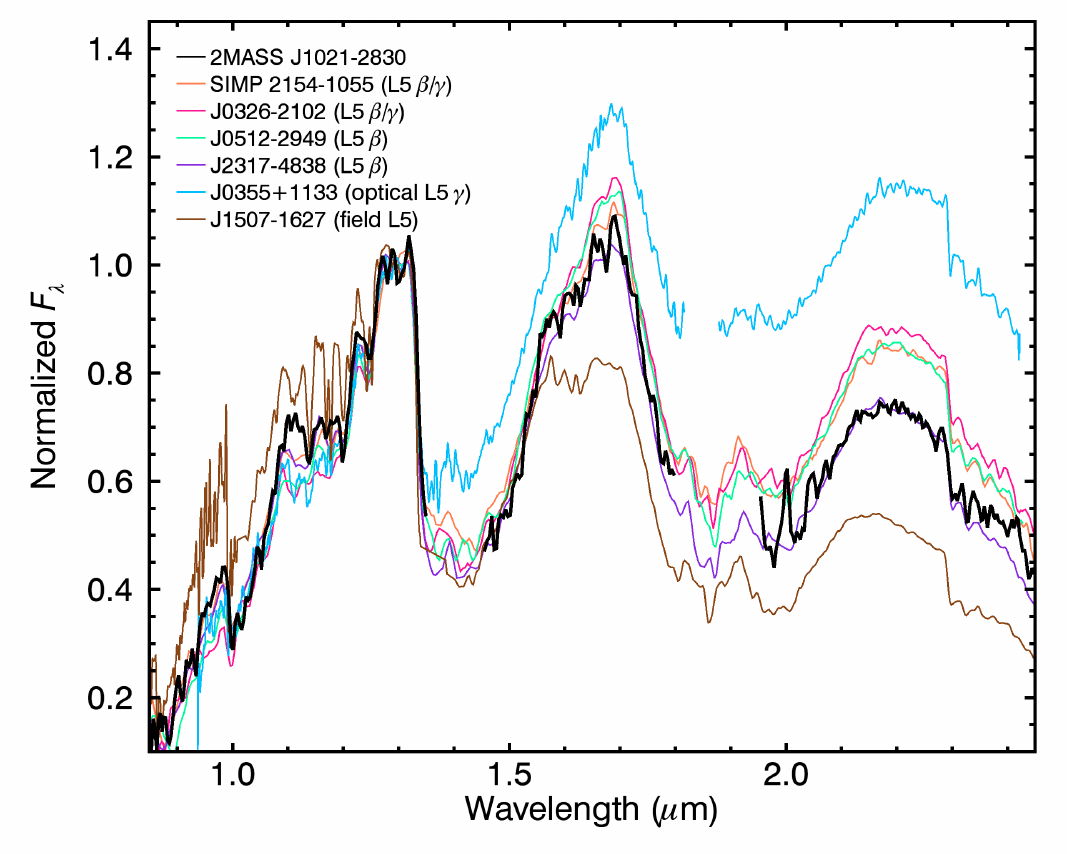}\label{fig:j1021}}
	\caption{NIR SpeX spectra of 2MASS~J09553336--0208403 and 2MASS~J10212570--2830427 compared with field and young substellar objects of the same spectral types. Both objects display a trinagular $H$-band continuum and red $J-K$ colors, which are telltale signs of a low surface gravity. All spectra were normalized by their median in the 1.27--1.33\,$\mu$m range. The full names of the reference spectra given in the legends are, from top to bottom; PSO~J318.5338--22.8603 \citep{2013ApJ...777L..20L}, WISEP~J004701.06+680352.1 \citep{2012AJ....144...94G}, SDSS~J102552.43+321234.0 \citep{2006AJ....131.2722C}, 2MASS~J23254530+4251488 \citep{2007AJ....133..439C}, SIMP~J21543454--1055308 \citep{2014ApJ...792L..17G}, 2MASS~J032642250--2102057 \citep{2003AJ....125.3302G}, 2MASS~J05120636--2949540 \citep{2003AJ....126.2421C}, 2MASS~J23174712--4838501 \citep{2008AJ....136.1290R}, 2MASS~J03552337+1133437 \citep{2008AJ....136.1290R}, and 2MASS~J150747690-1627386 \citep{2000AJ....119..369R}. See Section~\ref{sec:obs_spex} for more details.}
	\label{fig:j0955_1021}
\end{figure}

The low-resolution NIR spectrum of 2MASS~J10212570--2830427 (Figure~\ref{fig:j1021}) \replaced{shows}{displays} a triangular-shaped $H$-band continuum and a unusually red slope. Using the spectral templates of \cite{2015ApJS..219...33G} yields a spectral type of L5\,$\beta$. A higher-resolution spectrum will be necessary to confirm \replaced{if}{whether} these characteristics are clearly due to a young age.

\subsubsection{The Isolated Planetary-Mass Object 2MASS~J11472421--2040204}\label{sec:J1147}

As \replaced{pointed out}{mentioned} in Section~\ref{sec:buc}, 2MASS~J11472421--2040204 has been reported as a candidate member of TWA by \cite{2016ApJ...822L...1S}, using its sky position, proper motion, spectrophotometric distance and tentative indications of youth. The very red $J-K_S$ color ($2.57 \pm 0.03$) and triangular-shaped $H$-band continuum of 2MASS~J11472421--2040204 were used to determine that it is likely a young L7 substellar object. However, \replaced{other authors have pointed out that}{it has been discussed in the literature that} these characteristics could also be \added{potentially }caused by other effects, such as a high metallicity or unusual cloud thickness, \replaced{that is not due to}{without needing to invoke} a young age \citep{2013ApJ...772...79A,2014MNRAS.439..372M,2016AJ....151...46A}.

As one of the first few high-priority discoveries of the BASS-Ultracool survey, 2MASS~J11472421--2040204 (BASS-UC~56) was observed with FIRE in both the prism and echelle modes. The new FIRE-echelle spectrum allowed a radial velocity measurement that strengthened the TWA membership (see Section~\ref{sec:rv}), as well as a diagnosis of the surface gravity based on the strength of the \ion{K}{1} absorption lines at 1.168--1.179\,$\mu$m and 1.243--1.253\,$\mu$m. The relative strength of these absorption lines, along with the other characteristics mentioned above, can be used to safely determine whether 2MASS~J11472421--2040204 is a young L7 dwarf.

In Table~\ref{tab:J1147_K1}, the \ion{K}{1} equivalent widths as defined by \cite{2003ApJ...596..561M} are compared to those of other known young L7 dwarfs and to those of field L7 dwarfs \citep{2003ApJ...596..561M}. The weak \ion{K}{1} equivalent widths of 2MASS~J11472421--2040204 demonstrate that it has a low surface gravity, and is thus a young substellar object\added{, as suspected by \cite{2016ApJ...822L...1S}}.

Only a parallax measurement is \replaced{now}{still} needed before this object can be assigned \replaced{the status of a}{as a} bona fide member of TWA. Since its spectrophotometric distance matches its BANYAN~II kinematic distance\deleted{ very well }\citep{2016ApJ...822L...1S}, it is\deleted{ very }likely that this object is a\deleted{ real }member of TWA (see also \citealt{2016ApJS..225...10F}).

\begin{deluxetable}{lcccc}
\tabletypesize{\scriptsize}
\tablecolumns{5}
\tablecaption{$J$-Band \ion{K}{1} Equivalent Widths of L7-type Objects \label{tab:J1147_K1}}
\tablehead{\colhead{} & \multicolumn{4}{c}{EW(\ion{K}{1})\tablenotemark{a} (\AA)}\\
\cline{2-5}
\colhead{Name} & \colhead{1.169\,$\mu$m} & \colhead{1.177\,$\mu$m} & \colhead{1.243\,$\mu$m} & \colhead{1.254\,$\mu$m}}
\startdata
PSO~J318.5--22\tablenotemark{b} & $2 \pm 1$ & $3 \pm 1$ & $1.7 \pm 0.6$ & $1.3 \pm 0.7$\\
J1119--1137\tablenotemark{c} & $1.2 \pm 0.6$ & $3.9 \pm 0.6$ & $1.9 \pm 0.3$ & $3.1 \pm 0.3$\\
J1147--2040\tablenotemark{d} & $3.0 \pm 0.7$ & $4.4 \pm 0.7$ & $3.4 \pm 0.3$ & $3.0 \pm 0.3$\\
J0955--0208\tablenotemark{e} & $2.3 \pm 0.7$ & $4.6 \pm 0.7$ & $2.9 \pm 0.3$ & $3.7 \pm 0.3$\\
Field L7\tablenotemark{f} & 6.0--7.0 & 9.0--10.0 & 4.0--6.0 & 5.5--7.5\\
\enddata
\tablenotetext{a}{As defined by \cite{2003ApJ...596..561M}.}
\tablenotetext{b}{The complete PSO designation is PSO~J318.5338--22.8603; the FIRE spectrum of \cite{2016arXiv160507927F} was used.}
\tablenotetext{c}{The complete 2MASS designation is 2MASS~J11193254--1137466l; the FIRE spectrum of \cite{2016ApJ...821L..15K} was used.}
\tablenotetext{d}{The complete 2MASS designation is 2MASS~J11472421--2040204.}
\tablenotetext{e}{The complete 2MASS designation is 2MASS~J09553336--0208403.}
\tablenotetext{f}{Complete range of values for field L7 dwarfs obtained by \cite{2003ApJ...596..561M}}
\tablecomments{See Section~\ref{sec:J1147} for more details.}
\end{deluxetable}

\section{A COMPILATION OF THE TW~HYA MEMBERS AND CANDIDATES}\label{sec:candcomp}

In light of the new data presented in this work, an updated list of members and candidate members of TWA is compiled in this section. A discussion on the confusion between members of TWA and LCC is presented in Section~\ref{sec:lcc}. The final list of TWA objects is displayed in Tables~\ref{tab:kinematic} (kinematic properties), \ref{tab:spectro} (spectrophotometric properties) and \ref{tab:banyan} (final BANYAN~II membership probabilities). The final BANYAN~II probabilities listed in Table~\ref{tab:banyan} take into account positions, proper motions, radial velocities and/or trigonometric distances, when available. In addition to this, 2MASS and WISE photometry were used for all objects with spectral types $\geq$\,M5 to constrain the distance using two color-magnitude diagrams ($M_{W1}$ versus $J-K_S$, and $M_{W1}$ versus $H-W2$), as described by \cite{2014ApJ...783..121G}.

A proper motion was calculated from the 2MASS and AllWISE astrometr\replaced{y or}{ies for} all members and candidate members of TWA using the method of \cite{2015ApJ...798...73G}. The resulting measurements were adopted only when the proper motion was more accurate than those available in the literature. These measurements are reported in Table~\ref{tab:kinematic} along with all kinematic properties.

Objects that show all the necessary observational evidence for TWA membership (signs of youth, proper motion, radial velocity and trigonometric distance) and have Bayesian probabilities above 90\% are assigned the \emph{bona fide member} (BF) status (all have Bayesian membership probabilities above 98\%). Those that are missing only one of these measurements and have a Bayesian membership probability above 90\% are assigned the \emph{high-likelihood candidate member} (HM) status. Objects that are members of other groups, have properties that are inconsistent with the age of TWA, or have Bayesian membership probabilities below 1\% are rejected (R). All remaining objects are divided between \emph{candidate members} (CM; Bayesian probability $\geq$\,20\%) or \emph{low-likelihood candidate members} (LM; Bayesian probability $<$\,20\%). Two systems (TWA~6~AB and TWA~31) respect all of the observational criteria \replaced{of}{for} \emph{bona fide members}, but have a significantly lower Bayesian membership probability. These two objects are further discussed below.

In Table~\ref{tab:multiples}, a list of all known binaries in the present sample is reported, along with their projected separations. 2MASS~J12421948--3805064~B was discovered in this work as a visual binary in the GMOS follow-up, using the $i$-band acquisition image that\deleted{ immediately }preceded the spectral data acquisition. The two components were placed in the slit, which allowed for a separate data extraction and spectral typing of the two components.
\deleted{
A few objects that deserve further discussion are listed below.}

\textbf{TWA~6~AB} and \textbf{TWA~31} were identified as candidate members of TWA by \cite{2003ApJ...599..342S} and \cite{2011ApJ...727....6S}. At the present stage, they benefit from full kinematic measurements, but display slight discrepancies with other TWA members in $XYZ$ and $UVW$ spaces, resulting in relatively low Bayesian membership probabilities of 62.5\% and 68.4\%. These two objects are discussed in more detail\deleted{s} in Section~\ref{sec:lcc}.

\textbf{TWA~9~AB} has been identified as a candidate member of TWA by \cite{2003ApJ...599..342S}. \cite{2013ApJ...762..118W} and \cite{2014AA...563A.121D} noted that this object \replaced{seemed}{seems} slightly older ($38 \pm 18$\,Myr; \citealt{2014AA...563A.121D}) than other TWA members from a comparison with the \cite{1998AA...337..403B} and \cite{2000AA...358..593S} evolutionary models. \cite{2013ApJ...762..118W} used a trigonometric distance measurement to show that it is relatively discrepant to other TWA members in $UVW$ space. \cite{2013ApJS..208....9P} argued that all of these discrepancies were possibly explained by the Hipparcos distance if it is in error by $\approx$\,3$\sigma$, as adopting a distance of $\approx$\,70\,pc (rather than $46.8 \pm 5.4$\,pc) would solve both the isochronal age and kinematic discrepancy. A recent trigonometric distance measurement of $52.1 \pm 3.0$\,pc \citep{2014A&A...563A.121D} made this scenario seem unlikely\added{, however a more precise measurement from the \emph{Gaia} Data Release 1 \citep{2016arXiv160904303L} places TWA~9 at $75.7 \pm 1.7$\,pc, corroborating the hypothesis of \cite{2013ApJS..208....9P}. In this paper, we adopt the \emph{Gaia} distance measurement, which makes TWA~9 a bona fide member of TWA.}\deleted{ The BANYAN~II high membership probability of TWA~9 does not provide an additional argument for its membership to TWA, as it was included in the list of bona fide members of TWA that was used to build its spatial-kinematic model. This object will be further discussed in Section~\ref{sec:lcc} where membership to the LCC OB association is considered.}

\textbf{TWA~21} was identified \added{by \cite{2003ApJ...599..342S} }as a candidate member of TWA\deleted{by \cite{2003ApJ...599..342S}}. \cite{2014AA...563A.121D} noted that this object seemed slightly older ($25 \pm 7$\,Myr) than other TWA members from a comparison with the \cite{1998AA...337..403B} and \cite{2000AA...358..593S} evolutionary models. Using all available kinematic data from the literature, TWA~21 is instead an excellent match (98.2\%) to the Carina association ($45_{-7}^{+11}$\,Myr; \citealp{2008hsf2.book..757T,2015MNRAS.454..593B}), although its age seems slightly lower than that of the Carina association. A more detailed analysis of this object will be required to confirm whether it\deleted{ really }is a new bona fide member of Carina, however this scenario seems likely\added{ and it is therefore rejected from the census of TWA members and candidates}.

\textbf{TWA~22~AB} was identified as a candidate member of TWA by \cite{2003ApJ...599..342S}, a claim that was subsequently questioned by \cite{2005ApJ...634.1385M} using the method of convergent proper motion. \cite{2009AA...503..281T} used a trigonometric distance measurement to confirm that it is not a member of TWA, and suggested that it is rather a member of the young $\beta$~Pictoris moving group ($24 \pm 3$\,Myr; \citealp{2001ApJ...562L..87Z,2015MNRAS.454..593B}). This is consistent with the resulting BANYAN~II classification (99.7\% membership to $\beta$~Pictoris).

\begin{deluxetable*}{llllcccc}
\tablecolumns{8}
\tablecaption{Multiple Systems in TWA \label{tab:multiples}}
\tablehead{\colhead{2MASS} & \colhead{Other} & \colhead{Type\tablenotemark{a}} & \colhead{Host} & \colhead{$N_*$\tablenotemark{b}} & \colhead{$N_*$\tablenotemark{b}} & \colhead{Sep.\tablenotemark{c}} & \colhead{Ref.}\\
\colhead{Designation} & \colhead{Name} & \colhead{} & \colhead{Name} & \colhead{2MASS} & \colhead{AllWISE} & \colhead{($''$)} & \colhead{}}
\startdata
10023100--2814280~A & 1002--2814~A & P & $\cdots$ & 2 & 2 & $\cdots$ & $\cdots$\\
10023100--2814280~B & 1002--2814~B & C & 1002--2814~A & 2 & 2 & 0.56 & (1)\\
10120908--3124451~A & TWA~39~A & P & $\cdots$ & 2 & 2 & $\cdots$ & $\cdots$\\
10120908--3124451~B & TWA~39~B & C & TWA~39~A & 2 & 2 & 1.0 & (2)\\
10182870--3150029~A & TWA~6~A & P & $\cdots$ & 2 & 2 & $\cdots$ & $\cdots$\\
10182870--3150029~B & TWA~6~B & C & TWA~6~A & 2 & 2 & SB & (3)\\
11015191--3442170 & TWA~1 & P & $\cdots$ & 1 & 1 & $\cdots$ & $\cdots$\\
11020983--3430355 & TWA~28 & C & TWA~1 & 1 & 1 & 735.6 & (4)\\
11091380--3001398~A & TWA~2~A & P & $\cdots$ & 2 & 2 & $\cdots$ & $\cdots$\\
11091380--3001398~B & TWA~2~B & C & TWA~2~A & 2 & 2 & 0.56 & (3)\\
11102788--3731520~Aa & TWA~3~Aa & P & $\cdots$ & 3 & 3 & $\cdots$ & $\cdots$\\
11102788--3731520~Ab & TWA~3~Ab & C & TWA~3~Aa & 3 & 3 & SB & (5)\\
11102788--3731520~B & TWA~3~B & C & TWA~3~Aa & 3 & 3 & 1.4 & (6)\\
11211723--3446454~A & TWA~13~A & P & $\cdots$ & 2 & 2 & $\cdots$ & $\cdots$\\
11211723--3446454~B & TWA~13~B & C & TWA~13~A & 2 & 2 & 5 & (7)\\
11220530--2446393~Aa & TWA~4~Aa & P & $\cdots$ & 4 & 4 & $\cdots$ & $\cdots$\\
11220530--2446393~Ab & TWA~4~Ab & C & TWA~4~Aa & 4 & 4 & 0.2 & (8)\\
11220530--2446393~Ba & TWA~4~Ba & C & TWA~4~Aa & 4 & 4 & 0.8 & (9)\\
11220530--2446393~Bb & TWA~4~Bb & C & TWA~4~Ba & 4 & 4 & SB & (10)\\
11315526--3436272~Aa & TWA~5~Aa & P & $\cdots$ & 2 & 2 & $\cdots$ & $\cdots$\\
11315526--3436272~Ab & TWA~5~Ab & C & TWA~5~Aa & 2 & 2 & 0.066 & (11)\\
11315526--3436272~B & TWA~5~B & C & TWA~5~Aa & 1 & 1 & 2.0 & (12)\\
11324116--2652090 & TWA~8~B & C & TWA~8~A & 1 & 1 & 13.1 & (4)\\
11324124--2651559 & TWA~8~A & P & $\cdots$ & 1 & 1 & $\cdots$ & $\cdots$\\
11321822--3018316 & TWA~30~B & C & TWA~30~A & 1 & 1 & 80.2 & (13)\\
11321831--3019518 & TWA~30~A & P & $\cdots$ & 1 & 1 & $\cdots$ & $\cdots$\\
11392944--3725531 & HIP~56863~A & P & $\cdots$ & 1 & 2 & $\cdots$ & $\cdots$\\
11392960--3725538 & HIP~56863~B & C & HIP~56863~A & 1 & 2 & 3.34 & (14)\\
11482422--3728491 & TWA~9~A & P & $\cdots$ & 1 & 1 & $\cdots$ & $\cdots$\\
11482373--3728485 & TWA~9~B & C & TWA~9~A & 1 & 1 & 6.0 & (3)\\
12072738--3247002~Aa & TWA~23~A & P & $\cdots$ & 3 & 3 & $\cdots$ & $\cdots$\\
12072738--3247002~Ab & TWA~23~B & C & TWA~23~A & 3 & 3 & SB2 & (15)\\
12073346--3932539~A & TWA~27~A & P & $\cdots$ & 1 & 1 & $\cdots$ & $\cdots$\\
12073346--3932539~b & TWA~27~b & C & TWA~27~A & 1 & 1 & 0.78 & (17)\\
12074836--3900043 & TWA~40 & C & TWA~27~A & 1 & 1 & 1977.2 & (4)\\
12100648--4910505 & HIP~59315 & P & $\cdots$ & 1 & 1 & $\cdots$ & $\cdots$\\
12265135--3316124~A & TWA~32~A & P & $\cdots$ & 2 & 2 & $\cdots$ & $\cdots$\\
12265135--3316124~B & TWA~32~B & C & TWA~32~A & 2 & 2 & 0.656 & (15)\\
12313807--4558593~A & TWA~20~A & P & $\cdots$ & 2 & 2 & $\cdots$ & $\cdots$\\
12313807--4558593~B & TWA~20~B & C & TWA~20~A & 2 & 2 & SB & (18)\\
12345629--4538075~A & TWA~16~A & P & $\cdots$ & 2 & 2 & $\cdots$ & $\cdots$\\
12345629--4538075~B & TWA~16~B & C & TWA~16~A & 2 & 2 & 0.61 & (19)\\
12354893--3950245 & TWA~11~C & C & TWA~11~A & 1 & 1 & 174.9 & (4)\\
12360055--3952156 & TWA~11~B & C & TWA~11~A & 1 & 1 & 7.6 & (3)\\
12360103--3952102 & TWA~11~A & P & $\cdots$ & 1 & 1 & $\cdots$ & $\cdots$\\
12421948--3805064~A & 1242--3805~A & P & $\cdots$ & 2 & 2 & $\cdots$ & $\cdots$\\
12421948--3805064~B & 1242--3805~B & C & 1242--3805~A & 2 & 2 & 1.53 & (20)\\
\enddata
\tablenotetext{a}{Type of system component (P: Primary; C: Companion).}
\tablenotetext{b}{Number of known unresolved components in the 2MASS or AllWISE entry.}
\tablenotetext{c}{Separation to host component. SB indicates a spectroscopic binary for which the separation was not measured.}
\tablecomments{See Section~\ref{sec:candcomp} for more details.}
\tablerefs{(1)~\citealt{2012ApJ...754...44J}; (2)~\citealt{2014AJ....147...85R}; (3)~\citealt{1999ApJ...512L..63W}; (4)~2MASS catalog \citep{2006AJ....131.1163S}; (5)~\citealt{1997Sci...277...67K}; (6)~\citealt{2015AJ....149..145R}; (7)~\citealt{1972AA....18..341S}; (8)~\citealt{1999AstL...25..669T}; (9)~\citealt{2008hsf2.book..757T}; (10)~\citealt{2005ApJ...635..442B}; (11)~\citealt{2007AJ....133.2008K}; (12)~\citealt{2013ApJ...762..118W}; (13)~\citealt{2010AJ....140.1486L}; (14)~\citealt{2002AA...384..180F}; (15)~\citealt{2011ApJ...727....6S}; (16)~\citealt{Elliott:2016dd}; (17)~\citealt{2004AA...425L..29C}; (18)~\citealt{2006ApJ...648.1206J}; (19)~\citealt{2001ApJ...549L.233Z}; (20)~this paper.}
\end{deluxetable*}

\textbf{2MASS~J102825000-3959230} was identified as a candidate member of TWA as part of the PRE-BASS survey, which consists of objects that were selected and observed while the BASS survey \citep{2015ApJ...798...73G} selection criteria were still \replaced{evolving}{getting refined}, and was subsequently rejected (see \citealt{2015ApJS..219...33G} for more discussion on the PRE-BASS survey). A recent radial velocity measurement of $20 \pm 2$\,\kms\ \citep{2015MNRAS.453.2220M} rejects it as a possible member of TWA, making it instead a candidate member of Carina with a Bayesian membership probability of 51.4\%.

\subsection{The Confusion Between TW~Hya and the LCC Association}\label{sec:lcc}

Before a final list of TWA candidates and members can be properly constructed, it is necessary to address the possible confusion between TWA and the LCC OB association. This slightly older ($\sim$\,16\,Myr; \citealt{2002AJ....124.1670M}) and more distant association ($\sim$\,120\,pc\added{;} \citealt{1999AJ....117..354D}) is located in the same region \replaced{on}{of} the sky as TWA, which\deleted{ often }causes confusion between the members of \replaced{both}{the two} groups. Furthermore, the members of \replaced{the two}{both} associations have similar space velocities $UVW$.

It has been demonstrated, for example, that several objects that were \replaced{believed}{thought} to be members of TWA (TWA~12, 14, 15, 17, 18, 19 and 24) are probable members of LCC \citep{2001ASPC..244..104M,2005MNRAS.357.1399L,2005ApJ...634.1385M}.\deleted{In order to differentiate between the two groups, }\cite{2005ApJ...634.1385M} suggested a separation at a Solar distance of $\sim$\,85\,pc\added{ to differentiate between the two groups}, \replaced{which}{as this distance} delimitates two populations of stars with distinct compositions, ages and rotation periods \citep{2005MNRAS.357.1399L}.

A new version of the BANYAN~II tool, which will include a larger number of associations, is currently under construction. The final version of this tool, BANYAN~$\Sigma$, will be presented in a future publication, but at this stage it is already in a state where it can be used to flag likely LCC contaminants in our sample. All LCC members compiled by \cite{1999AJ....117..354D} and \cite{2008hsf2.book..235P} that have signs of youth, radial velocity and distance measurements were used to build a spatial-kinematic model that is similar to those of BANYAN~II (see \citealt{2014ApJ...783..121G}). The probabilities of this new tool are not yet calibrated to yield true contamination rates and the model of the field population is still incomplete, however it is already possible to calculate the Bayes factor between the LCC and TWA hypotheses.

Any TWA candidate in our sample that has an LCC/TWA Bayes factor above 1 is thus likely to be a contaminant from LCC. There are 44 such systems in our sample (including TWA~6 and TWA~31); they were flagged in Table~\ref{tab:banyan} and excluded from the sample for the remainder of this work. Only 9/44 of these objects would also have been rejected by the \cite{2005ApJ...634.1385M} criterion, 7/9 from their trigonometric distance and 2/35 from their TWA statistical distance. \replaced{The criterion of \cite{2005ApJ...634.1385M}}{A criterion based on only the distance separation between TWA and LCC} is therefore \replaced{only}{not} reliable \replaced{when a trigonometric distance is used}{to distinguish LCC contaminants when using a kinematic distance estimate instead of a trigonometric distance measurement}.

Objects that have been defined as high-likelihood candidates \replaced{of}{or} bona fide members but that have a non-negligible probability of being an LCC interloper (\replaced{we define the threshold}{the threshold is defined as} as an LCC/TWA Bayes factor above 0.1) were demoted to ambiguous candidate members (CM) until more information is available. This was the case for only \replaced{two}{one} object (not counting those with a LCC/TWA Bayes factor above 1):\deleted{ TWA~9 and }2MASS~J11152992--2954436\replaced{. Both have}{, which has an} LCC/TWA Bayes factor\added{ of} $\sim$\,$0.9$.\deleted{ This is consistent with the above discussion where TWA~9 was found as having slightly discrepant kinematics compared to TWA.}

\begin{figure}
	\centering
\includegraphics[width=0.488\textwidth]{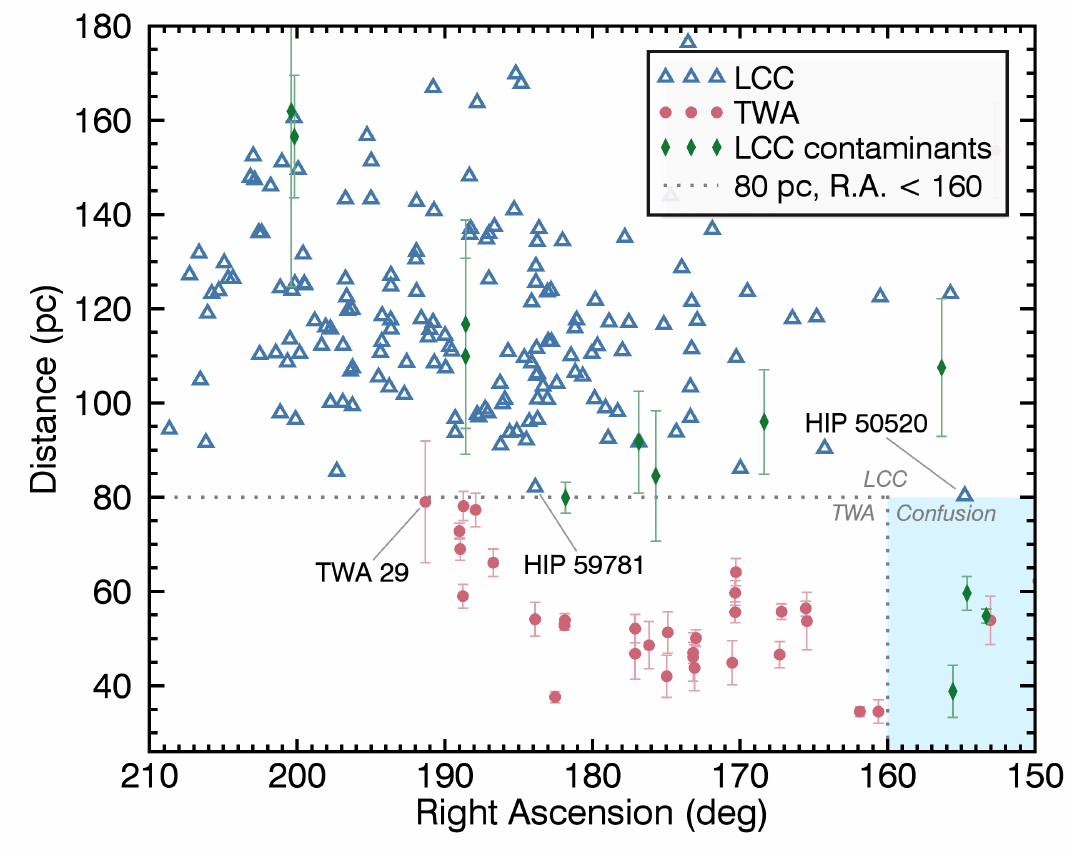}
	\caption{Distance as a function of right ascension for the current TWA \replaced{sample}{census} (red circles), and members of the LCC association (blue triangles). Likely LCC members that contaminated the current TWA sample before the application of BANYAN~$\Sigma$ are displayed with green diamonds. This figure is an update on Figure~3 of \cite{2005ApJ...634.1385M}. \cite{2005ApJ...634.1385M} suggest a distance threshold of 85\,pc to distinguish between members of LCC and TWA; this updated census seems to warrant a slight modification to a distance threshold of 80\,pc, which is\added{ only} valid at R.A.\,$>$\,160. It can be seen that East of this boundary (pale blue shaded region), the statistical distance cannot discriminate likely LCC contaminants from true TWA members. See Section~\ref{sec:lcc} for more details.}
	\label{fig:lcc}
\end{figure}

In Figure~\ref{fig:lcc}, an update on the Figure~3 of \cite{2005ApJ...634.1385M} is presented; two LCC members from \cite{1999AJ....117..354D} are located between 80--85\,pc; HIP~59781 and HIP~50520. Both objects have a 0\% BANYAN~II TWA membership probability. \deleted{It thus seems that a clear separation between TWA and LCC members lies at $\sim$\,80\,pc rather than $\sim$\,85\,pc.}\added{This figure also shows that three stars located at R.A.\,$<$\,10$^{\rm h}$40$^{\rm m}$ (R.A.\,$<$\,160\,\textdegree) and well within $\sim$\,85\,pc of the Sun are likely contaminants from LCC according to their LCC/TWA Bayes factor. It thus seems that a clear separation between TWA and LCC members lies at $\sim$\,80\,pc rather than $\sim$\,85\,pc, and that it is only valid for R.A.\,$>$\,160\,\textdegree; the members of LCC and TWA that are located East of this boundary cannot be discriminated with \replaced{this simple}{a simple distance} criterion.}

The TWA member that is closest to the $\sim$\,80\,pc boundary is TWA~29 at a distance of $79 \pm 13$\,pc \citep{2013ApJ...762..118W}; it would be useful to obtain a more precise distance measurement to clarify whether TWA~29 falls within a distance of 80\,pc. The \emph{Gaia} mission \citep{2001A&A...369..339P,Collaboration:2016cu} will likely answer this question, as TWA~29 is present in the Initial \emph{Gaia} Source List \citep{2013yCat.1324....0S} with a magnitude of $G = 16.8 \pm 0.4$\added{ (The \emph{Gaia} Data Release 1 does not provide its parallax measurement; \citealt{2016arXiv160904303L})}. Even when this large distance error bar is adopted, the LCC/TWA Bayes factor of TWA~29 remains very small at $\sim$\,3\%, hence it seems unlikely that it is a contaminant from LCC even if it were located above 80\,pc. \deleted{Despite the clear distance separation between TWA and LCC at most sky positions, it seems that the region at R.A.\,$<$\,10$^{\rm h}$40$^{\rm m}$ (160\textdegree) is subject to more confusion. It is therefore important to include a model of LCC in a membership analysis to avoid a high contamination in this region of the sky.}

\begin{figure}
	\centering
	\includegraphics[width=0.488\textwidth]{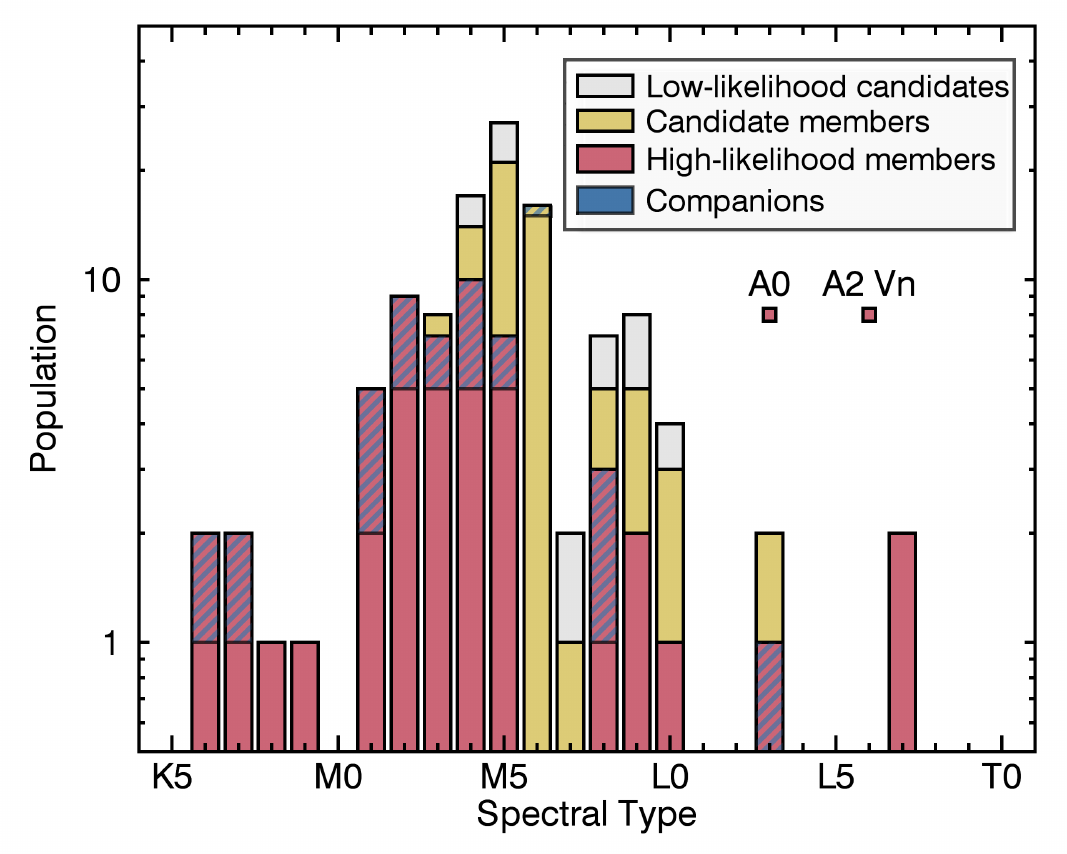}
	\caption{Spectral type histogram of members and candidate members of TWA. Optical spectral types are preferred when they are available. The bona fide members TWA~11~A and HIP~54477 with respective spectral types A0 and A2\,Vn are located outside of range, and are represented with square symbols. Known companions are represented with diagonal blue stripes, and are assumed to have the same spectral type as the primary star when the brightness ratio is close to unity. See Section~\ref{sec:candcomp} for more details.}
	\label{fig:spthist}
\end{figure}


\begin{figure*}[p]
	\centering
	\subfigure[TWA Spectral type--\teff\ relation]{\includegraphics[width=0.488\textwidth]{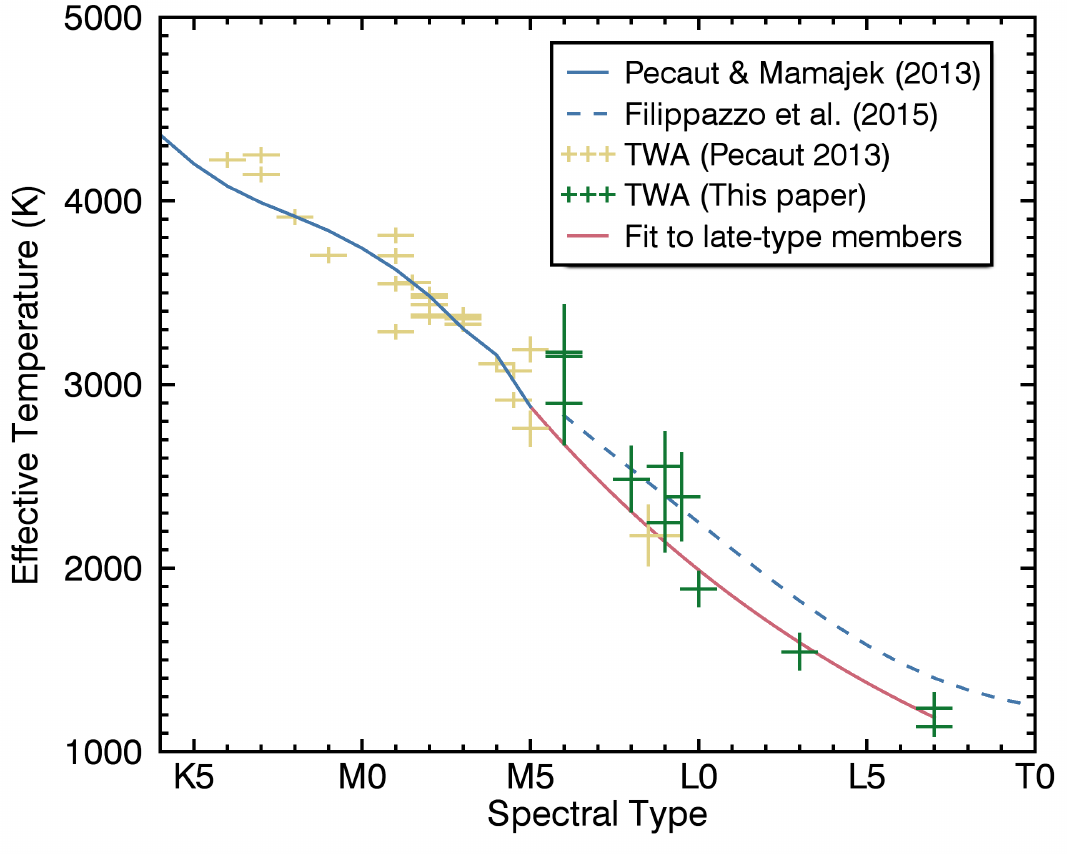}\label{fig:spttefftwa}}
	\subfigure[Model tracks at the age of TWA]{\includegraphics[width=0.488\textwidth]{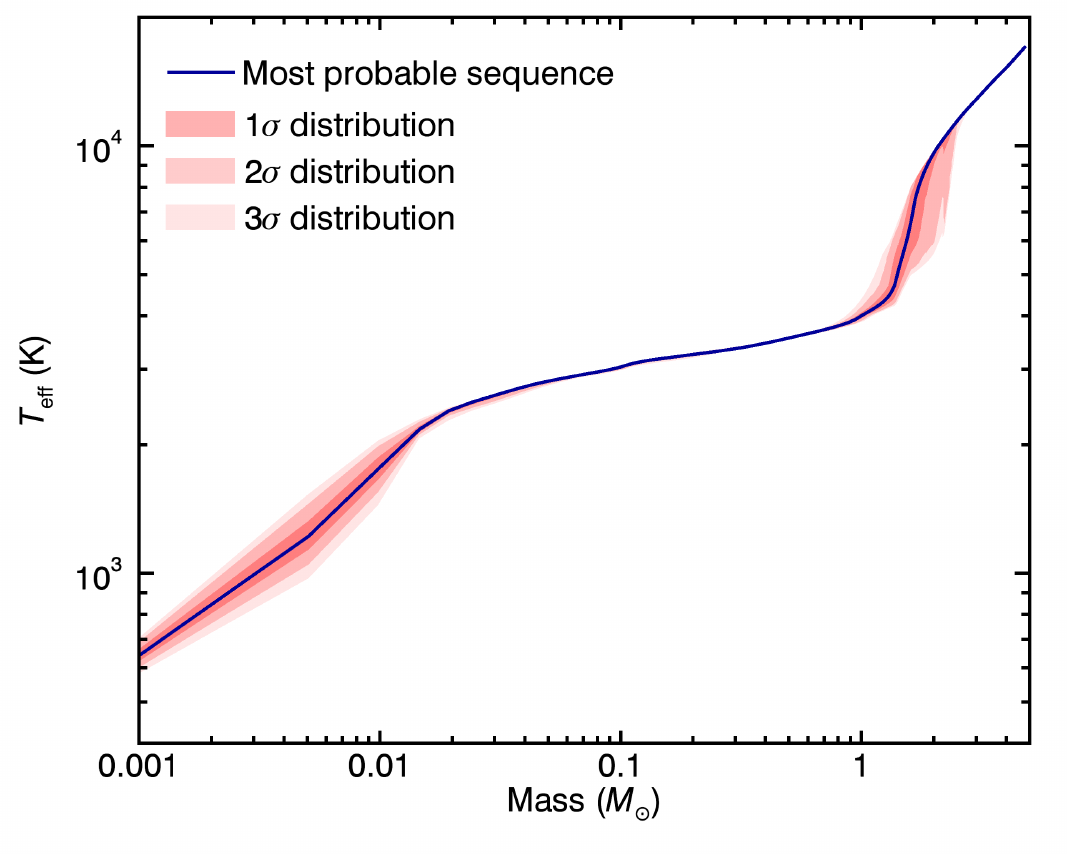}\label{fig:massteffmodels}}
	\subfigure[TWA Spectral type--mass relation]{\includegraphics[width=0.488\textwidth]{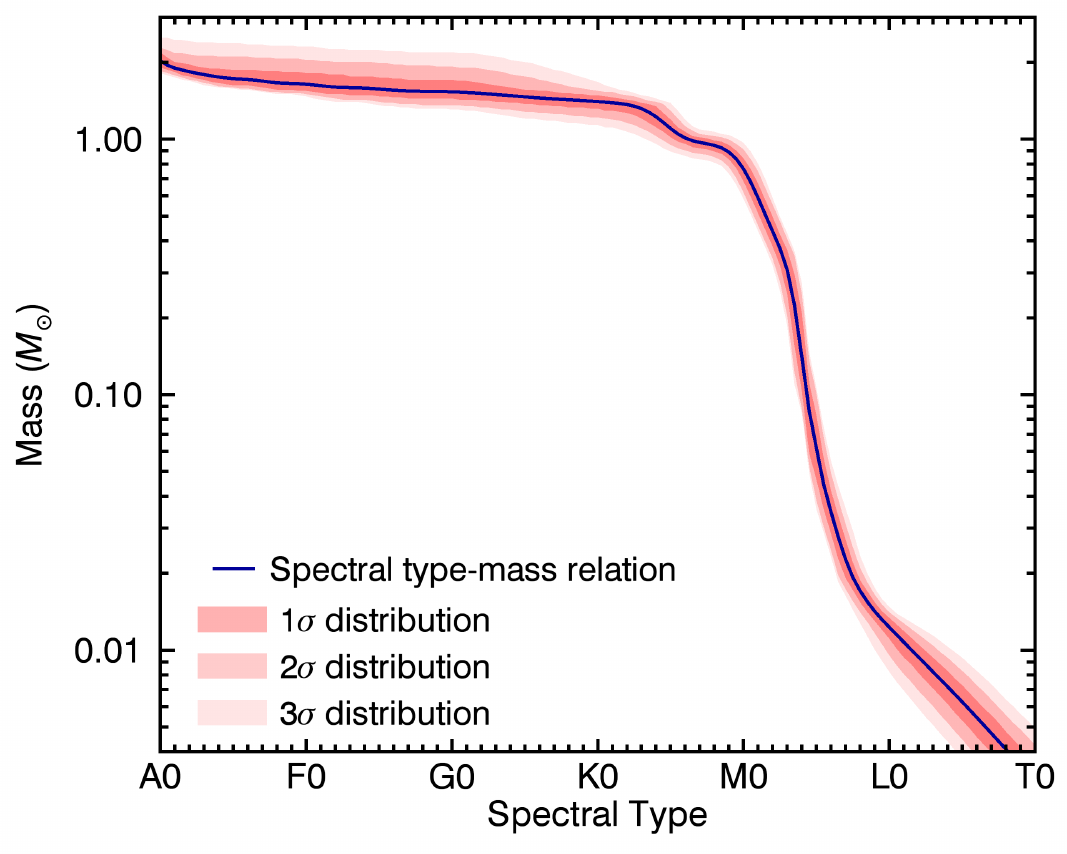}\label{fig:sptmasstwa}}
	\subfigure[Mass PDFs at the age of TWA]{\includegraphics[width=0.488\textwidth]{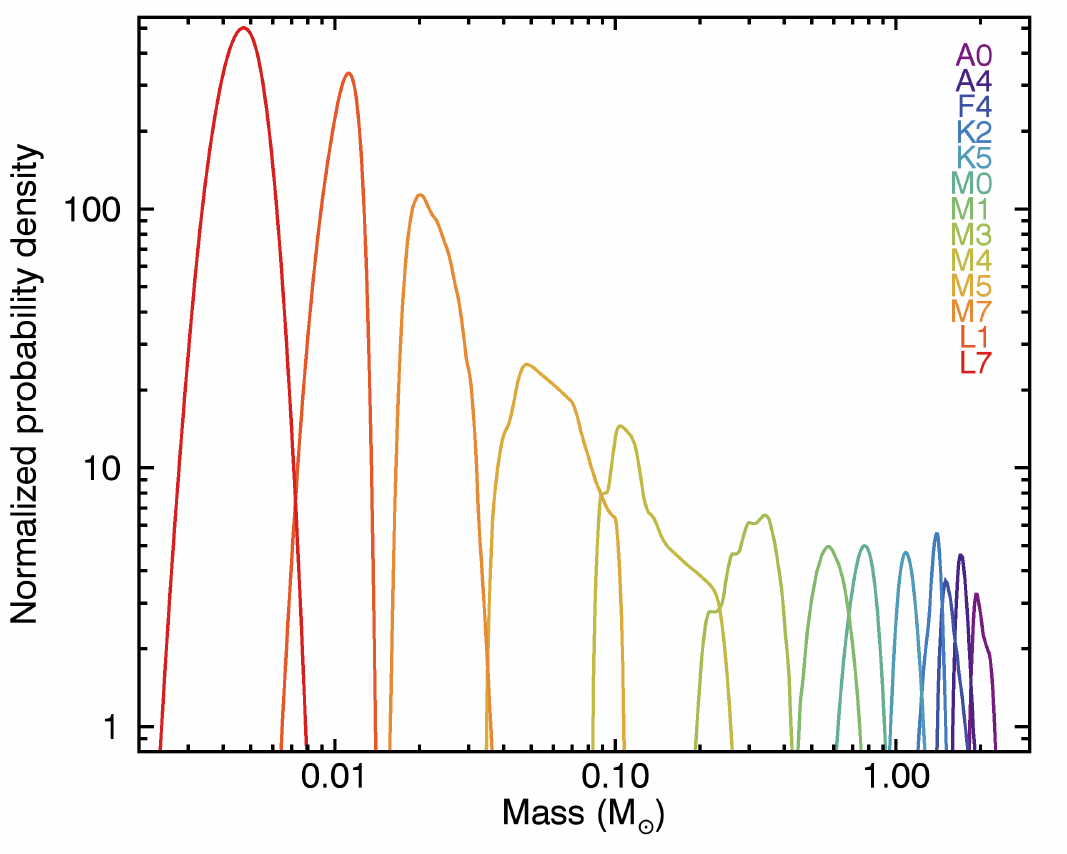}\label{fig:masspost}}
	\caption{Panel~a: Extension of the \cite{2013ApJS..208....9P} spectral type--\teff\ relation (blue line) to the $>$\,M5 regime (red line) using late-type members of TWA and the \teff\ determination method of \cite{2015ApJ...810..158F}. The dashed blue line is the spectral type--\teff\ relation of field stars calculated by \cite{2015ApJ...810..158F} at the age of the field. Young stars are expected to have a lower \teff\ at a fixed spectral type. It is possible that the sample of M6--M9 TWA stars is contaminated by unresolved binaries, which would explain that some objects lie above the young \teff\ sequence. \\
	Panel~b: Posterior PDFs for evolutionary model tracks obtained from the \citeauthor{2015A&A...577A..42B} (\citeyear{2015A&A...577A..42B}; $\leq$\,1.4\,\Msol) and \citeauthor{2000AA...358..593S} (\citeyear{2000AA...358..593S}; $>$\,1.4\,\Msol) models, at the age of TWA ($10 \pm 3$\,Myr; \citealp{2015MNRAS.454..593B}).\\
	Panel~c: Mass--\teff\ relations for TWA, obtained from a combination of the relations displayed in Panels~a and b.\\
	Panel~d: Mass posterior PDFs as a function of spectral type, obtained from the spectral type--temperature relation displayed in Panel~a and the model tracks presented in Panel~b. See Section~\ref{sec:physpar} for more details.}
	\label{fig:models_mult}
\end{figure*}

\subsection{Possible Common Proper-Motion Systems}\label{sec:cpm}

A few studies have uncovered potential common proper motion (CPM) systems in TWA. It has been suggested by \cite{2008AA...491..829K} that TWA~11~C and TWA~11~A are a CPM pair; by \cite{2005AA...430L..49S} that TWA~1 and TWA~28 are a possible CPM pair; by \cite{2010AJ....140.1486L} that TWA~30~A and TWA~30~B are a CPM pair; and \cite{Elliott:2016dd} have suggested 3 additional new potential CPM companions to known TWA members: 2MASS~J11130416-4516056 (TWA~14), 2MASS~J12073145-3310222 (TWA~23) and 2MASS~J12090628--3247453 (TWA~23). Since TWA~14 is a likely member of LCC (see Section~\ref{sec:lcc}), the former system is not discussed here. There is, however, a fundamental problem with the two latter CPM companions suggested by \cite{Elliott:2016dd}. Placing 2MASS~J12073145-3310222 at the distance of TWA~23 would make its absolute $K_S$-band magnitude ($7.35 \pm 0.06$) too faint for a K9-type dwarf by more than \replaced{an order of magnitude}{almost 4\,mag}. By comparison, the K9 bona fide member TWA~25 has an absolute $K_S$-band magnitude of $3.6 \pm 0.1$. Similarly, placing 2MASS~J12090628--3247453 at the distance of TWA~23 would mean that its absolute $K_S$-band magnitude is $8.32 \pm 0.06$, again much too faint for an M1 dwarf (the M1-type bona fide member TWA~13~A has an absolute $K_S$-band magnitude of $4.59 \pm 0.07$ after correcting for its binary nature). It is therefore very likely that both these objects are not related to TWA~23, or even to the TWA association. This is consistent with the fact that we measure an ESPaDOnS RV of $-9.0 \pm 0.2$\,\kms\ for 2MASS~J12073145-3310222, which safely rejects it from TWA.

In addition to these, \cite{2014ApJ...785L..14G} have discovered an isolated planetary-mass candidate member of TWA (2MASS~J12074836--3900043) that is located relatively close ($\sim$57$'$, corresponding to 181\,000\,AU at 52.8\,pc) to TWA~27 on the sky. While this fact has not been discussed in the discovery paper, it has \replaced{sparked}{triggered} discussions on the possibility that 2MASS~J12074836--3900043 is a CPM companion of TWA~27 (Niall Deacon, priv. comm. 2014).

Since the determination of false-positive probabilities of CPM discoveries in young associations requires a different approach than those in the field (e.g., see \citealt{2016MNRAS.457.3191D}), a determination of the false-positive probabilities of the possible CPM systems mentioned above is carried out in this section.

The development in this section aims at addressing the following question: \emph{What is the probability that any two members of TWA will have observables that are similar to a CPM system simply by chance ?}\deleted{ In order }To answer this, it is possible to model the distribution of the $XYZ$ \replaced{galactic}{Galactic} positions and $UVW$ space velocities of its members as a six-dimensional multivariate Gaussian distribution. If $\mathbf{Q}$ is a six-dimensional vector that contains the $XYZUVW$ coordinates\replaced{ of a TWA member}, the multivariate Gaussian distribution is obtained by calculating the covariance matrix $\mathbf{\Sigma}$ of the $XYZUVW$ positions of the members and their mean position $\mathbf{Q_0}$. The spatial density distribution of the TWA members is then described as:
\begin{align}
	P(\mathbf{Q}|\mathbf{Q_0},\mathbf{\Sigma}) &= N_{\mathrm{tot}}\frac{e^{-\frac{1}{2}\left(\mathbf{Q}-\mathbf{Q_0}\right)^T\mathbf{\Sigma}^{-1}\left(\mathbf{Q}-\mathbf{Q_0}\right)}}{\sqrt{\left(2\pi\right)^3\left|\mathbf{\Sigma}\right|}},\label{eqn:multivar}
\end{align}
\noindent where $N_{\mathrm{tot}}$ is the total number of members. Such a distribution is equivalent to the freely rotating Gaussian ellipsoids used in BANYAN~II \citep{2014ApJ...783..121G} if the spatial-dynamic terms of the covariance matrix are set to zero. The covariance matrix $\mathbf{\Sigma}$ and mean value $\mathbf{Q_0}$ of the $XYZUVW$ coordinates were calculated for all bona fide and high-likelihood members of TWA. Only primaries \replaced{are}{were} used in this calculation, to avoid introducing artificial biases in the structure of TWA. Kinematic distances or radial velocities were used when the\deleted{se} measurements were not available.

A set of $10^5$ artificial associations were then randomly drawn from the multivariate gaussian distribution of TWA described in Equation~\eqref{eqn:multivar}. These artificial associations were created with the same number (34) \replaced{or}{of} bona fide and high-likelihood systems as the current TWA census. The fraction of simulations that had any two members with proper motions, radial velocities, distances and sky positions \replaced{comparable to a}{at least as close together as a given} common proper motion system was then counted. 

Similar simulations were carried out for each potential CPM pair, the results of which are presented in Table~\ref{tab:cpm}. It can be seen that most of the proposed potential CPM pairs have relatively low probabilities of being located this close by pure chance, given the spatial structure of TWA. 2MASS~J12074836--3900043 has the largest probability (0.7\%) of a chance alignment. Obtaining a precise trigonometric distance would be helpful to determine whether it is a common proper motion companion of TWA~27, however it will be treated conservatively as an isolated object in the remainder of this work.

\begin{deluxetable}{llccc}
\tabletypesize{\scriptsize}
\tablecolumns{5}
\tablecaption{Potential Common Proper Motion Objects\label{tab:cpm}}
\tablehead{\colhead{Companion} & \colhead{Primary} & \colhead{Angular} & \colhead{Physical} & \colhead{$P_A$\tablenotemark{a}}\\
\colhead{Name} & \colhead{Name} & \colhead{Sep. ($''$)} & \colhead{Sep. (AU)} & \colhead{(\%)}}
\startdata
TWA~30~B & TWA~30~A & 80.2 & $3\,500 \pm 400$ & $< 10^{-6}$\\
TWA~11~C & TWA~11~A & 174.9 & $14\,400 \pm 300$ & $9\cdot 10^{-4}$\\
TWA~28 & TWA~1 & 735.6 & $160\,000 \pm 20\,000$ & $8\cdot 10^{-3}$ \\
J1207--3900\tablenotemark{d} & TWA~27 & 1977.2 & $181\,000 \pm 3\,000$ & 0.7\\
\enddata
\tablenotetext{a}{Chance alignment probability, assuming membership to TWA.}
\tablenotetext{b}{The complete 2MASS designation is 2MASS~J12073145--3310222.}
\tablenotetext{c}{The complete 2MASS designation is 2MASS~J12090628--3247453.}
\tablenotetext{d}{The complete 2MASS designation is 2MASS~J12074836--3900043.}
\tablecomments{See Section~\ref{sec:cpm} for more details.}
\end{deluxetable}

\subsection{An Update on TWA Names}\label{sec:twanames}

In light of the revisions to the list of TWA candidates and members presented here, it will be useful to assign new TWA names to the list that is currently available in the literature. The TWA numbers up to TWA~33 have been defined without ambiguity, however TWA~34 and TWA~35 have each been assigned to two distinct objects: TWA~34 stands for 2MASS~J12520989--4948280 \citep{2011PhDT.......245L} or  2MASS~J10284580--2830374 \citep{2015MNRAS.453.2220M}, whereas TWA~35 stands for 2MASS~J13265348-5022270 \citep{2011PhDT.......245L} or 2MASS~J12002750--3405371 \citep{2015MNRAS.453.2220M}. Here, the definitions of \cite{2015MNRAS.453.2220M} are adopted since it was the first refereed work to use both names. Furthermore, 2MASS~J12520989--4948280 was found here to be a likely LCC contaminant, and 2MASS~J13265348--5022270 has a 0\% BANYAN~II TWA membership probability; both objects were thus rejected from the current TWA sample. The names TWA~36 through 38 have been defined by \cite{2011PhDT.......245L} without ambiguity in the literature and refer to objects that are still credible candidate members; they were\added{ thus} adopted in this work. To our knowledge, no TWA name with a number above TWA~38 has been defined yet\added{ (with the exceptions of TWA~45 and TWA~46 discussed below)}.

There are seven high-likelihood candidates and bona fide members of TWA that currently do not have a TWA name. We therefore suggest assigning the names TWA~39~AB \replaced{for}{to} SCR 1012--3124 AB (2MASS~J10120908--3124451~AB; \citealt{2014AJ....147...85R}), TWA~40 \replaced{for}{to} 2MASS~J12074836--3900043 \citep{2014ApJ...785L..14G}, TWA~41 \replaced{for}{to} 2MASS~J11472421--2040204 \citep{2016ApJ...822L...1S}, TWA~42 \replaced{for}{to} 2MASS~J11193254--1137466 \citep{2016ApJ...821L..15K}, TWA~43 \replaced{for}{to} HIP~54477 (Section~\ref{sec:hip}), and TWA~44 \replaced{for}{to} 2MASS~J12175920--3734433 (Section~\ref{sec:rv2}).

\added{Two additional TWA names have been introduced for new candidate members of TWA reported by \cite{2016arXiv161001667D}: TWA~45 for 2MASS~J11592786--4510192 and TWA~46 for 2MASS~J12354615--4115531. One last high-likelihood candidate member of TWA was discovered by \cite{2016arXiv161003867R} while this paper was in review; we therefore assign the name TWA~47 to SCR~1237--4021 (2MASS~J12371238--4021480; \citealt{2016arXiv161003867R}).}

The distribution of spectral types \replaced{of}{for} the final TWA candidates and members is presented in Figure~\ref{fig:spthist}\added{.}\deleted{, and their positions in a $J-K$ versus absolute $J$-band color-magnitude diagram are displayed in Figure~\ref{fig:jjk}.}\added{ This figure demonstrates that there are only two massive ($<$\,K-type) members of TWA, and that all of its M6--M7 candidate members still remain to be confirmed as bona fide members. Furthermore, it is likely that several K-type members are still missing due to the lack of an all-sky survey targeting such members of TWA.}

\section{ESTIMATION OF PHYSICAL PARAMETERS}\label{sec:physpar}

\replaced{Before mass estimates can be obtained for candidate members of TWA,}{In order to obtain a mass estimate for the candidate members of TWA,} either their absolute magnitudes or effective temperatures must first be \replaced{obtained}{measured}, and then \replaced{transformed}{translated} to masses using evolutionary models. Since the multiplicity rate of TWA members is high, we choose to rely on effective temperature measurements over \replaced{spectral types}{absolute magnitudes} such that the effect of unknown binaries\deleted{ or bright companions }is minimized.

\subsection{Effective Temperatures of Early-Type Members}\label{sec:earlyteff}

The spectral types of all objects \replaced{earlier than M6}{with spectral types M5 or earlier} are translated to an effective temperature using the relations for young stars developed by \cite{2013ApJS..208....9P}. This is done by drawing a set of $10^7$ Gaussian random numbers distributed around each spectral type (error bars of 0.5 subtypes were assumed), and interpolating them into a set of temperatures with the aforementioned spectral type--\teff\ relations.

\begin{deluxetable*}{llccccc}
\tablecolumns{7}
\tablecaption{Empirical Bolometric Luminosity and Semi-Empirical \teff\ Measurements \label{tab:teffs}}
\tablehead{\colhead{2MASS} & \colhead{TWA} & \colhead{Spectral} & \colhead{Data} & \colhead{$\log _{10}\left(L_*/L_\odot\right)$} & \colhead{Radius}& \colhead{\teff}\\
\colhead{Designation} & \colhead{Name} & \colhead{Type} & \colhead{Used\tablenotemark{a}} & \colhead{} & \colhead{($R_\mathrm{Jup}$)} & \colhead{(K)}}
\startdata
\sidehead{\textbf{Bona Fide Members and High-Likelihood Candidate Members of TWA}}
12265135--3316124~A & TWA~32~A & M5 & di & $-1.71 \pm 0.04$ & $4.5 \pm 0.3$ & $3200 \pm 100$\\
12073346--3932539~A & TWA~27~A & M8\,pec & doi & $-2.59 \pm 0.02$ & $2.4 \pm 0.1$ & $2640 \pm 70$\\
11020983--3430355 & TWA~28 & M8.5\,$\gamma$ & di & $-2.54 \pm 0.03$ & $2.4 \pm 0.1$ & $2680 \pm 80$\\
11395113--3159214 & TWA~26 & M9\,$\gamma$ & doi & $-2.71 \pm 0.09$ & $2.2 \pm 0.2$ & $2600 \pm 200$\\
12451416--4429077 & TWA~29 & M9.5\,$\gamma$ & doi & $-2.9 \pm 0.1$ & $2.0 \pm 0.2$ & $2400 \pm 200$\\
12074836--3900043 & TWA~40 & L0\,$\gamma$ & oi & $-3.47 \pm 0.08$ & $1.67 \pm 0.04$ & $1890 \pm 90$\\
12073346--3932539~b & TWA~27~b & L3\,$\gamma$\,pec & di & $-3.59 \pm 0.02$ & $1.62 \pm 0.01$ & $1800 \pm 20$\\
11472421--2040204 & TWA~41 & L7\,pec(red) & i & $-4.51 \pm 0.07$ & $1.40 \pm 0.03$ & $1140 \pm 50$\\
11193254--1137466 & TWA~42 & L7\,pec(red) & i & $-4.3 \pm 0.1$ & $1.44 \pm 0.05$ & $1240 \pm 80$\\
\sidehead{\textbf{Candidate Members of TWA}}
10284580--2830374 & TWA~34 & M5\,$\gamma$e & oi & $-1.9 \pm 0.1$ & $4.0 \pm 0.5$ & $3100 \pm 300$\\
12175920--3734433 & TWA~44 & M5\,$\gamma$e & o & $-1.9 \pm 0.1$ & $3.9 \pm 0.4$ & $3100 \pm 200$\\
10585054--2346206 & $\cdots$ & M6\,$\gamma$e & o & $-1.7 \pm 0.1$ & $4.5 \pm 0.5$ & $3200 \pm 300$\\
11112820--2655027 & TWA~37 & M6\,$\gamma$e & oi & $-1.7 \pm 0.1$ & $4.4 \pm 0.5$ & $3200 \pm 300$\\
12574941--4111373 & $\cdots$ & M6\,$\gamma$ & i & $-2.2 \pm 0.1$ & $3.0 \pm 0.3$ & $2900 \pm 200$\\
12035905--3821402 & TWA~38 & M8\,$\gamma$ & i & $-2.80 \pm 0.09$ & $2.1 \pm 0.2$ & $2500 \pm 200$\\
11064461--3715115 & $\cdots$ & M9 & oi & $-3.1 \pm 0.1$ & $1.8 \pm 0.1$ & $2300 \pm 200$\\
11480096--2836488 & $\cdots$ & L3\,$\beta$ & i & $-3.9 \pm 0.1$ & $1.56 \pm 0.03$ & $1500 \pm 100$\\
\enddata
\tablenotetext{a}{Data used to derive the measurements presented in this table. d: trigonometric distance (otherwise kinematic distance assuming membership to TWA is used); i: NIR spectrum and photometry, o: optical spectrum and photometry.}
\tablecomments{See Section~\ref{sec:physpar} for more details.}
\end{deluxetable*}

\subsection{Effective Temperatures of Late-Type Members}\label{sec:lateteff}

In the case of later-type ($\gtrsim$\,M6) targets, estimating effective temperatures is more delicate, especially because there are relatively large deviations in spectral types ($\geq$\,1--2 subtypes) depending upon what wavelength regime and/or spectral typing method is used. This is especially true for young substellar objects, in part because too few of them are known, which has yet prevented the construction of definite and reliable spectral standards.

For this reason, \cite{2015ApJ...810..158F} developed an empirical method to estimate the bolometric luminosity of an object using all available spectrophotometric data. A model-dependent radius must then be derived from this bolometric luminosity measurement, assuming an age of $10 \pm 3$\,Myr for all TWA objects \citep{2015MNRAS.454..593B}. The inferred radius and empirical bolometric luminosity are then translated to an effective temperature using the Stefan-Boltzmann law. One major advantage of this method is that the result is weakly dependent on model systematics, as \teff\ depends on the square root of the assumed radius, and as the radii of $\sim$\,10\,Myr-old substellar objects only span a factor of $\sim$\,2 depending on the mass of the object.

The method of \cite{2015ApJ...810..158F} was therefore used to estimate the effective temperature of targets in the TWA sample for which resolved NIR and/or optical photometry and spectroscopy were available. The resulting bolometric luminosities, \teff\ and radii measurements are presented in Table~\ref{tab:teffs}. \replaced{Some of these measurements are slightly different (but within 1$\sigma$) compared to those listed by \cite{2016ApJS..225...10F} and \cite{2015ApJ...810..158F}}{All measurements in common with those presented by \cite{2016ApJS..225...10F} and \cite{2015ApJ...810..158F} agree within 1.3$\sigma$}. The differences are due to slightly different distance estimates (either because a different kinematic tool was used or because no radial velocity was available at the time), or because different spectra were used.

The resulting \teff\ estimates were assumed to be described by Gaussian PDFs. A linear power law was then adjusted to the derived \teff\ values as a function of spectral type, which yielded:
\begin{align}
	T_{\rm eff} = 10^{3.46-0.0321\,\left(x-5\right)},\label{eqn:teff}
\end{align}
\noindent where $x$ is the numerical spectral type (e.g., M6 = 6, L0 = 10). The resulting spectral type--\teff\ relation is displayed in Figure~\ref{fig:spttefftwa} and is associated with a temperature scatter of 140\,K. Note that this relation should only be used within the spectral type range M5--L7. It is likely that the slope of the spectral type--\teff\ relation will flatten at the L/T transition, however this remains to be demonstrated at such a young age.

\deleted{In order }To avoid the aforementioned problems (unknown unresolved binaries are of special concern given the high binary fraction in TWA), the spectral type--\teff\ relations of \citeauthor{2013ApJS..208....9P} (\citeyear{2013ApJS..208....9P}; $\leq$\,M5, see Section~\ref{sec:earlyteff}) and Equation~(\ref{eqn:teff}; $>$\,M5) are used in the remainder of this work to transform gaussian spectral type PDFs into temperature PDFs.

\subsection{Mass Estimates}\label{sec:mass}

The \cite{2015A&A...577A..42B} and \cite{2000AA...358..593S} model masses were separately interpolated on a regular $10^4\times 10^4$ logarithmic grid of \teff\ and ages that span 400--17\,000\,K and 1\,Myr--12\,Gyr, respectively. The two grids were combined by adopting the \cite{2015A&A...577A..42B} masses below $1.4$\,\Msol\ and the \cite{2000AA...358..593S} models otherwise. The \cite{2000AA...358..593S} models over-predict \teff\ by up to 500\,K ($\sim$\,14\%) compared to \cite{2015A&A...577A..42B} in the mass range where both models are available (0.1--1.4\,\Msol); the \cite{2015A&A...577A..42B} models were adopted in this range as they include a more thorough treatment of convection, which is particularly important in this range of masses and age. The resulting mass--\teff\ relations that are obtained from a random draw along a log-normal distribution at the age of TWA ($10 \pm 3$\,Myr; \citealt{2015MNRAS.454..593B}) are displayed in Figure~\ref{fig:massteffmodels}. A combination of these model tracks with the spectral type--\teff\ relations derived above allows calculating mass--SpT tracks at the age of TWA; these are displayed in Figure~\ref{fig:sptmasstwa}.

The model grid described above was used to transform the $10^7$ Monte Carlo \teff\ determinations of each target to a mass PDF, while assuming the same log-normal age prior as above. The resulting mass PDF functions are displayed for a selection of spectral types spanning A0 to L7 in Figure~\ref{fig:masspost}.

\section{THE COMPLETENESS OF THE CURRENT TWA CENSUS}\label{sec:twa_completeness}

\added{Measuring an accurate IMF for TWA requires computing the completeness as a function of mass for the census of TWA members. It is however not possible to determine it given current data; the set of TWA candidates and members presented here originates from 39 distinct surveys that are based on different selection criteria, several of which are still not completed or do not provide enough information to determine the survey completeness or the overlap between different surveys. For this reason, the IMF that is determined in Section~\ref{sec:imf} should be taken as preliminary until a single all-sky survey with a well defined completeness as a function of mass is carried out.}

\added{The \emph{Gaia} Data Release 1 \citep{2016arXiv160904303L} does not allow to build a sample of TWA candidate members for which a completeness as a function of mass can be determined, because of several systematic effects and the heterogeneous nature of the sample \citep{2016arXiv160904172G}. The full release of the \emph{Gaia} mission will however provide a unique opportunity to assemble a well behaved sample of TWA candidates and members and re-visit the IMF of TWA, while taking completeness into account. The full release will include the full astrometric solution of 33/40 of the bona fide members and high-likelihood candidate members of TWA compiled in Section~\ref{sec:candcomp}, as well as 35/44 of the candidate members and 12/16 of the low-likelihood candidate members. It will also likely uncover additional low-mass members of TWA.}

\added{In the low-mass regime, all current surveys for TWA members are at best limited by the 2MASS sensitivity. This is true even though the AllWISE survey is more sensitive to substellar objects than 2MASS, because proper motions derived from the WISE mission data alone are not precise enough to identify new TWA candidate members without relying on 2MASS (the typical precision is of 100--3000\,\masyr\ for $W1$\,$\sim$\,13--18; \citealt{2014ApJ...783..122K}). For this reason, the completeness limit of TWA candidates imposed by 2MASS is informative as it provides the best-case scenario for the completeness of low-mass TWA members. This completeness is determined in Section~\ref{sec:tmass_completeness}.}

\added{In Section~\ref{sec:hip_completeness}, the completeness of the Hipparcos survey that was presented in Section~\ref{sec:hip} will be examined to determine a mass regime within which completeness is constant as a function of mass.}

\subsection{The 2MASS Completeness of Low-Mass TWA Members}\label{sec:tmass_completeness}

Although the census of 2MASS-detectable members of TWA is \deleted{clearly }still\added{ likely} incomplete, it is possible to use the spatial distribution of TWA members\added{,} combined with the sensitivity limits of 2MASS\added{,} to derive the maximum \replaced{completion fraction}{completeness} of low-mass members that due to the 2MASS sensitivity limit.

This is especially important for TWA because its members span a wide range of distances ($\sim$\,25--75\,pc). Such a correction will be most important for very low-mass and cool members with spectral types in the L spectral class, as will be demonstrated \replaced{in this section}{here}. Such cool objects have very red $J-K_S$ colors at young ages, which makes them more easily detected in the $K_S$ band, despite the shallower $K_S$-band limiting 2MASS magnitude compared to $J$ and $H$. For this reason, the determination of the \replaced{completion}{completeness} fraction carried out in this section will be based on the published sensitivity limits of 2MASS in the $K_S$ band.

\begin{figure}
	\centering
	\includegraphics[width=0.48\textwidth]{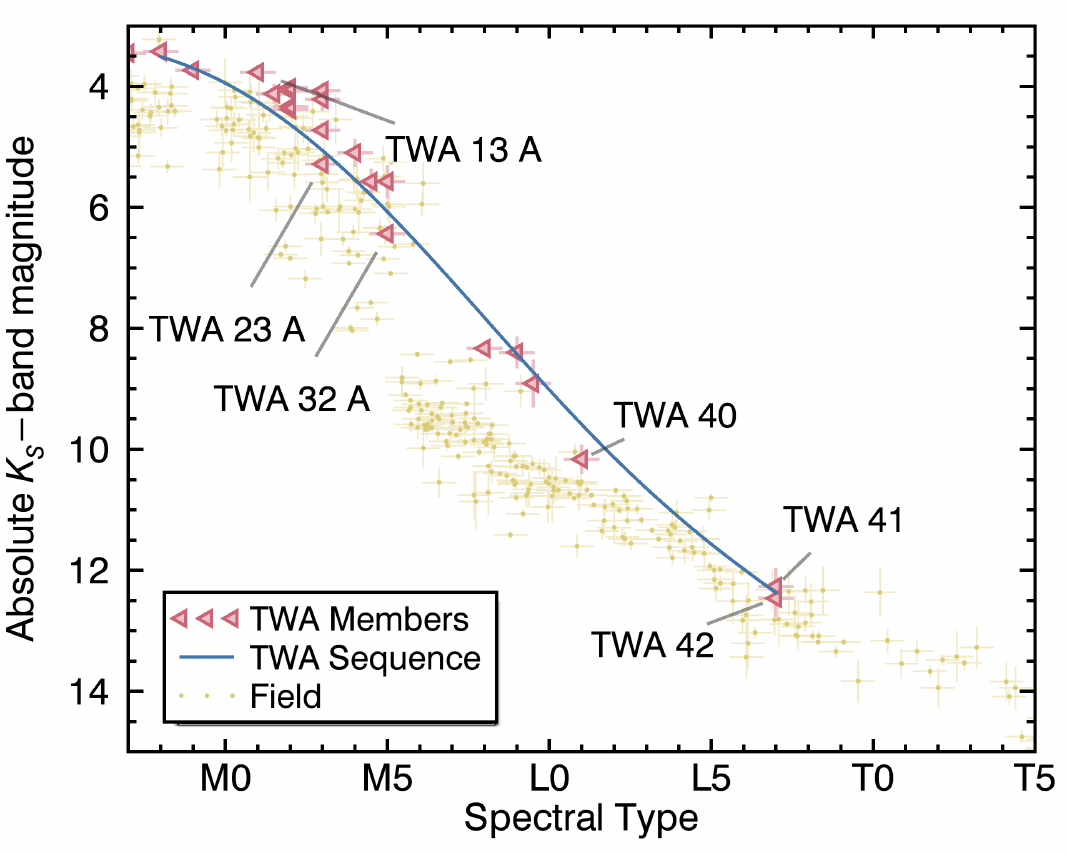}
	\caption{2MASS $K_S$-band spectral type--absolute magnitude sequence (blue line) of high-likelihood and bona fide members of TWA (leftwards red triangles), corrected for unresolved companions. Field objects are represented with small yellow dots. See Section~\ref{sec:tmass_completeness} for more details.}
	\label{fig:kabs}
\end{figure}

A photometric sequence in the $K_S$ band must first be constructed for TWA members\deleted{ in order }to perform this analysis. Such a sequence of absolute $K_S$-band magnitude as a function of spectral type is presented in Figure~\ref{fig:kabs}, using only high-likelihood candidates and bona fide members. Absolute magnitudes were corrected by adding a factor $2.5 \log _{10} N$ for $N$-components equal-luminosity multiples. Equal luminosities were assumed when individual magnitude measurements were not available.

\deleted{In order to derive a smooth spectral type--absolute $K_S$ magnitude sequence, the}\added{The} magnitudes of TWA objects were interpolated on a regular grid in spectral type \added{to obtain a smooth spectral type--absolute $K_S$ magnitude sequence}, to which a 7-order polynomial was fitted.

The $K_S$-band \replaced{completion}{completeness} fraction curve $f_C(K_S)$ of 2MASS\footnote{Available at \url{http://www.ipac.caltech.edu/2mass/releases/allsky/doc/sec6\_5a1.html}, Figure~5} was interpolated on a two-dimensional map $\mathcal{K}_{i,j}$ of apparent $K_S$-band magnitudes, where the two dimensions correspond to a grid of distances $\varpi_i$ (500 elements that range from 0.1 to 200\,pc) and spectral types $x_j$ (1000 elements that range from A0 to L8). This map of apparent $K_S$-band magnitudes can be described with the following equation:
\begin{align}
	\mathcal{K}_{i,j} = M_{K_S}(x_j) + 5\left(\log _{10}\varpi_i-1\right).
\end{align}

\begin{figure}
	\centering
	\includegraphics[width=0.48\textwidth]{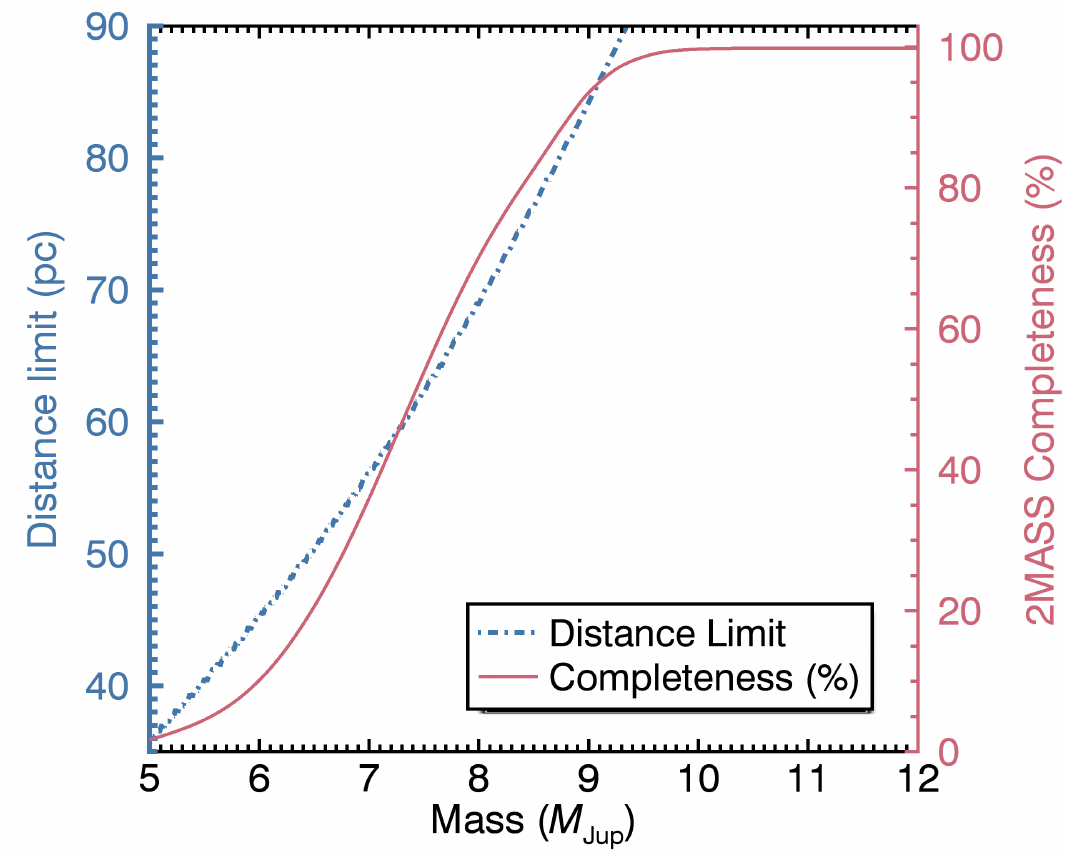}
	\caption{Limiting distance at which a TWA member of a given mass is detected \replaced{by}{in} 2MASS (dash-dotted blue line) obtained from the photometric sequence of Figure~\ref{fig:kabs} and the posterior mass distributions of Figure~\ref{fig:masspost}. The expected \replaced{completion}{completeness} fraction of TWA members, obtained from this limiting distance relation combined with the 2MASS $K_S$-band \replaced{completion}{completeness} limits and the BANYAN~II spatial model of TWA, is displayed as a red line. See Section~\ref{sec:tmass_completeness} for more details.}
	\label{fig:distlim_mult}
\end{figure}

The spectral type dimension was subsequently mapped to most probable masses using the PDFs derived in Section~\ref{sec:physpar} and displayed in Figure~\ref{fig:sptmasstwa}. The limiting distance at which TWA members can be detected in 2MASS\deleted{ then }corresponds to the value of $\varpi_i$ that yields a null \replaced{completion}{completeness} fraction $f_C(\mathcal{K}_{i,j}) = 0$. \replaced{This relation}{The solution to this constraint as a function of mass} is displayed in Figure~\ref{fig:distlim_mult}.

\begin{figure}
	\centering
	\includegraphics[width=0.48\textwidth]{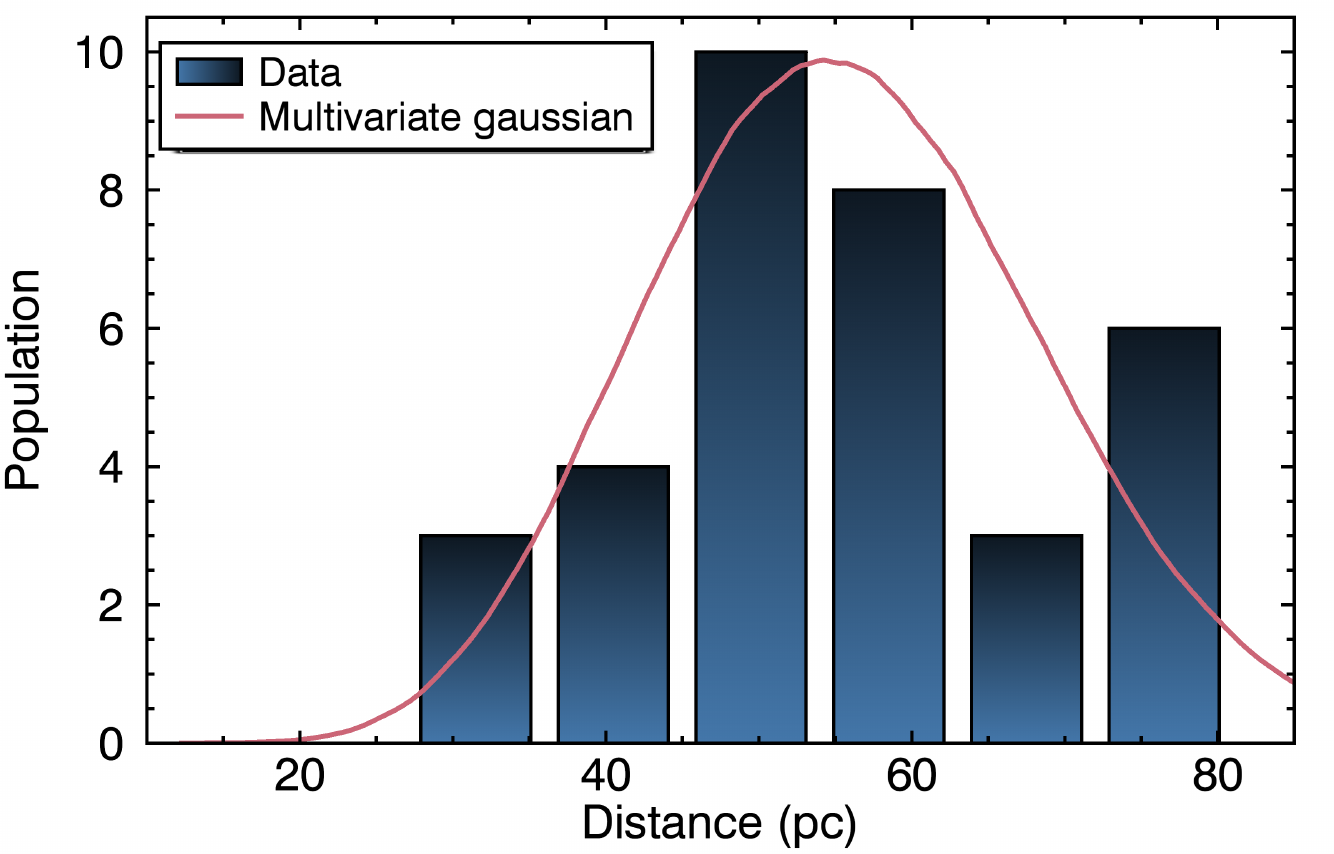}
	\caption{Distance histogram of TWA bona fide members and high-likelihood candidate members (dark blue bars), compared with a synthetic population drawn from a multivariate Gaussian PDF (red line). See Section~\ref{sec:tmass_completeness} for more details.}
	\label{fig:distpdf}
\end{figure}

\deleted{In order to determine the completeness limit of TWA members as a function of their mass, }It is necessary to invoke a model of \replaced{their}{the TWA} spatial distribution\added{ to determine the completeness limit of its members as a function of mass}. The multivariate Gaussian spatial model developed in Section~\ref{sec:cpm} was used to obtain the most up-to-date distance distribution of TWA members. A Monte Carlo simulation \replaced{of}{consisting in} $10^6$ random draws along this distribution was performed, and the distance to each of \replaced{these}{the $10^6$} synthetic objects was calculated. A distance histogram was \replaced{then reconstructed}{built from these} to serve as a distance probability density distribution $\mathcal{D}\left(\varpi_i\right)$, which is displayed in Figure~\ref{fig:distpdf}.

The \replaced{completion}{completeness} fraction as a function of mass $f_C^\prime\left(m_j\right)$ can then be obtained from the following expression:
\begin{align}
	f_C^\prime\left(m_j\right) = \frac{\sum_i f_C\left(\mathcal{K}_{i,j}\right) \mathcal{D}\left(\varpi_i\right)}{\sum_i \mathcal{D}\left(\varpi_i\right)}.
\end{align}

This quantity is also displayed in Figure~\ref{fig:distlim_mult}. It can be observed that \replaced{this correction only has a significant ($\geq$\,10\%) effect for masses of $\leq 9.5$\,\Mjup, which correspond to temperatures of $\lesssim 1\,700$\,K at the age of TWA, or to spectral types $\gtrsim$\,L2 (see Figure~\ref{fig:spttefftwa} and \citealt{2015ApJ...810..158F})}{more than 90\% of TWA members with masses $\geq 8.8$\,\Mjup\ should be detected in 2MASS, which correspond to temperatures of $\gtrsim 1\,600$\,K at the age of TWA, or to spectral types $\lesssim$\,L3 (see Figure~\ref{fig:spttefftwa} and \citealt{2015ApJ...810..158F})}.

\subsection{The Completeness of the Hipparcos Search for TWA members}\label{sec:hip_completeness}

\added{The Hipparcos-based search for new TWA members presented in Section~\ref{sec:hip} can be used to determine a range of masses for which the sample completeness is largest and constant as a function of mass.}

\added{Although there is no published completeness curve as a function of magnitude for the Hipparcos survey, the Hipparcos input catalog was constructed to be complete for G5 or earlier stars that are brighter than $V = 7.9 + 1.1\sin{|b|}$, where $b$ is the Galactic latitude \citep{1992BICDS..41....9T}. In the case of later-type stars, this limiting magnitude is given by $V = 7.3 + 1.1\sin{|b|}$. Averaging this limit over the spatial distribution of TWA members yields respective limiting magnitudes of $V = 8.4$ and $V = 7.8$ for early- and late-type stars.}

\added{A completeness curve for Hipparcos was determined using a Monte Carlo simulation: $10^4$ synthetic objects were drawn from the spatial distribution of TWA presented in Equation~\eqref{eqn:multivar} of Section~\ref{sec:cpm} at each point of a $10^3$ log-uniform array of masses, and the \cite{2016ApJ...823..102C} solar-metallicity isochrones at the age of TWA were used to determine their absolute $V$-band magnitudes and effective temperatures. The $XYZ$ coordinates of each synthetic star were used to determine their distances, galactic latitudes, and relative $V$-band magnitudes. The G5 spectral type threshold that is used to select the appropriate magnitude limit corresponds to a temperature of $5500$\,K at the age of TWA \citep{2013ApJS..208....9P}; the appropriate Hipparcos magnitude limit was used for each synthetic star to compute the fraction of stars that were detected in Hipparcos, while making the conservative assumption that no stars below the magnitude limits of \cite{1992BICDS..41....9T} were detected.}

\begin{figure}
	\centering
	\includegraphics[width=0.48\textwidth]{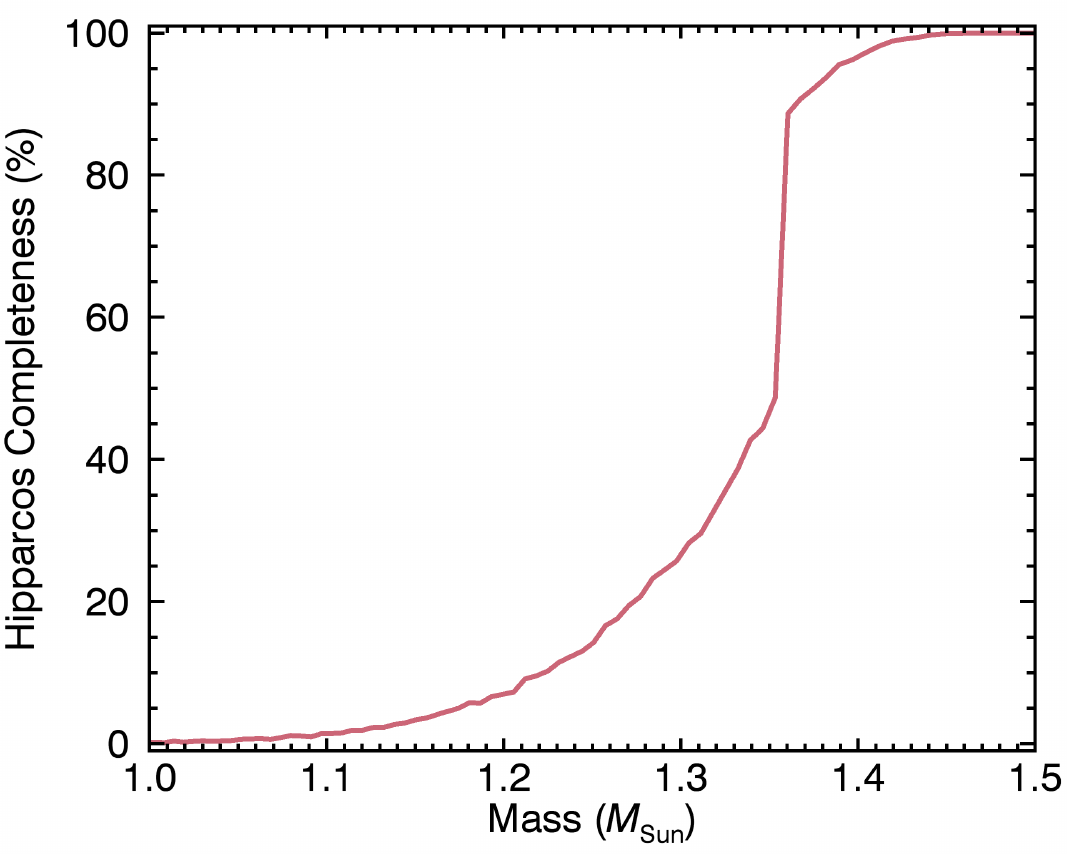}
	\caption{Sample completeness as a function of mass for TWA candidate members that are expected to be safely detected by the Hipparcos survey. The discontinuity at $\sim$\,1.35\,\Msol\ is due to a break in the Hipparcos detection limit for stars later than G5. Only members with masses above $\sim$\,1.43\,\Msol\ are all expected to be safely detected by Hipparcos. See Section~\ref{sec:hip_completeness} for more details.}
	\label{fig:hip_completeness}
\end{figure}

\added{This Monte Carlo simulation yielded a minimal completeness curve as a function of mass for the TWA members detectable in Hipparcos, which is presented in Figure~\ref{fig:hip_completeness}. This figure demonstrates that only TWA members with masses above $\sim$\,1.43\,\Msol\ have been detected with confidence in Hipparcos. At the age of TWA, this mass corresponds to a temperature of $\sim$\,5800\,K, or to the spectral type G2. Only two TWA members (TWA~11; A0 and TWA~43; A2) fall in this Hipparcos-complete regime. Hence, the Hipparcos survey does not provide a significant sample of bona fide TWA members to measure its IMF parameters in a regime of uniform completeness.}

\section{THE INITIAL MASS FUNCTION OF TW~HYA}\label{sec:imf}

\added{In this section, a continuous and empirical IMF of TWA is constructed. The Salpeter and log-normal functional forms are then fitted to this IMF using an MCMC statistical method that is independent on binning and accounts for small number statistics.}

\subsection{A Spatial Model of TWA}\label{sec:imf_spatmod}

\deleted{In order to express the IMF as a space density rather than a number of stars,} A model for the spatial extent of TWA must be considered \added{to express the IMF as a space density rather than a number of stars}. Such a model is available as part of the BANYAN~II tool \citep{2014ApJ...783..121G}, however it does not benefit from recent updates to the list of high-likelihood members of TWA (see Section~\ref{sec:candcomp}). For this reason, an updated spatial model of TWA is developed here.

\deleted{In order }To do this, a \replaced{3-dimensions}{3 dimensional} version of the multivariate gaussian model of TWA that was described in Equation~\eqref{eqn:multivar} was used, where only the spatial dimensions $XYZ$ were conserved. The density $n_{\mathrm{max}}$ of objects at the core of TWA \replaced{is then given by}{then follows}:
\begin{align}
	n_{\mathrm{max}} &= N_{\mathrm{tot}}/V_{\mathrm{eff}},\\
	V_{\mathrm{eff}} &= \left(2\pi\right)^{3/2}\sqrt{|\mathbf{\Sigma}|},
\end{align}
where $V_{\mathrm{eff}}$ is referred to in this work as the \emph{effective volume} of TWA. In the equation above, the covariance matrix $\mathbf{\Sigma}$ is a $3\times 3$ matrix that contains only the spatial coordinates.

The \replaced{galactic}{Galactic} coordinates $XYZ$ and measurement errors were calculated for all 31 high-likelihood and bona fide systems listed in Table~\ref{tab:kinematic}. In the cases where a trigonometric distance was not available, the BANYAN~II kinematic distance was used. The covariance matrix $\mathbf{\Sigma}$ and mean position $\mathbf{X_0}$ were calculated in a Monte Carlo simulation with $10^6$ cases that are normally distributed along measurements and errors of the $XYZ$ positions of TWA members. This yielded an effective volume of $V_{\mathrm{eff}} = 6\,200^{+690}_{-630}$\,pc$^3$. \replaced{An}{The} IMF of TWA can thus be divided by this \replaced{number}{volume} to obtain a space density IMF.

\begin{deluxetable*}{rccccc|cccccc}
\tabletypesize{\scriptsize}
\tablecolumns{12}
\tablecaption{Best-Fitting IMF Parameters\label{tab:imf}}
\tablehead{\colhead{Sample} & \colhead{$M_{\mathrm{tot}}$} & \multicolumn{3}{c}{Salpeter} & \colhead{} & \multicolumn{6}{c}{Log-normal}\\
\cline{3-5}
\cline{7-12}
\colhead{Name} & \colhead{($M_\odot$)} & \colhead{$\alpha$} & \colhead{$\phi_0$ (pc$^{-3}$)} & \colhead{$\rho\left(\alpha,\phi_0\right)$} & \colhead{} & \colhead{$m_c$ (\Msol)} & \colhead{$\sigma$ (dex)} & \colhead{$\phi_t$ (pc$^{-3}$)} & \colhead{$\rho\left(m_c,\sigma\right)$} & \colhead{$\rho\left(m_c,\phi_t\right)$} & \colhead{$\rho\left(\sigma,\phi_t\right)$}}
\startdata
\sidehead{\textbf{Bona Fide Members and High-Likelihood Candidate Members}}
Primaries + Companions & $19.0_{-0.6}^{+0.4}$ & $2.23^{+1.05}_{-0.45}$ & $3.55^{+0.64}_{-1.74}$ & $-0.96$ & & $0.21^{+0.11}_{-0.06}$ & $0.76^{+0.18}_{-0.13}$ & $9.8^{+1.5}_{-1.4}$ & $0.29$ & $0.85$ & $0.55$\\
Primaries only & $12.6 \pm 0.3$ & $1.92^{+1.15}_{-0.45}$ & $2.15^{+0.47}_{-1.02}$ & $-0.90$ & & $0.19^{+0.14}_{-0.06}$ & $0.88^{+0.25}_{-0.19}$ & $6.1^{+1.1}_{-1.0}$ & $0.35$ & $0.84$ & $0.51$\\
\sidehead{\textbf{All Candidate Members Except Low-Likelihood}}
Primaries + Companions & $20.8_{-0.4}^{+0.3}$ & $\cdots$ & $\cdots$ & $\cdots$ & & $0.08 \pm 0.02$ & $0.70^{+0.12}_{-0.09}$ & $15.6^{+2.2}_{-1.8}$ & $0.74$ & $0.86$ & $0.71$\\
Primaries only & $14.2_{-0.3}^{+0.5}$ & $\cdots$ & $\cdots$ & $\cdots$ & & $0.05 \pm 0.02$ & $0.63 \pm 0.12$ & $11.3^{+1.8}_{-1.2}$ & $0.72$ & $0.83$ & $0.70$\\
\enddata
\tablecomments{Results in the table section corresponding to all candidate members are not reported for samples that contain only high-likelihood candidates and\\ bona fide members. See Section~\ref{sec:imf} for more details.}
\end{deluxetable*}

\subsection{The Construction of a Continuous IMF}\label{sec:imf_const}
The mass probability functions that were derived in this work for individual TWA members (see Section~\ref{sec:physpar}) can be summed together to obtain a continuous version of the observed TWA IMF.\deleted{ In order to give more weight to the more likely candidate members,} Each candidate member's mass PDF was weighted by $1-C_B$, where $C_B$ is the BANYAN~II probability that a contaminant from the field imitates the properties of a given candidate member (see Section~5 of \citealt{2014ApJ...783..121G} for a detailed discussion)\added{, to account for the expected rate of false positives by assigning more weight to the more likely candidate members}.\deleted{ All bona fide members were assigned a weight of one.}

\replaced{Eight}{Four} distinct IMFs were constructed for TWA, using a variety of input data sets. The first class of \replaced{4}{two} data sets includes only high-likelihood candidate members and bona fide members of TWA, whereas the second class of\deleted{ remaining }data sets also includes all candidate members\added{ (but excludes low-likelihood candidate members)}. The \replaced{four}{two} subclasses are divided as follows: (1) primary and companion components of multiple systems \replaced{are considered as separate objects}{are included (and counted as separate objects)}; \added{and }(2) only primaries of multiple systems (or isolated objects) are counted\added{.} \deleted{; (3) only companions of multiple systems are counted; or (4) the IMF of systems is constructed, where each system has a mass that is equal to the total of its parts. In case (4), it is required to build a probability density distribution for the total mass of two stars from the PDFs of their individual masses.}\added{The total masses of these four samples were also calculated and are listed Table~\ref{tab:imf}.} \replaced{This is obtained from}{Calculating the total mass of \replaced{several}{multiple} objects necessitates} a special convolution-like combination of the \replaced{two}{individual} PDFs, which is detailed in Appendix~\ref{an:totalmass}.\deleted{ This method can also be used to derive the PDF for the total mass of TWA, which are listed in Table~\ref{tab:imf}.} A total mass of $19.0_{-0.6}^{+0.4}$\,\Msol\ \replaced{is}{was} obtained for the current census of TWA high-likelihood and bona fide members.

The compilation of TWA companions used to derive the companion IMF is likely incomplete and results in a compilation of heterogeneous literature searches for companions. It should therefore serve only as a rough estimation until a systematic search for TWA member companions is carried out.

\begin{figure*}
	\centering
	\subfigure[Continuous IMF of all system components]{\includegraphics[width=0.48\textwidth]{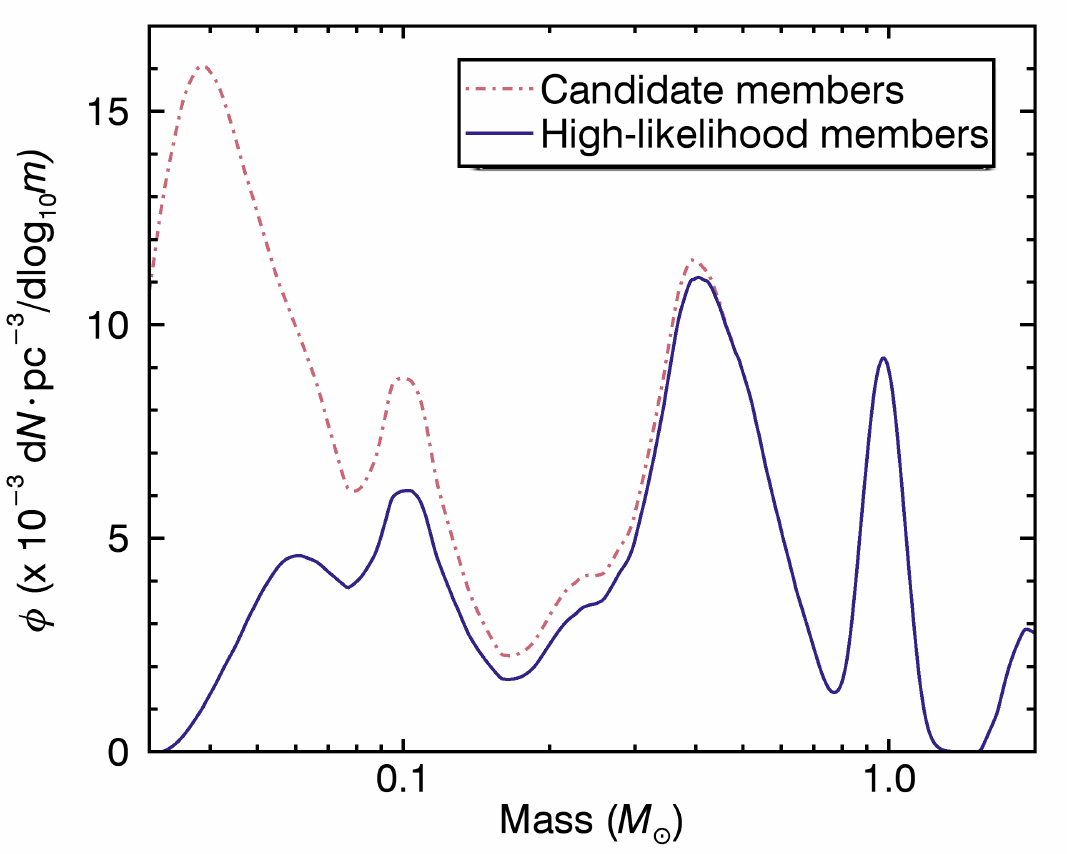}}
	\subfigure[Continuous IMF of primary stars only]{\includegraphics[width=0.48\textwidth]{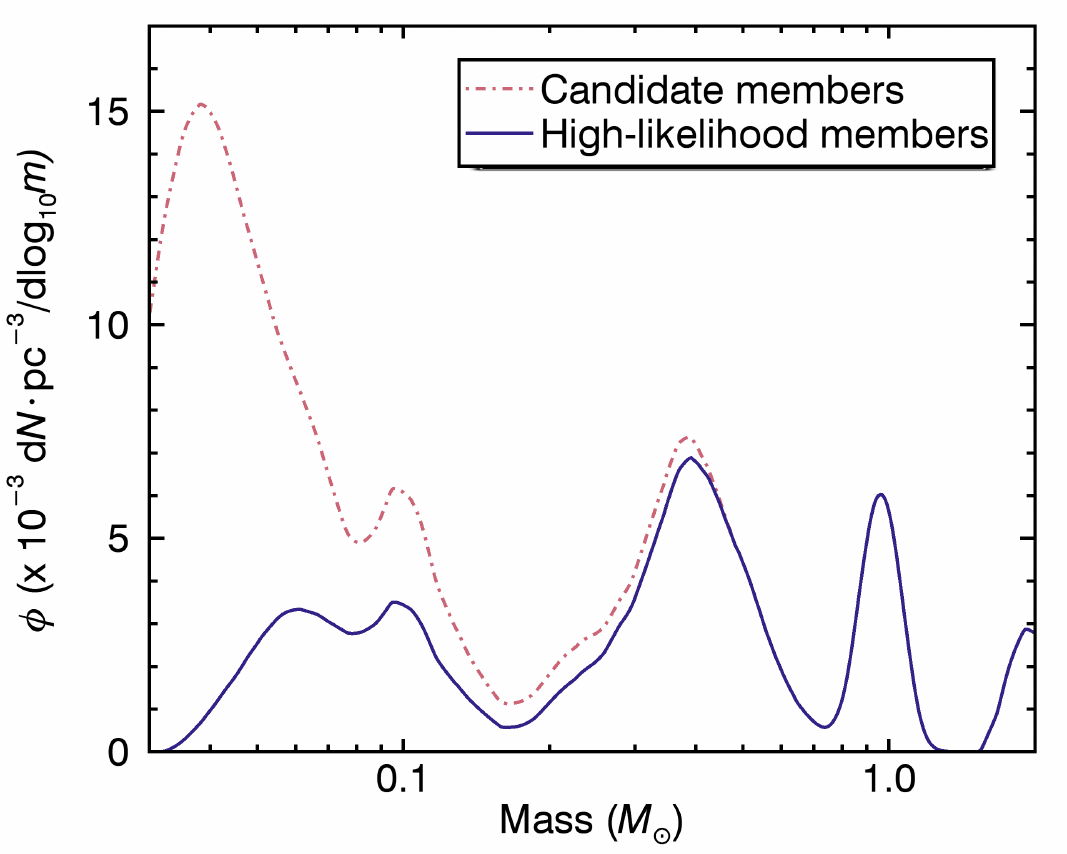}}
	\subfigure[Binned IMF of all system components]{\includegraphics[width=0.48\textwidth]{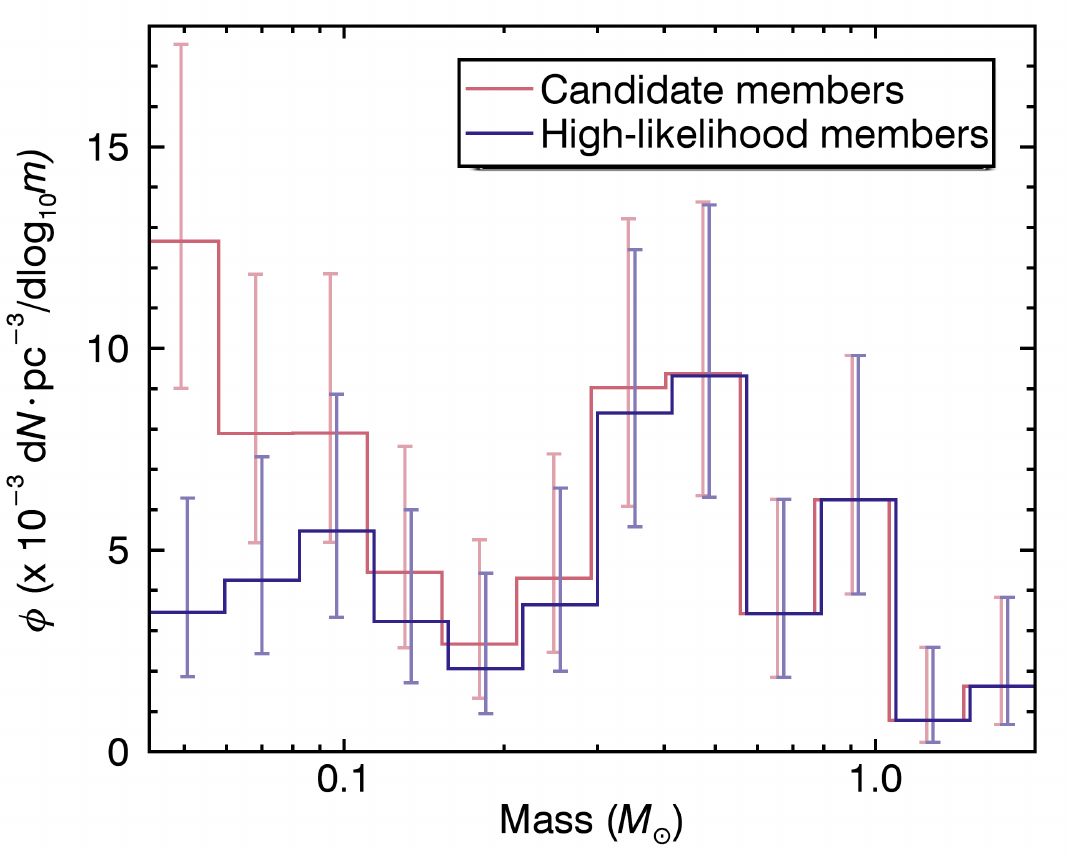}}
	\subfigure[Binned IMF of primary stars only]{\includegraphics[width=0.48\textwidth]{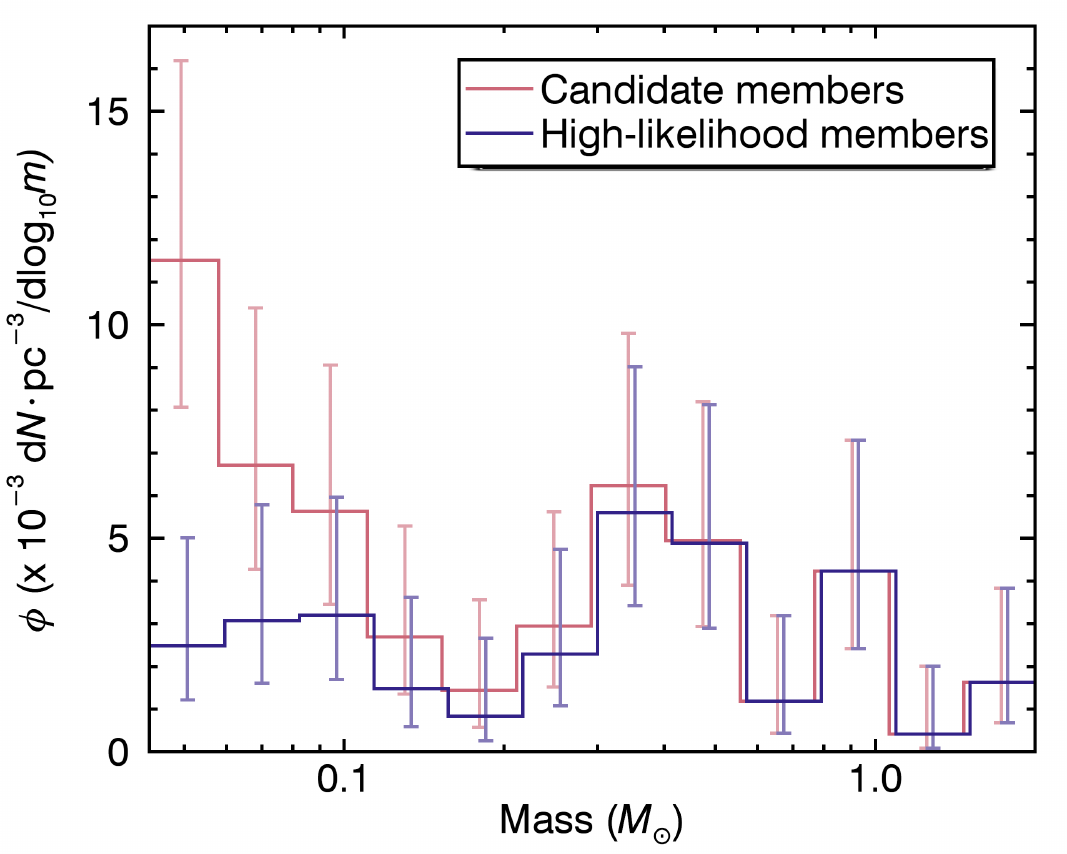}}
	\caption{Panels~a and b: Continuous IMFs of TWA candidate members (red dot-dashed line) and high-likelihood and bona fide members (blue line), for all system components (Panel~a) and primary stars only (Panel~b).\\
	Panels~c and d: Binned versions of the IMFs that include Poisson error bars and uncertainties on the volume of TWA. Bins of $\Delta\log{M_\odot} = 0.13$ were used, with the same color scheme as Panels~a and b. Histogram bars were shifted slightly to the left and right (by $\pm 6\times 10^{-3}$\,\Msol) for visibility. See Section~\ref{sec:imf} for more details.}
	\label{fig:pdf_data}
\end{figure*}

\replaced{In order to account for the effect of small number statistics in the calculation of the IMF, it must be assumed that}{Accounting for the effect of small number statistics in the calculation of the IMF requires to make the assumption that} the formation of $N$ stars in a given range of masses $\left[m_0,m_1\right]$ is a random process in which every star formation event is independent of the previous ones\added{ to account for the effect of small number statistics in the calculation of the IMF}. It follows that the PDF for the space density of objects $n$ is described by a Poisson distribution $\mathcal{P}(n|\lambda)$ parametrized with the mean number of star formation events $\lambda$, that is obtained from integrating the IMF $\phi$ over this mass range:
\begin{align}
	\lambda &= \int_{\log _{10}m_0}^{\log _{10}m_1} \phi\left(\log _{10} m\right)\,\mathrm{d}\log _{10}m, \\
	\mathcal{P}(n|\lambda) &= \frac{e^{-\lambda}\lambda^{nV_{\mathrm{eff}}}}{\Gamma\left(nV_{\mathrm{eff}}+1\right)},\label{eqn:poisson}
\end{align}
\noindent where $\Gamma(x)$ is the Euler Gamma function.

The parameter $\lambda$ can also be seen as the mean number of stars $n$ that would be formed in the mass range $\left[m_0,m_1\right]$ after a large number of simulations for the star formation of TWA members. The problem then consists of determining $P(\lambda|n)$, the PDF that describes the probable value of $\lambda$ given the measured space density $n$. This can be obtained using Bayes' formula:
\begin{align}
	P(\lambda|n) = \frac{\mathcal{P}(n|\lambda)\pi(\lambda)}{P_n(n)},
\end{align}
\noindent where $\pi(\lambda)$ is the prior distribution on the parameter $\lambda$, and:
\begin{align}
	P_n(n) = \int_0^\infty \mathcal{P}(n|\lambda)\pi(\lambda)\,\mathrm{d}\lambda.
\end{align}

The Jeffrey's non-informative prior \citep{Jeffreys:iy-ZBWsz} was chosen for $\pi(\lambda)$, which ensures in the case of 1-parameter PDFs that no prior knowledge on the value of this parameter is injected in the problem. This choice of a non-informative prior also ensures that the results will be independent under coordinate changes. The Jeffrey's prior of the Poisson distribution is given by \cite{Jaynes:1968tp}:
\begin{align}
	\pi(\lambda) = \lambda^{-1/2}.
\end{align}

It follows that:
\begin{align}
	P_n(n) &= \frac{\Gamma\left(nV_{\mathrm{eff}}+1/2\right)}{\Gamma\left(nV_{\mathrm{eff}}+1\right)},\\
	P(\lambda|n) &= \frac{e^{-\lambda}\lambda^{nV_{\mathrm{eff}}-1/2}}{\Gamma\left(nV_{\mathrm{eff}}+1/2\right)}.\label{eqn:imfposterior}
\end{align}

Hence, the value for the the cumulative IMF $\Phi$ within a given range of masses follows the PDF described by Equation~\eqref{eqn:imfposterior}, which is a continuous analog of the Poisson distribution $\mathcal{P}\left(k|\lambda\right)$ centered at $k-1/2$, where the roles of the variable $k$ and the parameter $\lambda$ have been swapped.

The resulting IMF PDFs are displayed in Figure~\ref{fig:pdf_data}, along with binned versions that include Poisson error bars. It can be noted that there is an unexpectedly large number of candidate members with masses in the range 0.04--0.1\,\Msol. Since this \added{effect }is not observed in the high-likelihood/bona fide members IMF, it could be caused by a larger number of contaminants at fainter magnitudes. This is similar to the relatively high fraction of M-type contaminants that were uncovered in Section~\ref{sec:data_an}, and would \replaced{coincide}{be consistent} with the \replaced{interpretation}{possibility} that the large number of brown dwarfs in Upper~Scorpius \citep{2007MNRAS.374..372L} \replaced{might}{could} be artificial.

\begin{figure*}[p]
	\centering
	\subfigure[High-likelihood members, all components]{\includegraphics[width=0.49\textwidth]{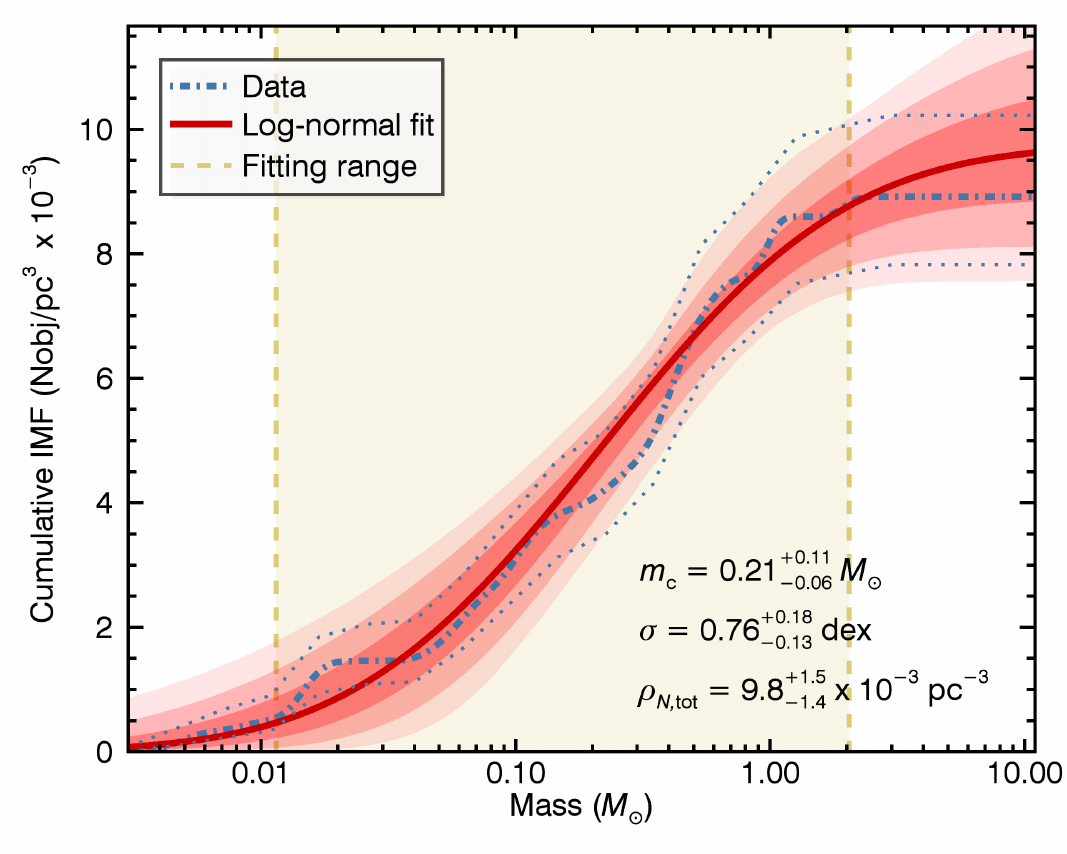}}
	\subfigure[High-likelihood members, primaries only]{\includegraphics[width=0.49\textwidth]{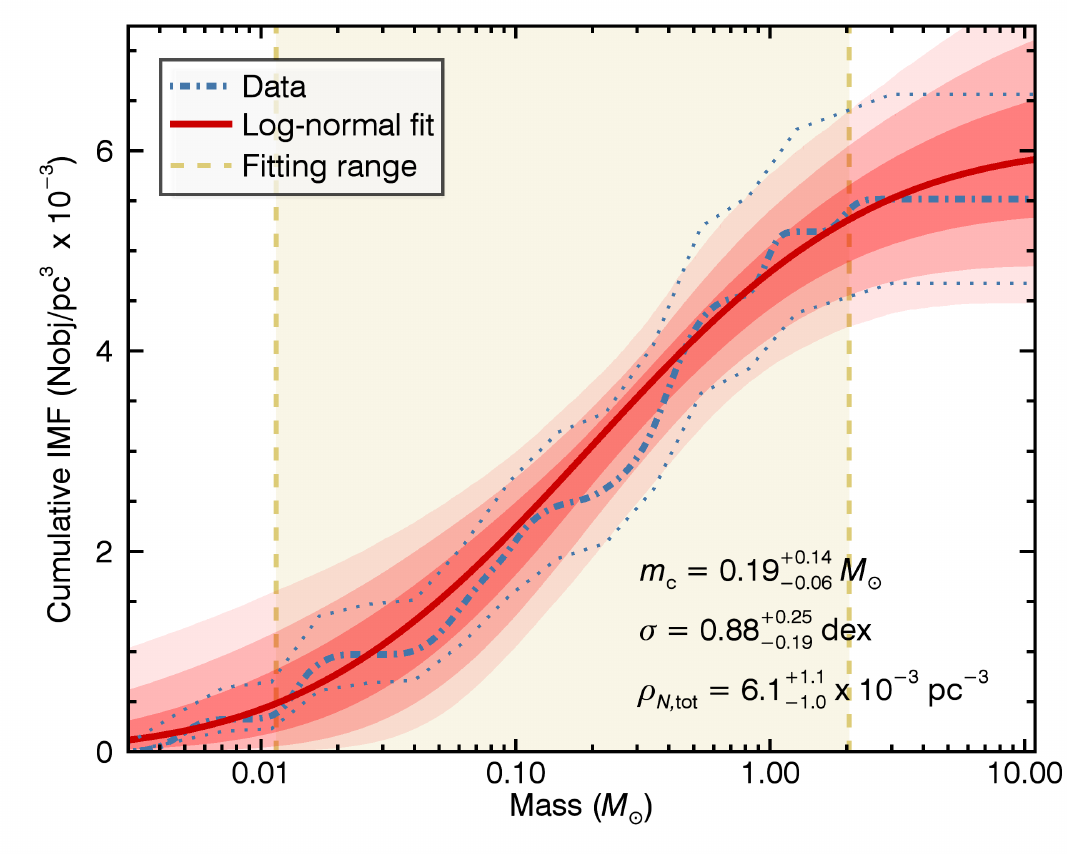}}
	\subfigure[Members and Candidate members, all components]{\includegraphics[width=0.49\textwidth]{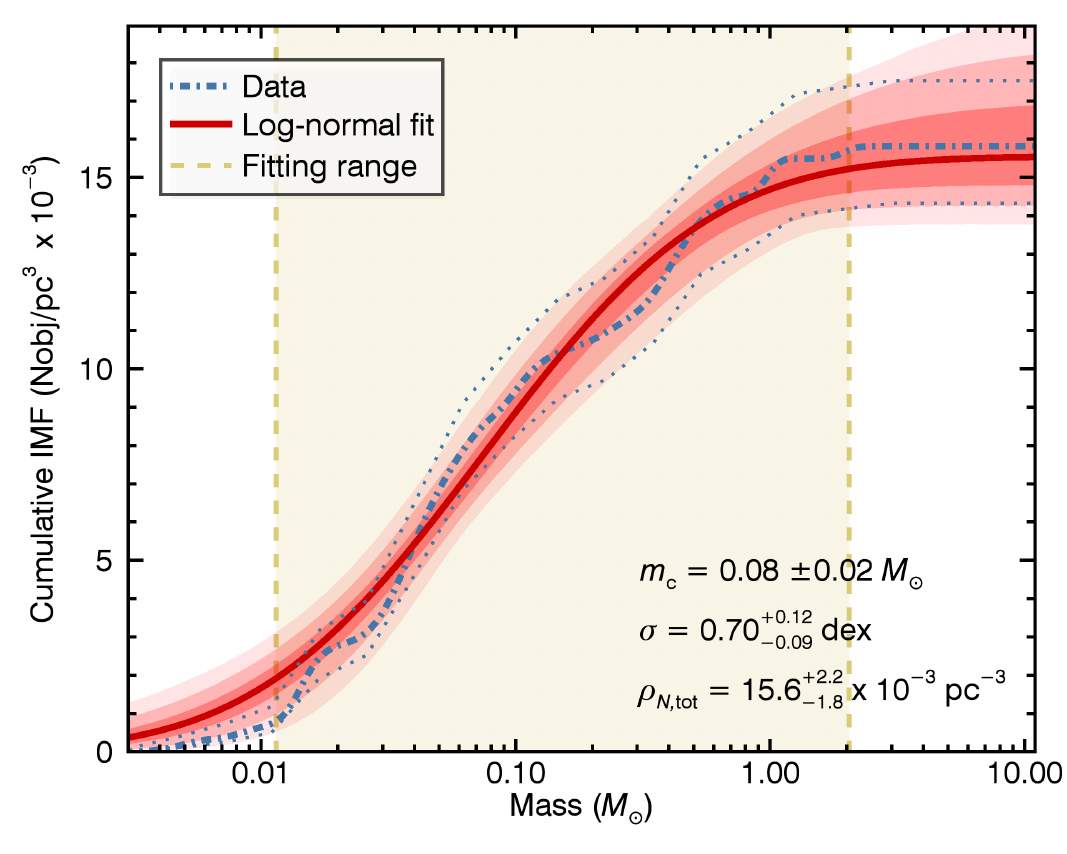}}
	\subfigure[Members and Candidate members, primaries only]{\includegraphics[width=0.49\textwidth]{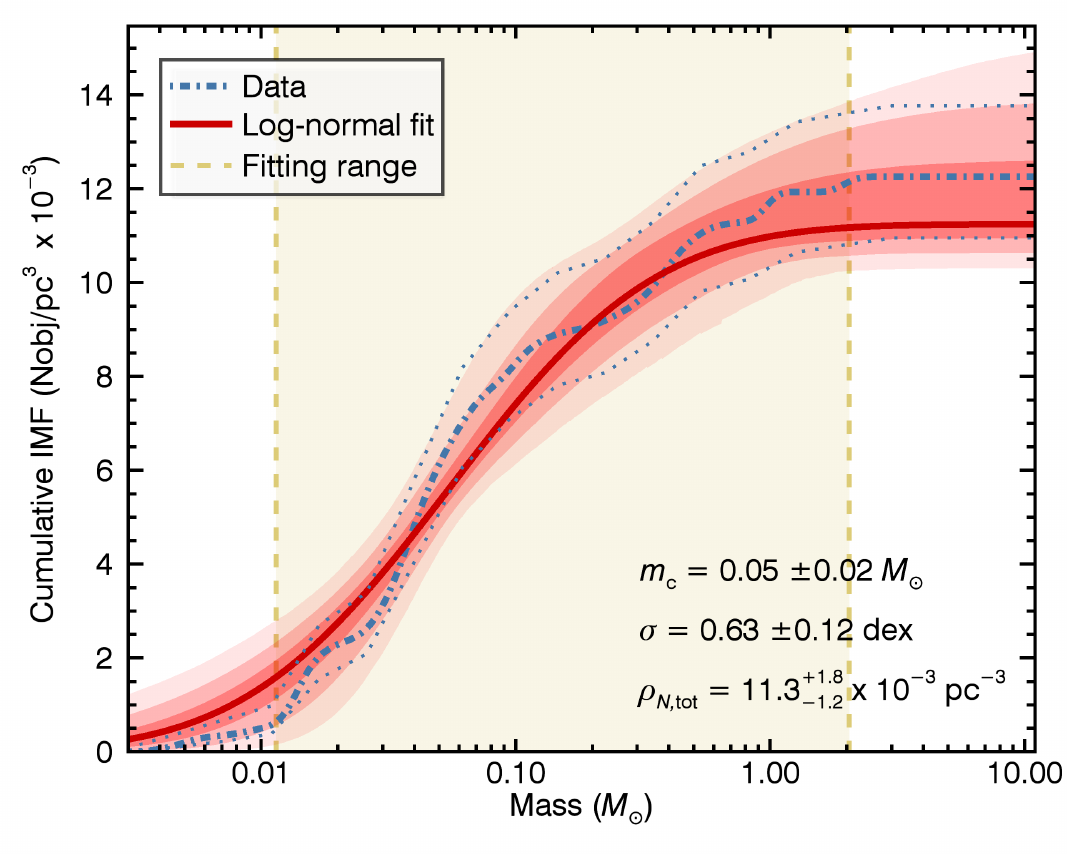}}
	\caption{Fitted log-normal cumulative IMFs (thick red line) to \added{the }observed un-binned TWA cumulative IMFs (blue dash-dotted line). The thin blue doted lines represent the $\pm$\,1\,$\sigma$ range due to small number statistics based on a Poisson distribution as well as errors on the TWA effective volume estimate. The yellow region represents the mass range used to perform the fit, and the red shaded regions indicate 1--3$\sigma$ random draws from the MCMC solutions. See Section~\ref{sec:imf} for more details.}
	\label{fig:IMF_fit}
\end{figure*}

\begin{figure*}
	\centering
	\subfigure[High-likelihood members, all components]{\includegraphics[width=0.49\textwidth]{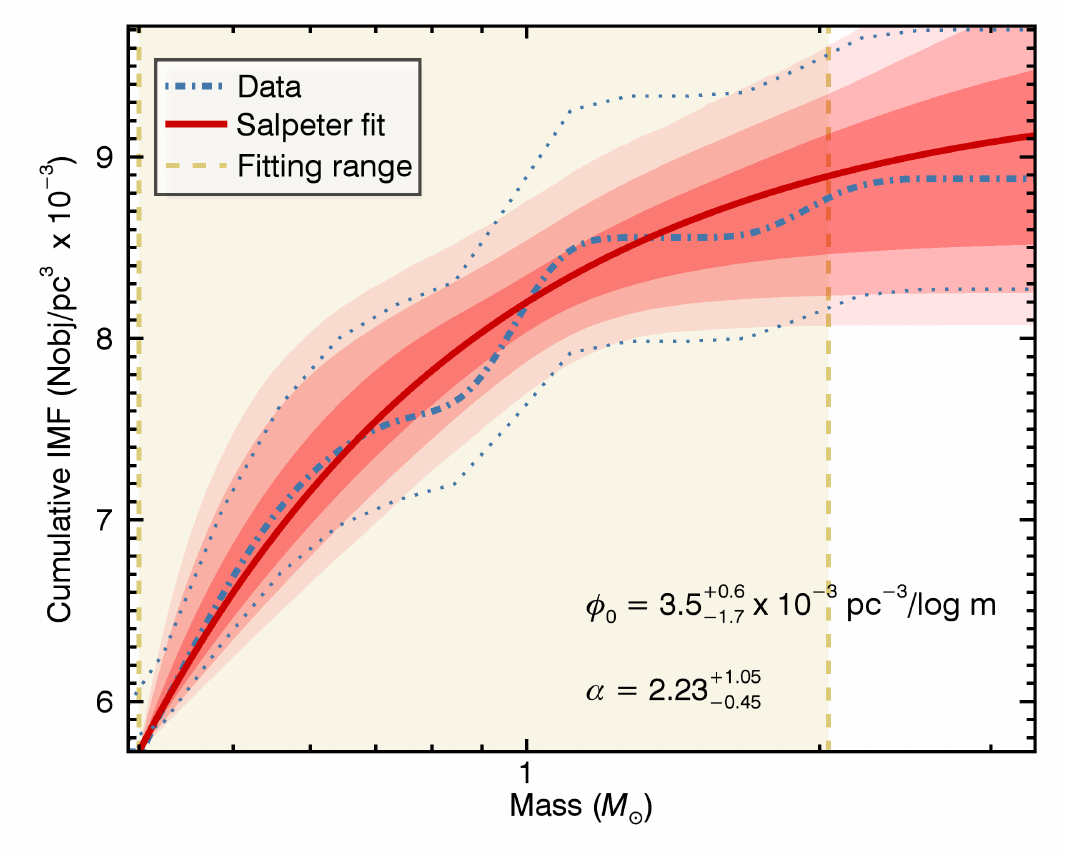}}
	\subfigure[High-likelihood members, primaries only]{\includegraphics[width=0.49\textwidth]{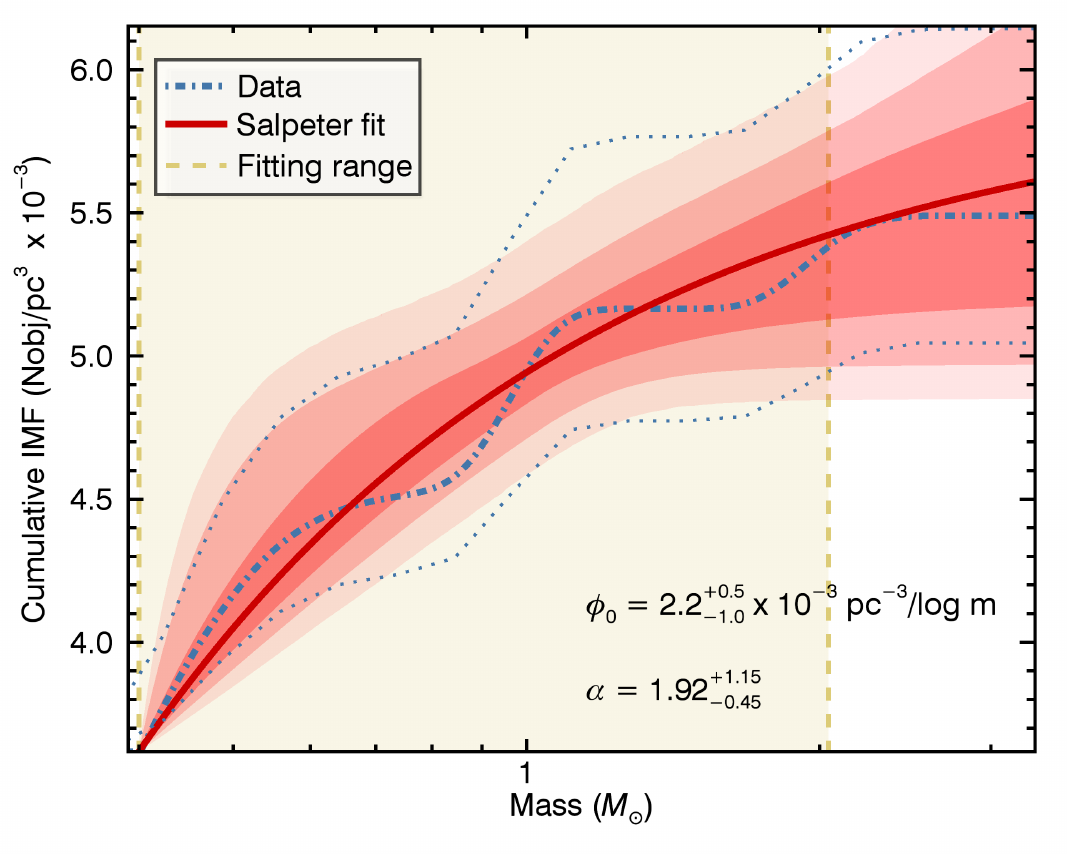}}
	\caption{Fitted Salpeter cumulative IMFs to observed un-binned TWA cumulative IMFs. Colors and curves are similar to those of Figure~\ref{fig:IMF_fit}. The cumulative integral of the IMFs used for the Salpeter fits have a lower bound of $0.4$\,\Msol. See Section~\ref{sec:imf} for more details.}
	\label{fig:salpeterIMF_fit}
\end{figure*}

\subsection{IMF Markov Chain Monte Carlo Fitting}\label{sec:imf_mcmc}
The DREAM(ZS) Markov Chain Monte Carlo algorithm \citep{terBraak:2008iw} was used to determine the best parameters that fit the measured TWA IMFs. Initial parameter estimates for $\phi_t$, $\sigma$ and $m_c$ were chosen by adjusting an error function to the cumulative IMF with a Levenberg-Marquardt least-squares fit (using the \emph{mpfitfun.pro} IDL routine). In the case of a Salpeter functional form, the initial value for $\alpha$ was set to the Salpeter slope $\alpha = 2.35$, and that of $\phi_0$ was chosen as the value of the adjusted error function at $m = 1$\,\Msol. These initial parameter choices did not affect the posterior PDFs.

The DREAM(ZS) algorithm requires building an array of parameters for an initial set of synthetic samples; this was done by randomly selecting parameter values for 10$D$ samples, where $D$ is the number of parameters (2 when fitting a Salpeter form or 3 when fitting a log-normal form). These random values were centered around the initial parameter estimates with a scatter of one tenth of the initial estimates. 

In all cases, we used $2D+1$ chains, a conservative $10^3$ samples for the burn-in phase, and let the MCMC run for a total of $10^5$ samples, with a thinning interval $K = 10$. The jumping scale factor $\gamma$ of the algorithm was set to eight times the default value of $2.38/\sqrt{2D}$ suggested by \cite{terBraak:2008iw}, to obtain mean acceptance rates below $\sim$\,90\%. The Snooker update probability was set to the default value of 10\%, and the jumping scale factor during a Snooker update was set to $\gamma_S = \sqrt{D}\gamma$. The value of $\gamma$ was set to unity once every 10 iterations, as suggested by \cite{terBraak:2008iw}. The improvement suggested by \cite{2016ApJ...819..133A} was used, where $\gamma$ is inflated by a uniform random small number that is bound between $-0.05$ and $0.05$. During Snooker updates, the value of $\gamma_s$ was allowed to randomly vary by a factor $1.3^{\pm 1}$, which corresponds to the same fractional random scatter that was used by \cite{terBraak:2008iw} with their value of $\gamma_s$.

The fits were performed directly on the cumulative IMFs to avoid the necessity of binning in accounting for small number statistics. The cumulative distribution function of the Salpeter IMF has the form:
\begin{align}
	\Phi\left(\log _{10}m\right) = \frac{\phi_0}{\left(1-\alpha\right)\ln{10}}\left(m_0^{1-\alpha}-m^{1-\alpha}\right),
\end{align}
\noindent where $m_0$ is the lower mass bound, and the cumulative distribution function of the log-normal distribution has the form:
\begin{align}
	\Phi\left(\log _{10}m\right) = \frac{\phi_t}{2}\erfc{\left(\frac{\log _{10}m-\log _{10}m_c}{\sigma\sqrt{2}}\right)},
\end{align}
\noindent where $\erfc{x}$ is the conjugate error function.

There is an additional complication in fitting a model to a cumulative distribution function. Using the classical approach of minimizing $\chi^2$ would be mathematically inconsistent, as can be illustrated by the fact that the results would depend on the sampling of the cumulative IMF (e.g., a denser sampling would artificially yield smaller error bars on the best-fitting IMF parameters). The question that must be asked at a given step of the MCMC solver is the following: \emph{What is the probability that the observed IMF was drawn from a given modelled cumulative IMF ?} One way to answer this question, as suggested by \cite{2010ARA&A..48..339B}, is to use a two-sided Kolmogorov-Smirnov (K-S) test \citep{Kolmogorov:RN6BgzTl}, which uses the maximal distance between two cumulative distribution functions to quantify the probability that their difference is significant.

In addition to the K-S test, a Poisson likelihood can be used to ensure that the value of the parameter $\phi_0$ or $N_{\mathrm{tot}}$ is consistent with the observed total number of TWA members. As a consequence, the likelihood function that the MCMC algorithm will explore in the Salpeter case can be written as:
\begin{align}
	\mathcal{L}(\mathcal{D}|\phi_0,\alpha) = \mathcal{K}_S\left(\mathcal{D}|\alpha\right)\mathcal{P}\left(N_{\mathrm{obs}}|\phi_\mathrm{tot}V_\mathrm{eff}\right),\label{eqn:mcmc_likelihood}
\end{align}
\noindent where $\mathcal{D}$ represents the data (the observed IMF), $N_\mathrm{obs}$ is the observed total number of TWA objects, $\phi_\mathrm{tot}$ is the total space density of objects (integrated over all masses) predicted from the model IMF, $\mathcal{K}_S\left(\mathcal{D}|\alpha\right)$ is the probability returned by the K-S test given the data and model IMF, and $\mathcal{P}(k|\lambda)$ is a Poisson distribution. In the log-normal case, $\phi_0$ is replaced with $\phi_t$ and $\alpha$ is replaced with $\left\{m_c,\sigma\right\}$ in the equation above.

The prior PDFs $\pi_1\left(\phi_0,\alpha\right)$ and $\pi_2\left(\phi_t,m_c,\sigma\right)$ were chosen such that no information is injected in the algorithm. In the present case where the likelihood depends on more than one parameter, the Jeffrey's priors do not correspond to the non-informative case, and the more general ``reference priors'' must be used instead (e.g., see \citealt{Bernardo:1979uq}). This choice of priors also ensures (1) that the problem is invariant under change of parameter variables, and (2) that the available data maximizes the difference between the prior and posterior distributions.

The determination of the reference prior associated with a likelihood is generally complicated to calculate, which is especially true in the present situation since the likelihood function includes a Kolmogorov-Smirnov test. \cite{Berger:2009wx} presented a way to circumvent this with a numerical algorithm to compute the reference priors on a grid of parameters. The reference priors for the present problem were derived separately for the Salpeter and log-normal IMFs, and are discussed in more detail\deleted{s} in Appendix~\ref{an:refpriors}. Their applications did not significantly affect the resulting shapes of the posterior distributions.

\subsection{IMF Results}\label{sec:imf_results}
The MCMC fitting algorithm described above was applied on the \replaced{eight}{four} distinct sets of data that were previously mentioned. On each of these data sets, two fitting steps were performed : a log-normal IMF as described in Equation~\eqref{eqn:lognormalimf} was first fitted in the mass range 12\,\Mjup--2\,\Msol and a Salpeter IMF as described in Equation~\eqref{eqn:salpeterimf} was then fitted in the mass range 0.1--2.0\,\Msol. \replaced{In some specific cases}{For the Salpeter fitting range}, there were no candidate members in the TWA sample that were not also high-likelihood candidates or bona fide members, thus making the ``members'' or ``candidates'' IMFs identical.\deleted{ This was the case for all Salpeter fits except the one treating systems as wholes, and for the log-normal fit to the IMF of companions only. }In \replaced{those cases}{this case}, only a fit to the high-likelihood and bona fide members is \added{thus }reported.

A visual inspection of the chains revealed that the burn-in phase remained well within $10^3$ samples in all cases. The autocorrelation length of the parameter chains were found to be in the range of 1--28 samples, meaning that the number of independent samples were in the range $\sim$\,3\,500--10\,000. Acceptance fractions in the range $\sim$\,28--75\% were obtained, depending on the data samples that were fit. The central parameter values reported in Table~\ref{tab:imf} were chosen as the peak locations of the marginalized PDFs, and the asymmetrical error bars were chosen as the regions that encompass 34\% of the total area under its curve on each side.

\begin{figure*}
	\centering
	\subfigure[IMF of all system components]{\includegraphics[width=0.488\textwidth]{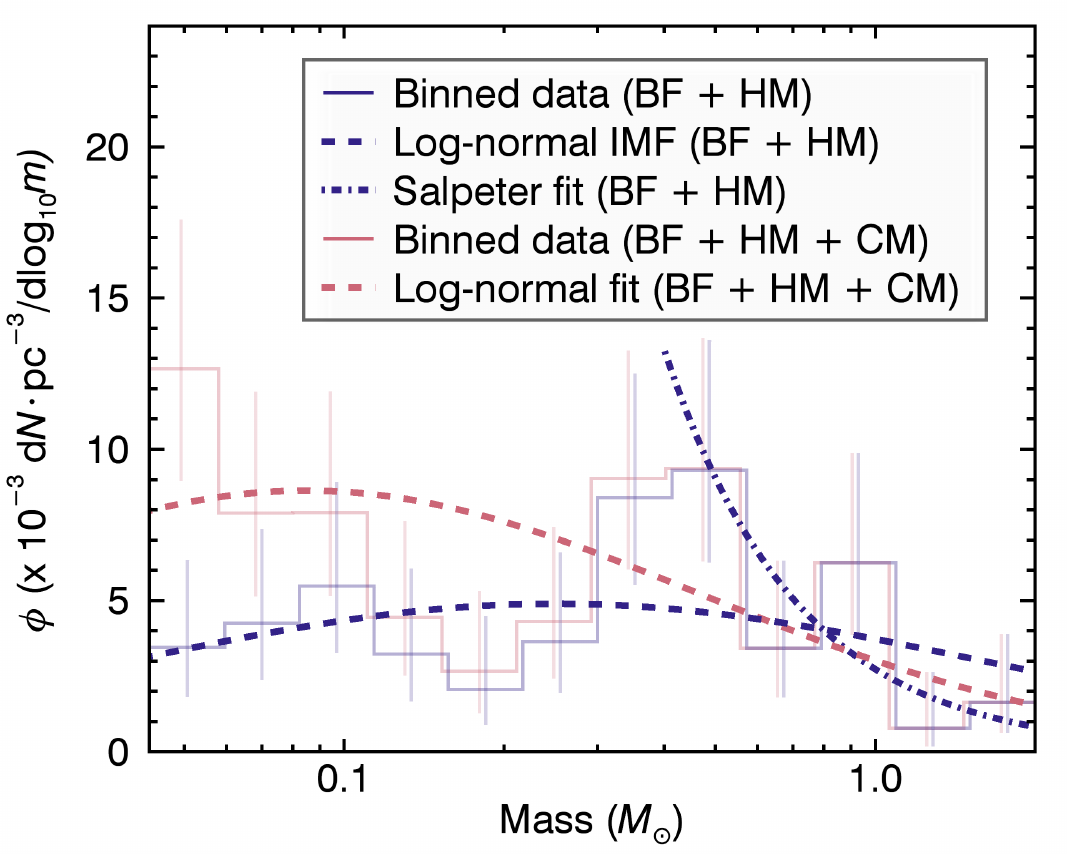}\label{fig:imfall}}
	\subfigure[IMF of primary stars only]{\includegraphics[width=0.488\textwidth]{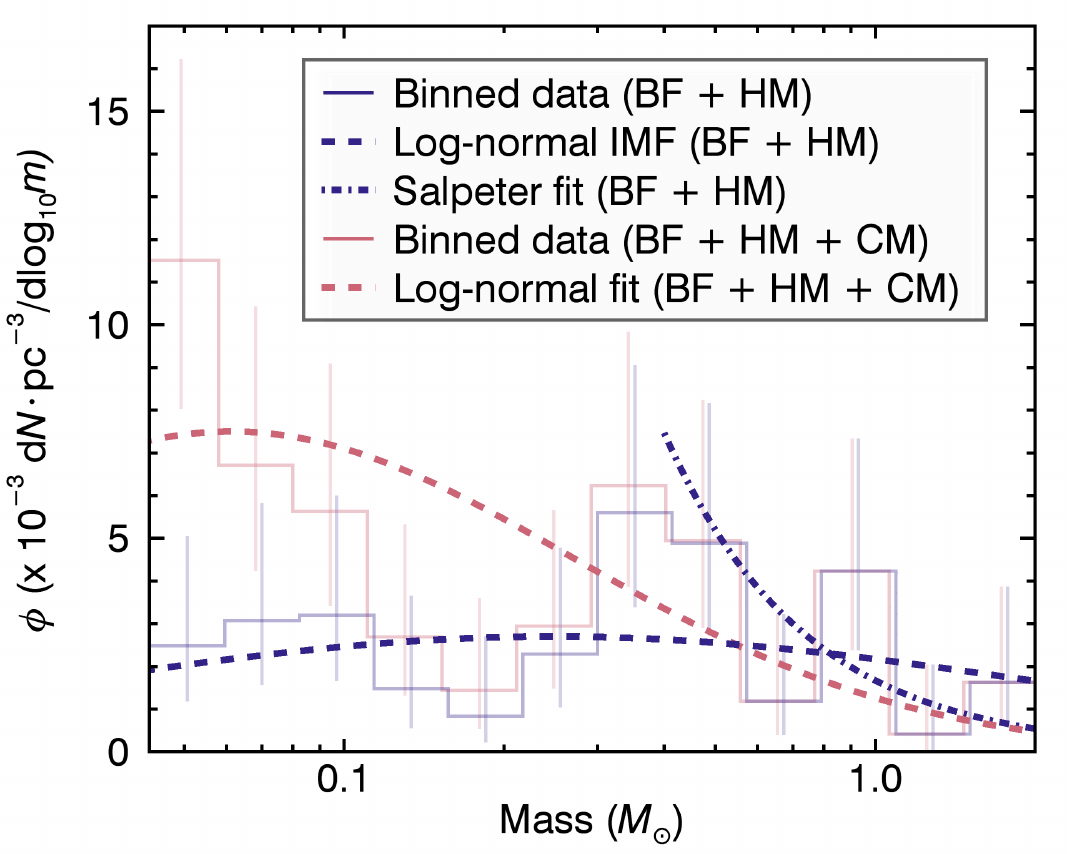}\label{fig:imfprim}}
	\caption{Best log-normal and Salpeter fits \added{compared }to the observed binned IMFs of TWA objects, for the \replaced{eight}{four} input samples described in the text. \emph{BF+HM} indicates bona fide members and high-likelihood candidate members only, whereas \emph{CM} indicates all currently known candidate members. Binned versions of the empirical IMFs are displayed here to make the comparison easier\deleted{, especially} because Poisson error bars cannot be determined for a continuous un-binned IMF. See Section~\ref{sec:imf} for more details.}
	\label{fig:imfs}
\end{figure*}

The resulting cumulative IMFs are displayed in Figures~\ref{fig:IMF_fit} and \ref{fig:salpeterIMF_fit} and the best-fitting parameters are listed in Table~\ref{tab:imf} along with their error bars and the Pearson correlation coefficients $\rho\left(x,y\right)$ of two given parameters $x$ and $y$. In most cases, correlations between the fitted parameters are significant. The error bars on the volume of TWA were added in quadrature to those of the space density parameters $\phi_0$ and $\phi_t$. The best-fitting IMF curves are displayed in Figure~\ref{fig:imfs}.

All posterior PDFs are unimodal, but are not always well represented by Gaussian distributions, even if asymmetrical error bars are used. The marginalized PDFs for $\alpha$ (Salpeter fit) in particular are heavy-tailed, with residual kurtosis values in the range $\kappa \simeq$\,2--16. This is also true to a lesser extent for the central mass ($\kappa \simeq$\,0.1--2.5) and the characteristic width ($\kappa \simeq$\,0.3--2) of the log-normal fits. All other cases have $0 < \kappa < 1$.

\replaced{In order to}{The Kullback-Leibler divergence $D_{\rm KL}\left(P||G\right)$ between the true PDF $P$ and an asymmetrical Gaussian PDF $G$ (see \citealt{Kullback:1951va}) was calculated to} characterize how much information is lost when representing the true posterior PDFs using asymmetrical Gaussians with the values provided in Table~\ref{tab:imf}\replaced{, the Kullback-Leibler divergence $D_{\rm KL}\left(P||G\right)$ between the true PDF $P$ and an asymmetrical Gaussian PDF $G$ was calculated (see \citealt{Kullback:1951va})}{. }This divergence characterizes the entropy increase when representing $P$ with $G$. This value was then compared with the Shannon entropy \citep{1949mtc..book.....S} $H$ of the true PDF, to obtain the fractional amount of entropy $f_E$ that is gained when representing $P$ with $G$, where $f_E = D_{\rm KL}/H$. Low values of $f_E < $6\% were obtained for all parameters except $\alpha$ (Salpeter), meaning that the loss of information is small when approximating their posterior PDFs with asymmetrical Gaussians. In the case of $\alpha$, fractional entropy gains were found to be in the range 7--52\%.

A Salpeter slope of $\alpha = 2.2^{+1.1}_{-0.5}$ is obtained in the generic case of high-likelihood and bona fide members, where multiple system components are treated as separate objects. This value is similar to the Salpeter slope of field stars ($\alpha = 2.35$; \citealt{1955ApJ...121..161S}), despite the exclusion of 2--10\,\Msol\ objects in the present analysis -- including this mass range is impossible since there are no such known members of TWA.\added{ It is possible that the Salpeter slope derived here is biased towards a shallower (lower) value due to the incompleteness of $<$\,F6 stars in the current TWA sample.}

The log-normal fit for the same case scenario yields a central mass of $m_c = 0.21^{+0.11}_{-0.06}$\,\Msol\ and a characteristic width of $0.8^{+0.2}_{-0.1}$\,dex. The central mass derived here is consistent with typical values obtained for the field ($m_c = $0.15--0.25\,\Msol; \citealt{2005ASSL..327...41C}), and is smaller than the previous estimation of the TWA IMF (0.4--0.6\,\Msol; \citealt{2011PhDT.......245L}), which was carried out when fewer TWA brown dwarf members were known. Including candidate members in the IMF calculations drives the central mass to a much lower value of $m_c = 0.08 \pm 0.02$\,\Msol, slightly below the brown dwarf/low-mass star boundary. This is related to the previously mentioned large number of TWA candidate members in the 0.04--0.1\,\Msol\ range, which are possibly due to significant contamination from interlopers unrelated to TWA in the sample. It can therefore be expected that the true central mass of TWA members will be located between these two values, and may still agree with IMF determinations based on field stars.

The characteristic widths that are obtained here are high compared to most determinations based on the field or other young associations (0.3--0.55\,dex; \citealp{2011AJ....141...98B,2012EAS....57...45J}), whether candidate members are included in the analysis or not. This \replaced{indicates}{may indicate} that completing the census of low-mass members of TWA might not remove this discrepancy\added{, however this is not definitive as the completeness of low-mass stars of the current sample of TWA candidates and members is not known}. Such a large characteristic width is consistent with the observation that the IMF of \added{the current }TWA\added{ census} is flatter than that of the field or other young associations \citep{2011PhDT.......245L}.

In comparison to these previous \replaced{results}{determinations of the TWA IMF}, the \replaced{different case scenarios where multiple system components (primaries or companions) are fitted separately}{\emph{primaries only} and \emph{primaries + companions} case scenarios} display similar log-normal shape parameters.\deleted{ This is also true of the IMF for total system masses.}\replaced{The companion IMF should be seen as tentative}{This should however be seen as a tentative result} since the binary fraction of very low-mass objects in TWA is largely unexplored. It is therefore subject to change when future surveys identify additional low-mass companions of TWA members.

\section{THE SPACE DENSITY OF ISOLATED PLANETARY-MASS OBJECTS IN TW~HYA}\label{sec:density}

The recent discovery of two isolated high-likelihood candidate members of TWA with estimated masses in the range $\sim$\,5--7\,\Mjup\ prompts \replaced{for an}{an} estimation of their total population. Since such objects are located near the 2MASS detection limit, only those at the nearest end of TWA \replaced{have}{should have} been discovered to date. Adopting a Poisson probability distribution described by Equation~\eqref{eqn:imfposterior}, one can estimate $\lambda$, the value of the measured IMF integrated over this specific mass range. This yields an estimate of $\lambda = 1.5^{+2.0}_{-0.7}$ objects. At a \replaced{completion}{completeness} limit of $\sim$\,15\% (obtained from the statistical expectation of the relation displayed in Figure~\ref{fig:distlim_mult} over the mean mass PDF of TWA~41 and TWA~42), this corresponds to an estimated total of $10^{+13}_{-5}$ similar isolated objects in TWA\replaced{, with masses estimated at $\sim$\,5--7\,\Mjup.}{.} \replaced{These mass estimates}{The mass estimates of these objects} are strongly model-dependent and rely on a hot-start formation mechanism,\added{ but it can be noted that this predicted population is not model-dependent if viewed as that of TWA members with absolute magnitudes $K_S \sim$\,12.2--12.5.}\replaced{and it must be noted that these estimates are}{ This expected population is} based on the assumption that TWA~41 and TWA~42 are members of TWA. Only trigonometric distances remain to be measured for this to be confirmed, and the case for their membership is strengthened by their spectrophotometric distances (accounting for their young age) that are consistent with their TWA kinematic distances.

Using the effective volume of TWA that was estimated in Section~\ref{sec:imf} ($V_{\mathrm{eff}} = 6\,200^{+690}_{-630}$\,pc$^3$), this population of $\sim$\,5--7\,\Mjup\ objects can be translated to a space density around the core of TWA. This calculation yields a space density of $1.7^{+2.1}_{-0.8}\times 10^{-3}$\,objects\,pc$^{-3}$. This is remarkably high in comparison to the space density of field stars ($93 \pm 20 \times 10^{-3}$\,stars\,pc$^{-3}$; \citealt{2005ASSL..327...41C}), as \replaced{the latter}{it} would account for one such isolated planetary-mass object for every $24^{+42}_{-9}$ main-sequence stars in the field. This is not simply due to TWA being denser than the field: this estimate corresponds to one expected $\sim$\,5--7\,\Mjup\ TWA object for every $1.9^{+3.1}_{-0.6}$ currently known main sequence ($\geq$\,75\,\Mjup) member\deleted{s} of TWA.

\replaced{This space density estimate}{This estimated fraction of planetary-mass to stellar TWA members} is larger than recent estimates \replaced{for the field space density of objects near the deuterium burning limit (12.5--14.0\,\Mjup) predicted from the IMF of the Tucana-Horologium Association (THA; $5.3_{-2.9}^{+3.8}\times 10^{-3}$\,objects\,pc$^{-3}$ or one object per $17.5_{-5.0}^{+6.6}$ field main-sequence stars; \citealt{2015ApJS..219...33G} -- which are refined to $4.6^{+2.0}_{-1.2}\times 10^{-3}$\,objects\,pc$^{-3}$ or one object per $16.6^{+4.8}_{-3.2}\times 10^{-3}$\,objects\,pc$^{-3}$ using the Bayesian statistical method presented here), however it}{based on the lowest-mass candidate members of the Tucana-Horologium Association (THA). \cite{2015ApJS..219...33G} estimated this fraction at one 12.5--14.0\,\Mjup\ object per $17.5_{-5.0}^{+6.6}$ main-sequence member of THA. Using the Bayesian formalism presented here, we revise this to one low-mass object per $16.6^{+4.8}_{-3.2}\times 10^{-3}$\,objects\,pc$^{-3}$ main-sequence THA member. This estimate} remains high in comparison to predictions based on a log-normal IMF of TWA. Such an estimation for TWA, based on the \emph{primary + companions} IMF of high-likelihood and bona fide members (Section~\ref{sec:imf}), would predict a total of only $0.5_{-0.2}^{+0.4}$ $\sim$\,5--7\,\Mjup\ objects in the whole TWA association.

It is possible to make a prediction for the \added{field }space density of $\sim$\,5--7\,\Mjup\ objects by assuming that the low-mass end of the field IMF is identical to that of TWA. This was done by assuming that the field is well represented by a fiducial log-normal IMF anchored on the stellar space density of \cite{2005ASSL..327...41C}, and anchoring it in turn on the stellar population of TWA. Since TWA has a notable lack of massive, early-type stars (no A3--G9 high-likelihood candidates or members are currently known), the field IMF was anchored on K0--K9 members of TWA, corresponding to a mass range of $\sim$\,0.7--1.5\,\Msol. This calculation based on TWA yields a predicted space density of $26_{-15}^{+29}\times 10^{-3}$\,objects\,pc$^{-3}$ for field $\sim$\,5--7\,\Mjup\ objects, most of which would be faint, Y-type dwarfs ($\lesssim$\,300\,K; \citealt{2014ApJ...786L..18L}). This estimate should be seen as an upper limit because it is likely that there are still K dwarfs that are missing in the current TWA census. We avoided anchoring this estimation on the two A-type members of TWA (A0 and A2) because these spectral types span a\deleted{ very }small range of masses (1.84--2.00\,\Msol) at the age of TWA (see Figure~\ref{fig:sptmasstwa}). A similar calculation based on the low-mass members of THA would predict a lower space density of $4.6^{+2.0}_{-1.2}\times 10^{-3}$\,objects\,pc$^{-3}$ for objects at the deuterium-burning limit (12.5--14.0\,\Mjup), which would correspond to early Y-type dwarfs ($\sim$\,300--400\,K; \citealt{2011ApJ...743...50C}) at the age of the field.

A calculation similar to the one carried out in Section~\ref{sec:imf} yields an effective volume of $15\,200^{+900}_{-700}$\,pc$^3$ for THA. This allows an estimation of the THA density of stars ($23.1_{-1.7}^{+1.8}\times 10^{-3}$\,objects\,pc$^{-3}$) and objects at the deuterium-burning limit ($1.3 \pm 0.4 \times 10^{-3}$\,objects\,pc$^{-3}$). This demonstrates that \replaced{the THA association}{THA} is denser than TWA (\replaced{which}{the latter} has a core stellar density of $7.2_{-1.2}^{+1.4} \times 10^{-3}$\,objects\,pc$^{-3}$), but still much sparser than the population of field stars.

The recent discovery of WISE~J085510.83--071442.5 (W0855 hereafter; \citealt{2014ApJ...786L..18L}), which is an isolated $\sim$\,3--10\,\Mjup\ object unrelated to TWA and located at a distance of $2.23 \pm 0.04$\,pc \citep{2016arXiv160506655L}, also hints at the possibility that isolated objects in the planetary-mass regime may be more numerous in the field than predictions from a fiducial log-normal IMF. Deriving the space density PDF associated with one such object in a spherical volume with a radius of $2.31 \pm 0.08$\,pc while accounting for Poisson statistics yields a space density estimate of $25_{-17}^{+30}\times 10^{-3}$\,objects\,pc$^{-3}$ for objects similar to W0855. This estimate is associated with large error bars because it is based on only one object, however even the 99.7\% (3$\sigma$) interval of this PDF would be consistent with a minimal space density of $3.3\times 10^{-4}$\,objects\,pc$^{-3}$.

\begin{figure*}
	\centering
	\includegraphics[width=0.78\textwidth]{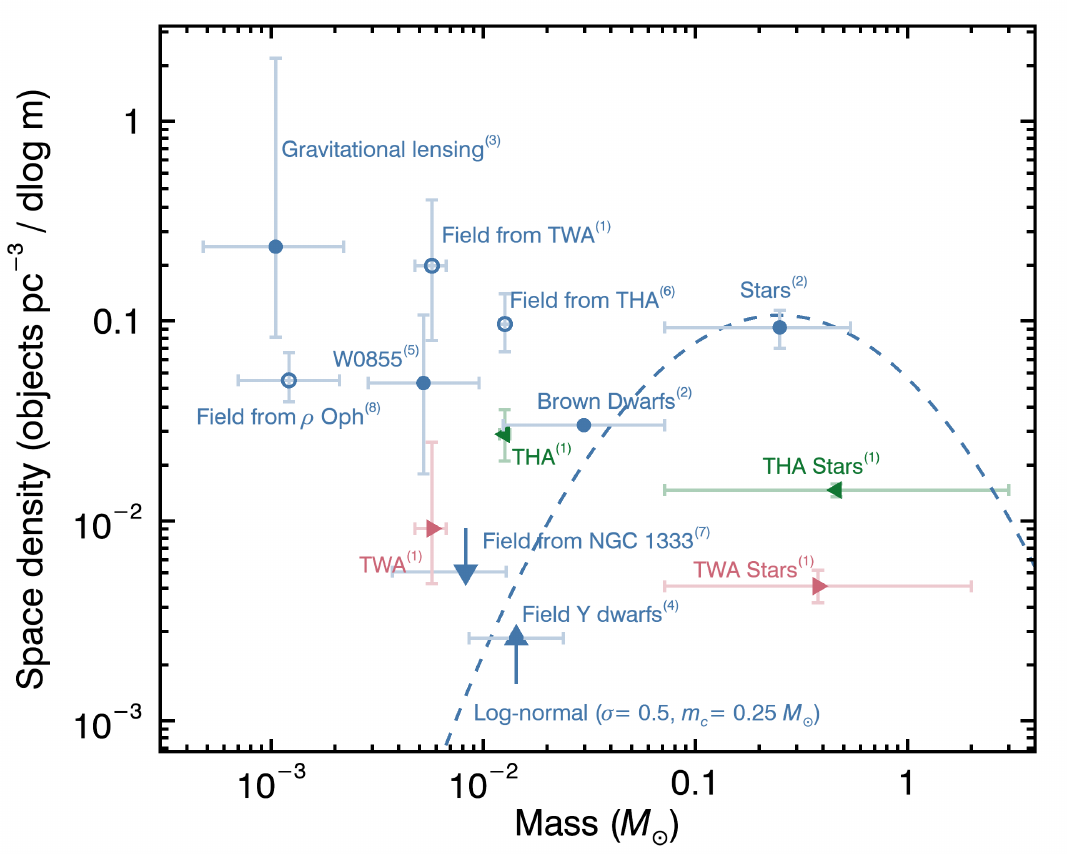}
	\caption{Estimates of space densities per unit logarithmic mass from various works, compared to a fiducial log-normal IMF ($\sigma = 0.5$, $m_c = 0.25$\,\Msol) anchored on the space density of stars (blue dashed line). All blue circles represent estimates for field objects; filled circles are direct measurements, \replaced{whereas}{and} open circles are predictions for the field based on the IMFs of TWA and THA anchored on the stellar space density estimated by \cite{2005ASSL..327...41C}. ``Field from TWA'' indicates the estimated field space density based on the ratio of low-mass to main-sequence stars in TWA and the \cite{2005ASSL..327...41C} space density of main-sequence stars. The space density of Y dwarfs (blue upwards triangle) is a lower limit, the space density translated to the field from NGC~1333 (blue downwards triangle) is an upper limit, and the predicted field space density based on TWA should be seen as an upper limit, due to the incomplete census of low-mass stars in TWA. Red rightwards triangles are estimates for TWA and green leftwards triangles are estimates for THA. References in this figure are as follows: (1)~This paper; (2)~\citealt{2005ASSL..327...41C}; (3)~\citealt{2011Natur.473..349S}; (4)~\citealt{2011ApJS..197...19K}; (5)~\citealt{2014ApJ...786L..18L}; (6)~\citealt{2015ApJS..219...33G}; (7)~\citealt{2012ApJ...756...24S}; (8)~\citealt{2010ApJ...719..550M}. See Section~\ref{sec:density} for more details.}
	\label{fig:spacedensities}
\end{figure*}

\added{The results of \cite{2012ApJ...756...24S} indicate that there may be as few as one 5--15\,\Mjup\ object for every 20--50 stars in the young ($\sim$\,1\,Myr) association NGC~1333, in strong contradiction with other results mentioned above. Similar but less constraining results have been obtained by \cite{2015ApJ...810..159M} for the $\sim$\,2\,Myr-old Chamaeleon~I region, and by \cite{2011A&A...531A..33C} and \cite{2015ApJ...810..159M} for the $\sim$\,1\,Myr-old Lupus~3 region. Results by \cite{2010ApJ...719..550M} yielded estimates of planetary-mass to main-sequence population ratios in the $\sim$\,1\,Myr $\rho$~Oph cloud core region that are more in line with our findings for TWA (see Figure~\ref{fig:spacedensities}). Detailed studies of completed young moving group censuses in the Solar neighborhood will be needed to assess whether young moving groups have fractions of isolated planetary-mass objects that are fundamentally different from NGC~1333.}

In Figure~\ref{fig:spacedensities}, space densities per logarithmic mass intervals from different works are compared with a typical log-normal IMF anchored on the stellar space density of \cite{2005ASSL..327...41C}. This figure demonstrates how the current planetary-mass space-density estimates of THA and TWA are higher than the predictions from a typical log-normal IMF anchored on the space density of main-sequence stars in the field, even though the stellar densities of both associations are much sparser than those of field stars. Predictions for the field space densities of planetary-mass (mostly Y-type) dwarfs are also displayed from the data available for both associations.

\section{SUMMARY AND CONCLUSIONS}\label{sec:conclusion}

New optical and near-infrared spectra were presented for several candidate members of TWA, in addition to 17 new radial velocity measurements. These new data allowed us to secure four new high-likelihood candidate members (2MASS~J10284580--2830374 or TWA~34; 2MASS~J12074836--3900043 or TWA~40; 2MASS~J11472421--2040204 or TWA~41; and 2MASS~J12175920--3734433 or TWA~44; i.e., objects with only a radial velocity, parallax or signs of youth left to measure) and three new bona fide members (TWA~28, TWA~29 and TWA~33).

The updated census of TWA objects contains 13 high-likelihood candidate members (11 systems) and 42 bona fide members (23 systems) with spectral types in the range A0--L7 and estimated masses in the range $\sim$\,5\,\Mjup--2\,\Msol. A determination of the initial mass function of TWA is presented using this updated census and a statistically robust method. A log-normal distribution was found to reproduce well the observed IMF of TWA, with a characteristic width that is larger than typical values for the field and other young associations.\added{These results are however possibly biased by the unknown completeness in the current sample of TWA candidates and members, which was constructed from a heterogeneous set of surveys. It is possible that a significant incompleteness in the low-mass star regime could entirely explain the unusually flat IMF.}

The recent discoveries of two new, isolated $\sim$\,5--7\,\Mjup\ high-likelihood members of TWA at the nearby end ($\sim$\,29 and $\sim$\,31\,pc) of its spatial structure are indicative that several more such members might remain to be discovered. We argue that only the nearest of these objects have been discovered yet because of the limited sensitivity of 2MASS, and that accounting for the spatial structure of TWA and Poisson statistics based on these two detections, a total of $10^{+13}_{-5}$ TWA members with similar properties could be expected.

This is much higher than what would be expected based on a log-normal IMF that is anchored on the \replaced{more massive}{higher-mass} population of TWA. This possible over-density of objects in the planetary-mass regime is surprising, but consistent with recent estimates for the space density of objects at the deuterium-burning limit in THA \citep{2015ApJS..219...33G}, the recent discovery of a cold, planetary-mass Y dwarf at only 2\,pc from the Sun \citep{2014ApJ...786L..18L}, as well as results from micro-lensing surveys \citep{2011Natur.473..349S}.

\deleted{However, the results of \cite{2012ApJ...756...24S} indicate that there may be as few as one 5--15\,\Mjup\ object for every 20--50 stars in the young ($\sim$\,1\,Myr) association NGC~1333, in strong contradiction with the results mentioned above. Similar but less constraining results have been obtained by \cite{2015ApJ...810..159M} for the $\sim$\,2\,Myr-old Chamaeleon~I region, and by \cite{2011A&A...531A..33C} and \cite{2015ApJ...810..159M} for the $\sim$\,1\,Myr-old Lupus~3 region.}More studies will be needed to further assess \replaced{this discrepancy}{the apparent discrepancy between the aforementioned results and those of \cite{2012ApJ...756...24S} for the NGC~1333 association, which seems to have less than one 5--15\,\Mjup\ object for every 20--50 stars}. For example, the James Webb Space Telescope will open the doors to a detailed study of a large number of more distant young clusters, which their IMFs to be characterized with a high significance down to $\sim$\,1\,\Mjup. The \emph{Gaia} mission will also allow completing the stellar census of TWA and other young moving groups, which will allow determining a more precise ratio of their BDs or planetary-mass objects to stellar members.

Our results indicate that many isolated planetary-mass members of TWA might still be hiding in deep large-area surveys such as VHS and AllWISE. It will however be challenging to identify them due to the relatively short temporal baseline between these surveys ($\sim$\,1\,yr), which makes it impossible to derive proper motions at the $\sim$\,10\,\masyr\ precision using the survey data alone. Future surveys such as LSST \citep{2008SerAJ.176....1I} and MaxWISE \citep{2015arXiv150501923F} will be able to reveal the population of TWA members well into the planetary-mass regime at distances up to 80\,pc.

\acknowledgments

The authors would like to thank the anonymous referee for comments and suggestions that significantly improved the quality of this manuscript, as well as Tri L. Astraatmadja for help with the \emph{Gaia} DR1, and Robert Simcoe for help with the FIRE data reduction. We thank Sarah Schmidt, Isabelle Baraffe, Gilles Chabrier, Peter Plavchan, Niall Deacon, Brendan Bowler, Michael Betancourt and Thierry Bazier-Matte for useful comments. We thank Kelle Cruz, Haley Fica, Victoria Ditomasso, Alan Munazza and Carolina Galindo for help with the observing. We thank Katelyn~N. Allers and Michael~C. Liu for sharing data. We thank the Gemini queue mode observers for data that was included in this work: Eder Martioli, Robert Bassett, K. Scott, Kathy Roth, Vinicius Placco, James Turner, Pablo Prado, Rachel Mason, Tom Geballe, Peter Pessev, German Gimeno, Mischa Schirmer, Eduardo Marin, Erich Wenderoth, David A. Krogsrud, Andrew Cardwell, Joanna E. Thomas-Osip, Andr\'e-Nicolas Chen\'e, Percy L. Gomez. This work has benefitted from the \emph{Best practices for reporting the results of MCMC analyses} document, created by Peter K.~G.~Williams and located at \url{https://github.com/pkgw/mcmc-reporting}.

This work was supported in part through grants from the Natural Science and Engineering Research Council of Canada. \emph{EEM} acknowledges support from National Science Foundation (NSF) award \emph{AST}-1313029 and the NASA NExSS program, and \emph{SL} acknowledges support from NSF grant \emph{AST}~09-08419. This research has benefited from the SpeX Prism Spectral Libraries, maintained by Adam Burgasser at \url{http://pono.ucsd.edu/\textasciitilde adam/browndwarfs/spexprism}. This document has benefitted from technical report SRON/EPS/TN/09-002 prepared by Paul Tol on color blind-friendly color schemes. This research made use of: the SIMBAD database and VizieR catalog access tool, operated at the Centre de Donn\'ees astronomiques de Strasbourg, France \citep{2000A&AS..143...23O}; data products from the Two Micron All Sky Survey (\emph{2MASS}; \citealp{2006AJ....131.1163S,2003yCat.2246....0C}), which is a joint project of the University of Massachusetts and the Infrared Processing and Analysis Center (IPAC)/California Institute of Technology (Caltech), funded by the National Aeronautics and Space Administration (NASA) and the National Science Foundation \citep{2006AJ....131.1163S}; data products from the \emph{Wide-field Infrared Survey Explorer} (\emph{WISE}; \citealp{2010AJ....140.1868W}), which is a joint project of the University of California, Los Angeles, and the Jet Propulsion Laboratory (JPL)/Caltech, funded by NASA; the VISTA Hemisphere Survey, ESO Progam, 179.A-2010 (PI: McMahon); the NASA/IPAC Infrared Science Archive (IRSA), which is operated by JPL, Caltech, under contract with NASA; and the Infrared Telescope Facility (IRTF), which is operated by the University of Hawaii under Cooperative Agreement NNX-08AE38A with NASA, Science Mission Directorate, Planetary Astronomy Program. This work has made use of data from the European Space Agency (ESA) mission {\it Gaia} (\url{http://www.cosmos.esa.int/gaia}), processed by the {\it Gaia} Data Processing and Analysis Consortium (DPAC, \url{http://www.cosmos.esa.int/web/gaia/dpac/consortium}). Funding for the DPAC has been provided by national institutions, in particular the institutions participating in the {\it Gaia} Multilateral Agreement. Part of this research was carried out at the JPL, Caltech, under a contract with NASA.

This paper includes data gathered with the 6.5 meter Magellan Telescopes located at Las Campanas Observatory, Chile (CNTAC program CN2013A-135). Based on observations obtained as part of the VISTA Hemisphere Survey, ESO Progam, 179.A-2010 (PI: McMahon). Based on observations obtained at the Gemini Observatory through programs number GN-2013A-Q-106, GN-2014A-Q-94, GS-2012B-Q-70, GS-2013A-Q-66, GS-2014A-Q-55, GS-2015A-Q-85 and GS-2015A-Q-60. The Gemini Observatory is operated by the Association of Universities for Research in Astronomy, Inc., under a cooperative agreement with the National Science Foundation (NSF) on behalf of the Gemini partnership: the NSF (United States), the National Research Council (Canada), CONICYT (Chile), the Australian Research Council (Australia), Minist\'{e}rio da Ci\^{e}ncia, Tecnologia e Inova\c{c}\~{a}o (Brazil) and Ministerio de Ciencia, Tecnolog\'{i}a e Innovaci\'{o}n Productiva (Argentina). All data were acquired through the Canadian Astronomy Data Center. This material is based upon work supported by AURA through the National Science Foundation under AURA Cooperative Agreement AST 0132798 as amended. This publication uses observations obtained at IRTF through program number 2015B091. The authors recognize and acknowledge the very significant cultural role and reverence that the summit of Mauna Kea has always had within the indigenous Hawaiian community. We are most fortunate to have the opportunity to conduct observations from this mountain.\\

\emph{JG} wrote the manuscript, generated figures, tables and led most of the analysis; obtained and reduced the Flamingos-2 data; obtained the GMOS data; obtained parts of the SpeX and FIRE data; reduced the SpeX data and parts of the FIRE data; and led the BASS-Ultracool survey. \emph{JKF} obtained parts of the FIRE and SpeX data; reduced parts of the FIRE data; provided help with observing and writing; and ideas in designing the BASS-Ultracool survey; \emph{EEM} performed the isochrone analysis of the Hipparcos candidates, wrote most of Section~\ref{sec:hip} and performed the related analysis, and generated Figure~\ref{fig:hip}; \emph{RD} led the kinematic re-analysis of Hipparcos data; \emph{JCF} calculated bolometric luminosities and effective temperatures for substellar objects; \emph{AJW} and \emph{JKD} provided parallax data and ideas for the IMF analysis; \emph{LM} obtained, reduced and analyzed ESPADONS data and performed the identification of TWA candidates from SUPERBLINK-South; \emph{DL} and \emph{\'EA} provided useful comments and help with parts of the observing; \emph{AJB} and \emph{DL} provided SpeX data and additional information on some TWA members; \emph{YB} observed parts of the FIRE data; \emph{AB} led the TWA~37 WISE excess analysis; \emph{SL} led the SUPERBLINK-South survey; \emph{CB} and \emph{GA} reduced and analyzed the GMOS data; and \emph{SC} observed parts of the SpeX data.

\facility{Gemini-South (Flamingos-2); Gemini-South (GMOS); Gemini-North (GMOS); IRTF (SpeX); Magellan:Baade (FIRE)}
\software{IDL by Harris Geospatial, Notability by Ginger Labs, Texpad by Valletta Ventures LLP}

\clearpage
\begin{deluxetable*}{llccccccc}
\tabletypesize{\tiny}
\tablecolumns{9}
\tablecaption{Kinematic Properties of Candidates and Members of TWA \label{tab:kinematic}}
\tablehead{\colhead{2MASS} & \colhead{Other} & \colhead{RA} & \colhead{DEC} & \colhead{$\mu_\alpha\cos\delta$} & \colhead{$\mu_\delta$} & \colhead{RV} & \colhead{Distance} & \colhead{Refer-}\\
\colhead{Designation} & \colhead{Names} & \colhead{J2000} & \colhead{J2000} & \colhead{(\masyr)} & \colhead{(\masyr)} & \colhead{(\kms)} & \colhead{(pc)} & \colhead{ences\tablenotemark{a}}}
\startdata
\sidehead{\textbf{Bona Fide Members of TW~Hydrae}}
10120908-3124451 & TWA 39 AB & 10:12:09.089 & -31:24:45.19 & $-74.8 \pm 1.1$ & $-9.4 \pm 1.0$ & $14.6 \pm 0.6$ & $53.9 \pm 5.1$ & 1,2,3,2\\
10423011-3340162 & TWA 7 & 10:42:30.113 & -33:40:16.21 & $-114.4 \pm 0.8$ & $-19.1 \pm 0.8$ & $11.9 \pm 0.4$ & $34.5 \pm 2.5$ & 4,5,6,5\\
11015191-3442170 & TWA 1 & 11:01:51.917 & -34:42:17.00 & $-68.2 \pm 0.1$ & $-13.9 \pm 0.1$ & $12.4 \pm 0.5$ & $59.5 \pm 1.0$ & 7,8,6,8\\
11020983-3430355 & TWA 28 & 11:02:09.833 & -34:30:35.53 & $-65.2 \pm 5.6$ & $-15.6 \pm 6.1$ & $9 \pm 3$ & $56.4 \pm 1.6$ & 9,10,10,11\\
11084400-2804504 & TWA 43; HIP 54477; HR 4334 & 11:08:44.002 & -28:04:50.44 & $-72.8 \pm 0.4$ & $-22.2 \pm 0.5$ & $16 \pm 5$ & $55.7 \pm 1.6$ & 12,13,14,13\\
11091380-3001398 & TWA 2 AB & 11:09:13.807 & -30:01:39.85 & $-91.1 \pm 0.8$ & $-21.0 \pm 0.8$ & $10.98 \pm 0.03$ & $46.6 \pm 2.8$ & 7,15,6,15\\
11210549-3845163 & TWA 12 & 11:21:05.498 & -38:45:16.34 & $-68.3 \pm 2.7$ & $-12.1 \pm 1.5$ & $10.9 \pm 0.3$ & $64.1 \pm 2.9$ & 16,15,6,15\\
11211723-3446454 & TWA 13 A & 11:21:17.135 & -34:46:45.77 & $-66.4 \pm 2.4$ & $-12.5 \pm 1.8$ & $10.8 \pm 0.4$ & $55.6 \pm 2.2$ & 16,15,6,15\\
\enddata
\end{deluxetable*}

\begin{deluxetable*}{lccccccccc}
\tabletypesize{\tiny}
\tablecolumns{10}
\tablecaption{Spectrophotometric Properties of Candidates and Members of TWA \label{tab:spectro}}
\tablehead{\colhead{2MASS} & \multicolumn{2}{c}{Spectral Type\tablenotemark{a}} & \colhead{} & \multicolumn{3}{c}{2MASS\tablenotemark{b}} &\colhead{} & \multicolumn{2}{c}{AllWISE\tablenotemark{b}}\\
\cline{2-3}
\cline{5-7}
\cline{9-10}
\colhead{Designation} & \colhead{Optical} & \colhead{NIR} & \colhead{Ref.\tablenotemark{c}} & \colhead{$J$} & \colhead{$H$} & \colhead{$K_S$} & \colhead{} & \colhead{$W1$} & \colhead{$W2$}}
\startdata
\sidehead{\textbf{Bona Fide Members of TW~Hydrae}}
10120908--3124451 A & M4\,Ve & $\cdots$ & 1, & [$9.60 \pm 0.00$] & [$9.01 \pm 0.04$] & [$8.75 \pm 0.03$] & & [$8.51 \pm 0.03$] & [$8.25 \pm 0.02$]\\
10120908--3124451 B & [M4\,Ve] & $\cdots$ & 1, & [$9.60 \pm 0.00$] & [$9.01 \pm 0.04$] & [$8.75 \pm 0.03$] & & [$8.51 \pm 0.03$] & [$8.25 \pm 0.02$]\\
10423011--3340162 & M3\,IVe & M4\,pec & 2,3 & $7.79 \pm 0.01$ & $7.13 \pm 0.03$ & $6.90 \pm 0.02$ & & $6.79 \pm 0.04$ & $6.66 \pm 0.02$\\
11015191--3442170 & K8\,IVe & M3\,e\,pec & 2,3 & $8.22 \pm 0.02$ & $7.56 \pm 0.04$ & $7.30 \pm 0.02$ & & $7.01 \pm 0.05$ & $6.88 \pm 0.02$\\
11020983--3430355 & M8.5\,$\gamma$ & M9\,$\gamma$ & 4,5 & $13.03 \pm 0.02$ & $12.36 \pm 0.02$ & $11.89 \pm 0.02$ & & $11.45 \pm 0.02$ & $10.81 \pm 0.02$\\
11084400--2804504 & A2\,Vn & $\cdots$ & 6, & $5.31 \pm 0.03$ & $5.35 \pm 0.03$ & $5.24 \pm 0.02$ & & $5.2 \pm 0.2$ & $5.04 \pm 0.06$\\
11091380--3001398 A & M1.5\,IVe & $\cdots$ & 2, & [$8.38 \pm 0.03$] & [$7.68 \pm 0.04$] & [$7.46 \pm 0.02$] & & [$7.41 \pm 0.07$] & [$7.24 \pm 0.02$]\\
11091380--3001398 B & M2 & $\cdots$ & 7, & [$8.38 \pm 0.03$] & [$7.68 \pm 0.04$] & [$7.46 \pm 0.02$] & & [$7.41 \pm 0.07$] & [$7.24 \pm 0.02$]\\
\enddata
\end{deluxetable*}

\begin{deluxetable*}{llccccc}
\tabletypesize{\scriptsize}
\tablecolumns{7}
\tablecaption{BANYAN~II Probabilities for Candidates and Members of TWA\label{tab:banyan}}
\tablehead{\colhead{2MASS} & \colhead{Other} & \colhead{BANYAN~II} & \colhead{Bayesian} & \colhead{Contamination} & \colhead{Statistical} & \colhead{Statistical}\\
\colhead{Designation} & \colhead{Names} & \colhead{Inputs\tablenotemark{a}} & \colhead{Probability (\%)} & \colhead{Probability (\%)\tablenotemark{b}} & \colhead{Distance (pc)} & \colhead{RV (\kms)}}
\startdata
\sidehead{\textbf{Bona Fide Members of TW~Hydrae}}
10120908-3124451 & TWA 39 AB & $\mu$yrt & $99.4$ & $< 0.1$ & $\cdots$ & $\cdots$\\
10423011-3340162 & TWA 7 & $\mu$yrt & $99.9$ & $< 0.1$ & $\cdots$ & $\cdots$\\
11015191-3442170 & TWA 1 & $\mu$yrt & $100.0$ & $< 0.1$ & $\cdots$ & $\cdots$\\
11020983-3430355 & TWA 28 & $\mu$yprt & $99.9$ & $< 0.1$ & $\cdots$ & $\cdots$\\
11084400-2804504 & TWA 43; HIP 54477; HR 4334 & $\mu$yrt & $98.7$ & $< 0.1$ & $\cdots$ & $\cdots$\\
11091380-3001398 & TWA 2 AB & $\mu$yrt & $100.0$ & $< 0.1$ & $\cdots$ & $\cdots$\\
11210549-3845163 & TWA 12 & $\mu$yrt & $100.0$ & $< 0.1$ & $\cdots$ & $\cdots$\\
11211723-3446454 & TWA 13 A & $\mu$yrt & $100.0$ & $< 0.1$ & $\cdots$ & $\cdots$\\
\enddata
\tablenotetext{a}{All observables that were included in BANYAN~II to derive membership probabilities; $\mu$: proper motion, y: signs of youth (i.e., age securely younger than $\sim$\,1\,Gyr), p: 2MASS and AllWISE photometry, r: radial velocity, t: trigonometric distance. Photometric measurements were not used for $\geq$\,L5--type objects, because the young photometric sequences used in BANYAN~II are not well calibrated at these late spectral types.}
\tablenotetext{b}{Probability that a synthetic object drawn from a field Galactic model obtains an equal or higher Bayesian probability. See \cite{2014ApJ...783..121G} for more detail (their Section 5).}
\tablenotetext{c}{Possible LCC contaminant.}
\tablecomments{See Section~\ref{sec:candcomp} for more details.}
\end{deluxetable*}

\appendix

\section{NEW OPTICAL AND NEAR-INFRARED SPECTRA}\label{an:spt}

\begin{figure*}[p]
	\centering
	\includegraphics[width=0.995\textwidth]{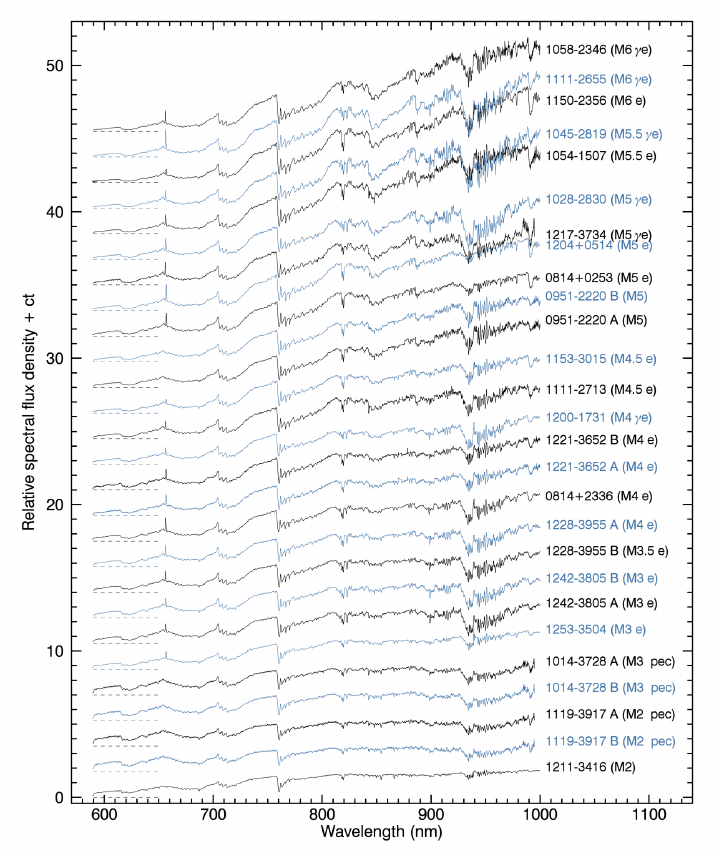}
	\caption{New GMOS spectra presented in this work. The individual zero flux levels are represented with horizontal dashed lines. See Section~\ref{sec:gmos} for more details.}
	\label{fig:gmos_seq}
\end{figure*}

\begin{figure*}[p]
	\centering
	\includegraphics[width=0.995\textwidth]{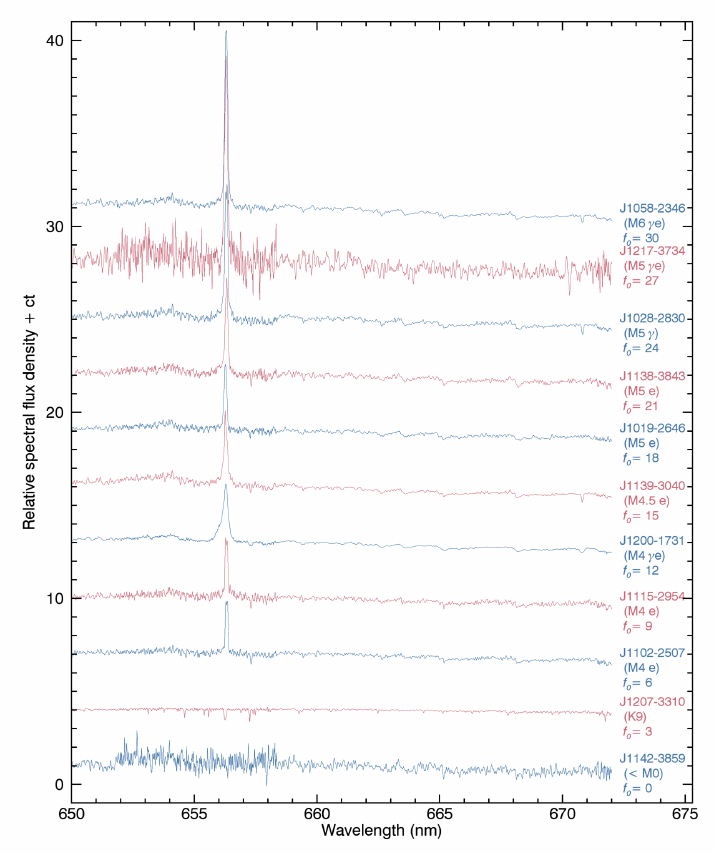}
	\caption{New ESPaDOnS spectra presented in this work. Zero levels in relative flux are indicated below object names and spectral types. See Section~\ref{sec:obs_espadons} for more details.}
	\label{fig:esp_seq}
\end{figure*}

\begin{figure*}[p]
	\centering
	\includegraphics[width=0.995\textwidth]{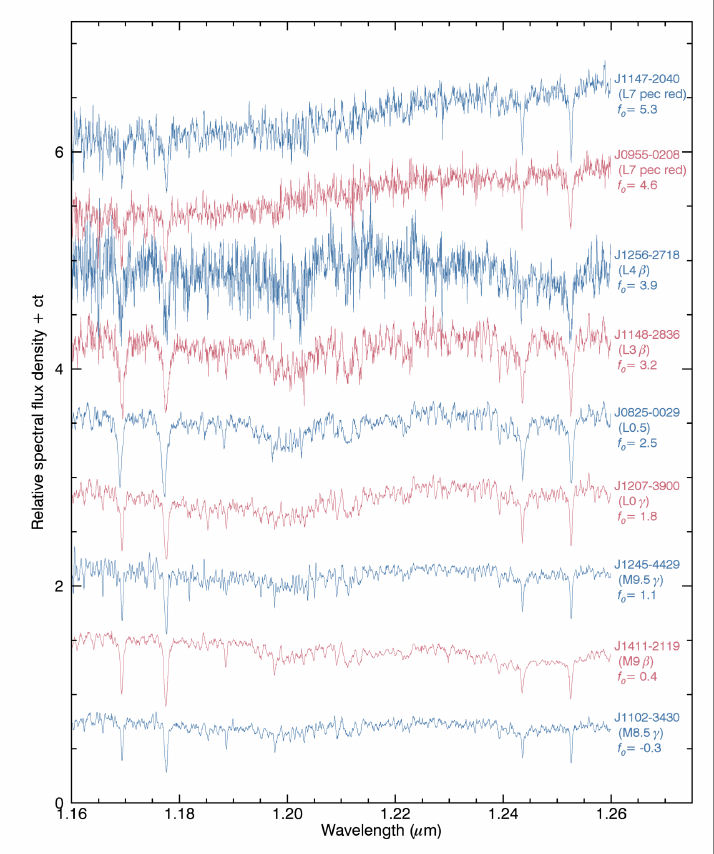}
	\caption{New $J$-band NIR spectra from FIRE obtained in echelle mode. Zero levels in relative flux are indicated below object names and spectral types. See Section~\ref{sec:obs_fire} for more details.}
	\label{fig:fire_ech_seq}
\end{figure*}

\begin{figure*}[p]
	\centering
	\includegraphics[width=0.995\textwidth]{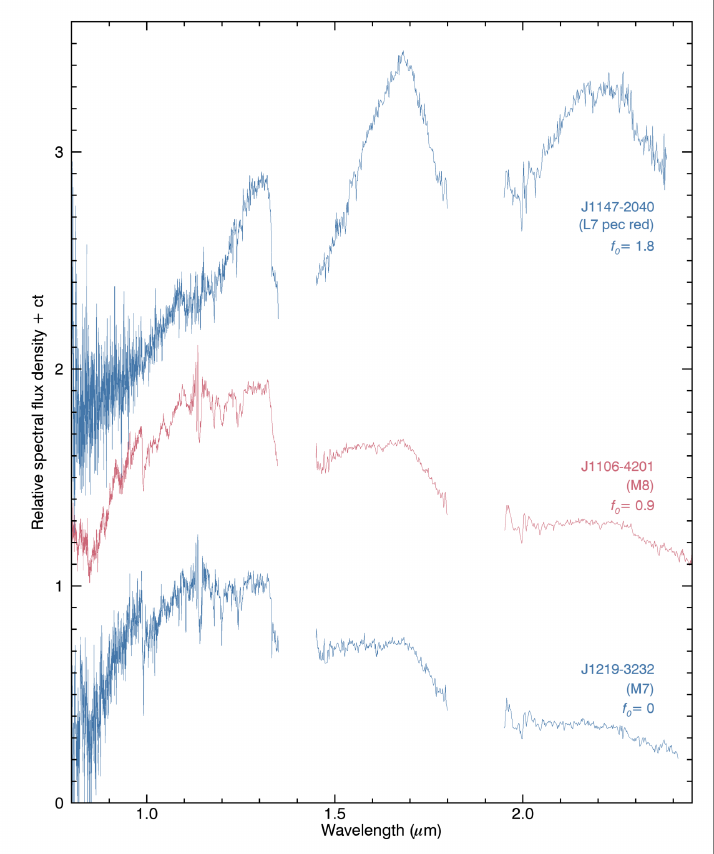}
	\caption{New NIR spectra from FIRE obtained in prism mode. Zero levels in relative flux are indicated below object names and spectral types. The slopes of all spectra were corrected using 2MASS photometry, as described in Section~\ref{sec:spt}. 2MASS~J12194846--3232059 and 2MASS~J11063147--4201251 are normal low-mass stars with no signs of youth and were thus rejected as TWA candidates. 2MASS~J11472421--2040204 (TWA~41) displays telltale signatures of youth such as a very red continuum and a triangular-shaped $H$ band ($\sim$\,1.5--1.8\,$\mu$m). See Section~\ref{sec:obs_fire} for more details.}
	\label{fig:fire_prz_seq}
\end{figure*}

\begin{figure}
	\centering
	\includegraphics[width=0.68\textwidth]{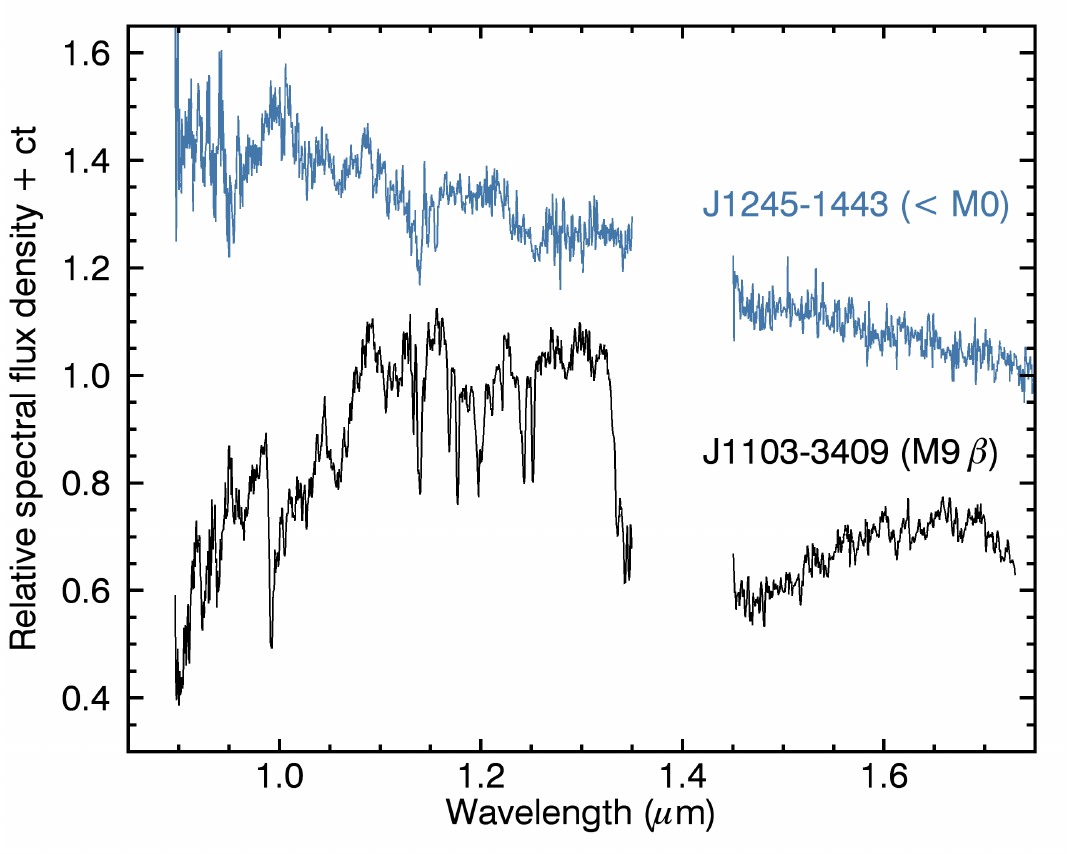}
	\caption{New Flamingos-2 spectra presented in this work. 2MASS~J12451035--1443029 has a spectral type earlier than M0 and is likely a background contaminant, whereas 2MASS~J11034950--3409445 is an M9\,$\beta$ dwarf with tentative signs of a low surface gravity. The relative flux of 2MASS~J11034950--3409445 was not offset, and that of 2MASS~J12451035--1443029 was offset by 0.3. See Section~\ref{sec:obs_f2} for more details.}
	\label{fig:f2_spectra}
\end{figure}

In this Section, new optical and near-infrared spectra described in Section~\ref{sec:obs} are displayed for reference (Figures~\ref{fig:gmos_seq}, \ref{fig:esp_seq}, \ref{fig:fire_ech_seq}, \ref{fig:fire_prz_seq}, and \ref{fig:f2_spectra}). The two new SpeX spectra (Section~\ref{sec:obs_spex}) are presented and discussed individually in Section~\ref{sec:spt}.

\section{THE TOTAL MASS PROBABILITY DENSITY FUNCTION OF A BINARY SYSTEM}\label{an:totalmass}

In this Appendix, the mathematical development leading to the total mass probability density function (PDF) of a system of two stars is carried out. In Section~\ref{sec:physpar}, a PDF $\mathcal{P}(\ell)$ for the logarithm of the mass ($\ell = \log _{10} m$) of each star in the sample is calculated. The determination of a systemic initial mass function (Section~\ref{sec:imf}) thus requires that the PDF of the total mass of binary systems be calculated. Let the two stars have masses $m_1$ and $m_2$, or $\ell_1$ and $\ell_2$ in logarithm space, and let $m_t$ be the total mass of the system. It follows that:
\begin{align}
	\ell_t = \log _{10} m_t = \log _{10} \left(10^{\ell_1}+10^{\ell_2}\right).\label{eqn:lt}
\end{align}
The problem is thus to determine $\mathcal{P}_t(\ell_t)$ from $\mathcal{P}_1(\ell_1)$ and $\mathcal{P}_2(\ell_2)$. \deleted{In order }To do this, the joint PDF $\mathcal{P}_j(\ell_1,\ell_2)$ is introduced, and followed by a change of variable:
\begin{align}
	\mathcal{P}_j(\ell_1,\ell_2) &= \mathcal{P}_1(\ell_1)\mathcal{P}_2(\ell_2),\\
	\mathcal{P}_j^\prime(\ell_1,\ell_t)\,\mathrm{d}\ell_1\mathrm{d}\ell_t &= \mathcal{P}_j(\ell_1,\ell_2)\,\mathrm{d}\ell_1\mathrm{d}\ell_2.
\end{align}
It follows that :
\begin{align}
	\mathcal{P}_j^\prime(\ell_1,\ell_t) = \mathcal{P}_j(\ell_1,\ell_2)\,|\mathbf{J}|,
\end{align}
\noindent where $|\mathbf{J}|$ is the determinant of the Jacobian matrix of this transformation:
\begin{align}
	\mathbf{J} = \left[\begin{array}{cc}
		 \frac{\partial\ell_1}{\partial\ell_1} & \frac{\partial\ell_1}{\partial\ell_t}\\
		 \frac{\partial\ell_2}{\partial\ell_1} & \frac{\partial\ell_2}{\partial\ell_t}\\
	\end{array}\right].
\end{align}

It follows from Equation~\eqref{eqn:lt} that:
\begin{align}
	|\mathbf{J}| = \frac{10^{\ell_t}}{10^{\ell_t}-10^{\ell_1}}
\end{align}

The final PDF $\mathcal{P}_t(\ell_t)$ can then be obtained by marginalizing $\mathcal{P}_j^\prime(\ell_1,\ell_t)$ over $\ell_1$, in the domain $\ell_1 < \ell_t$ where the joint PDF is defined:
\begin{align}
	\mathcal{P}_t(\ell_t) &= \int_{-\infty}^{\ell_t} \mathcal{P}_j^\prime(\ell_1,\ell_t)\,\mathrm{d}\ell_1,\\
	 &= \int_{-\infty}^{\ell_t} \mathcal{P}_j(\ell_1,\ell_2)\,|\mathbf{J}|\,\mathrm{d}\ell_1,\\
	 &= \int_{-\infty}^{\ell_t} \mathcal{P}_1(\ell_1)\,\mathcal{P}_2\Big(\log _{10} \left(10^{\ell_t}-10^{\ell_1}\right)\Big)\frac{10^{\ell_t}}{10^{\ell_t}-10^{\ell_1}}\,\mathrm{d}\ell_1.
\end{align}

This convolution-like integral representation of $\mathcal{P}_t(\ell_t)$ is then solved numerically.

\section{REFERENCE PRIORS FOR THE MCMC FIT OF CUMULATIVE INITIAL MASS FUNCTIONS}\label{an:refpriors}

The Markov Chain Monte Carlo (MCMC) fitting algorithm described in Section~\ref{sec:imf} requires the choice of a prior distribution on the parameters of the initial mass function (IMF) models to be fitted to the data. The numerical method described by \cite{Berger:2009wx} was used to obtain non-informative priors, i.e., priors distributions that do not inject any information on the values of the IMF parameters that are not informed by the data.

In the case of a Salpeter IMF, this was done by defining a 50\,$\times$\,50 grid of parameters $N = \phi_0 V_\mathrm{eff}$ and $\alpha$ defined in the ranges 10--100 and 0.3--4, respectively. For each values of $N_i$ and $\alpha_j$ on the grid, a random number of total TWA members $N_i^\prime$ was drawn from a Poisson distribution $\mathcal{P}\left(x|N_i\right)$. Each of these objects were attributed a mass following the IMF distribution described in Equation~\eqref{eqn:salpeterimf}, with a parameter $\alpha_j$. The cumulative IMF was then constructed, and compared to the model with a Kolmogorov-Smirnov test. The likelihood probability density $\mathcal{L}_{ij}(\mathcal{D}_{ij}|N_i,\alpha_i)$ was then calculated  following Equation~\eqref{eqn:mcmc_likelihood} at every point of the grid, where $\mathcal{D}_{ij}$ is the simulated data.

\begin{figure*}
	\center
	\includegraphics[width=0.688\textwidth]{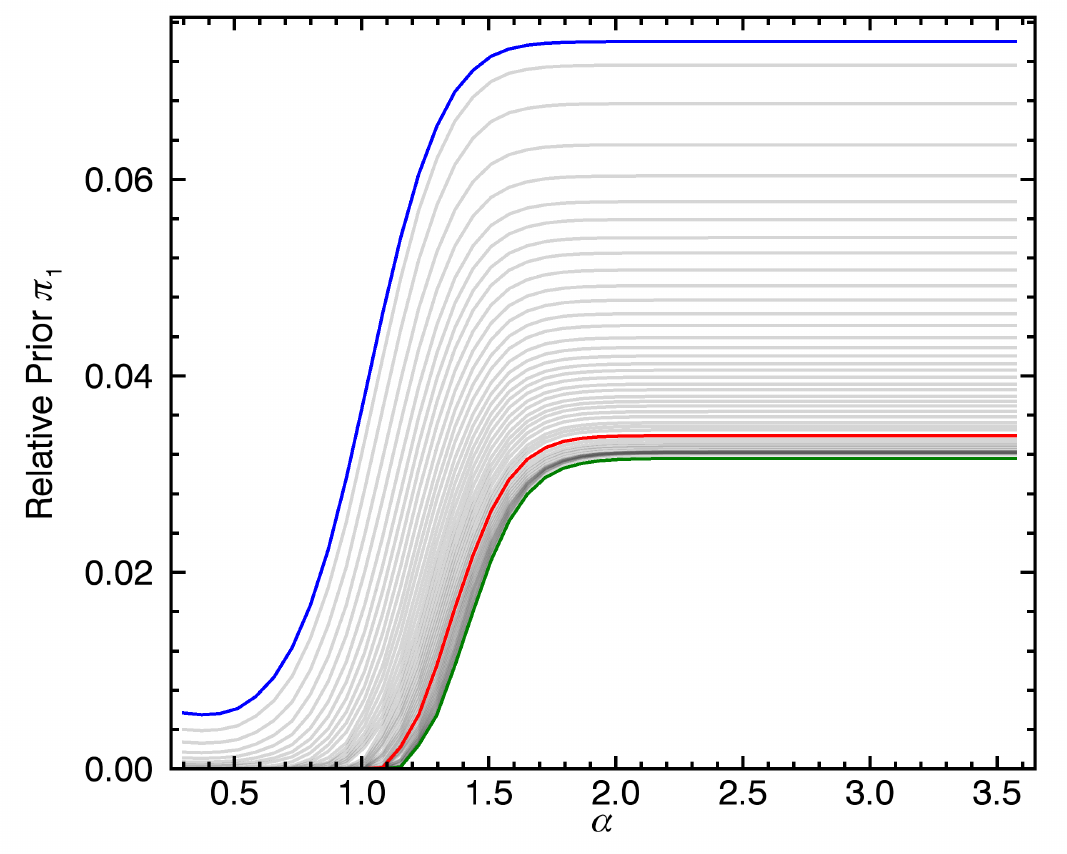}
	\caption{Reference priors in the Salpeter IMF case. The red curve corresponds to the case where $N$ has been fixed to a typical value of 70, the blue and green curves correspond to minimal and maximal values of $N$ respectively, and the grey curves to other values. All priors disfavour $\alpha \lesssim$\,1.7, but become uniform above this value. Smaller numbers $N$ of TWA members tend to be favoured.}
	\label{fig:ref_salpeterpriors}
\end{figure*}

\cite{Berger:2009wx} define the value of the reference prior $\pi_{ij}\left(\left\{\theta\right\}\right)$ (where $\left\{\theta\right\}$ is a set of parameters) as the mean value of $\mathcal{L}_{ij}(\mathcal{D}_{ij}|\left\{\theta\right\})$ over a large number of trials. Obtaining a properly normalized prior distribution requires an additional step at each trial that was ignored in the present case; the resulting prior distribution is thus not normalized to unity, but this has no effect on the MCMC algorithm because it only relies on the relative value of the likelihood and prior at different steps in the parameter space. A total of  1\,000 trials were performed, and the resulting prior array was smoothed in logarithmic space with a running box of 6\,$\times$\,6 elements. The resulting priors are displayed in Figure~\ref{fig:ref_salpeterpriors}.

\begin{figure*}[p]
	\centering
\subfigure[$\pi_2(\sigma)$ for varying $\log _{10} m_c$]{\includegraphics[width=0.488\textwidth]{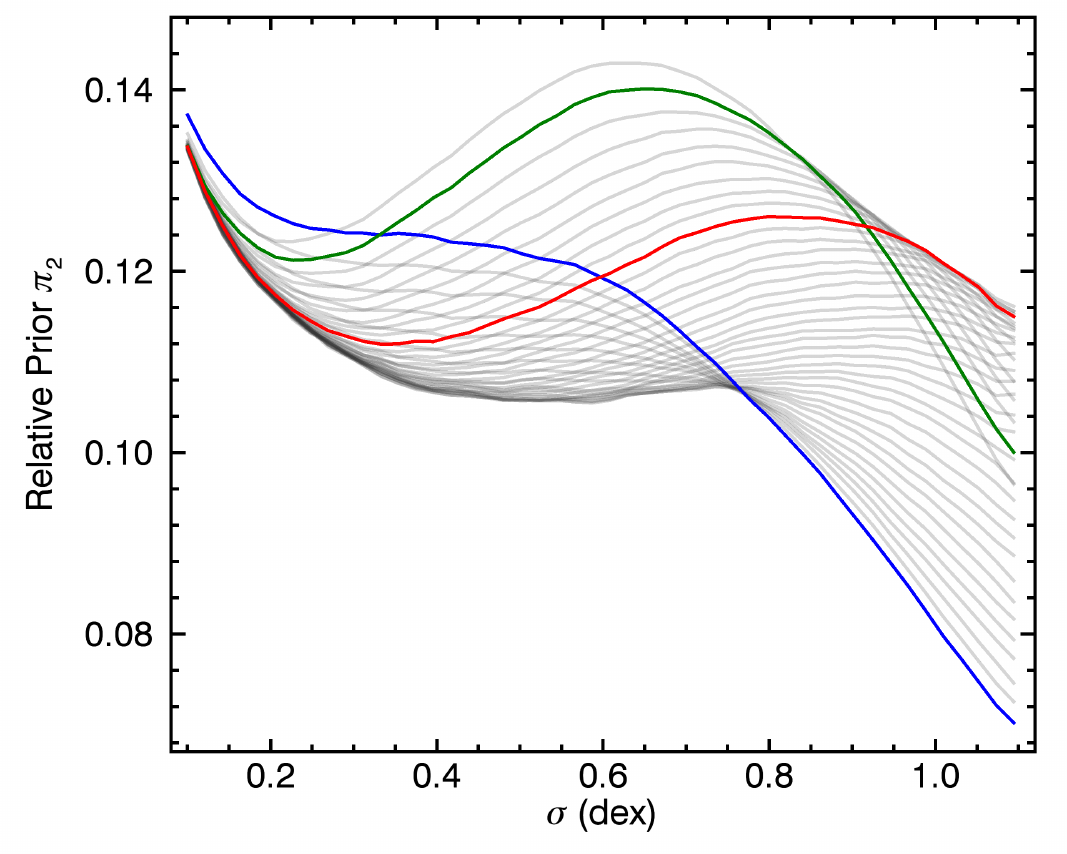}}
\subfigure[$\pi_2(\sigma)$ for varying $N$]{\includegraphics[width=0.488\textwidth]{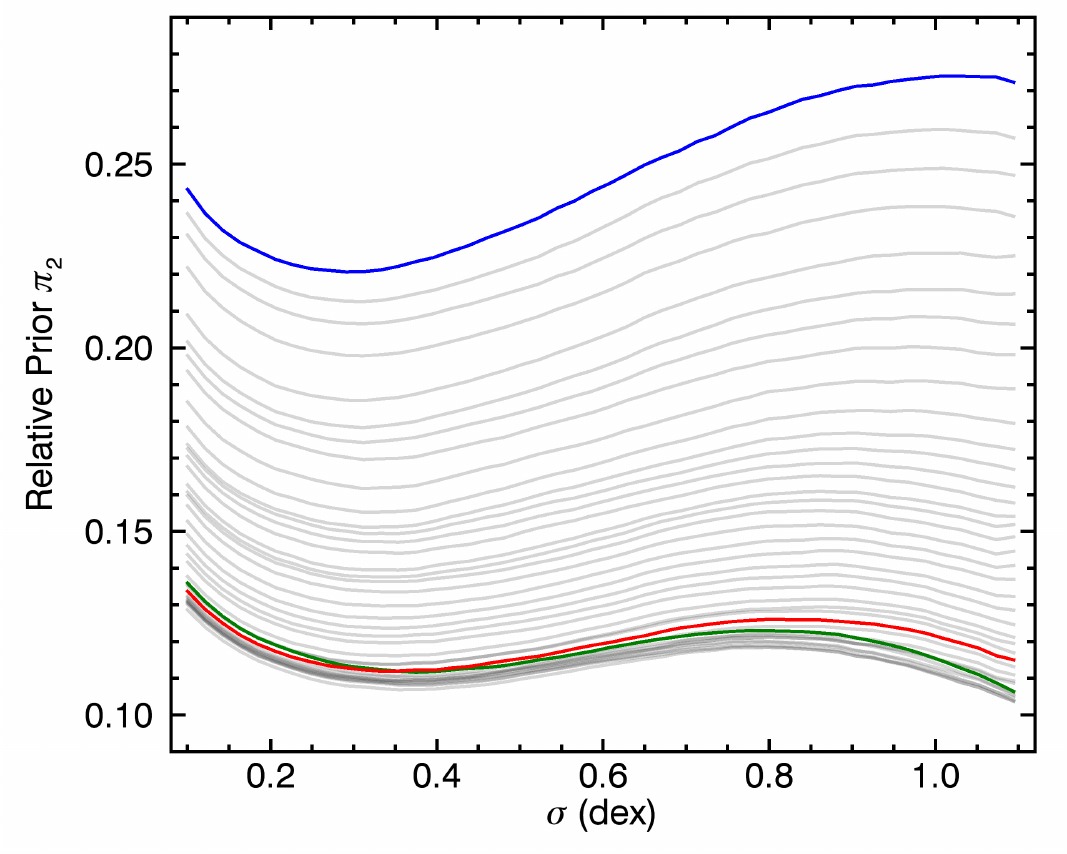}}
\subfigure[$\pi_2(m_c)$ for varying $\sigma$]{\includegraphics[width=0.488\textwidth]{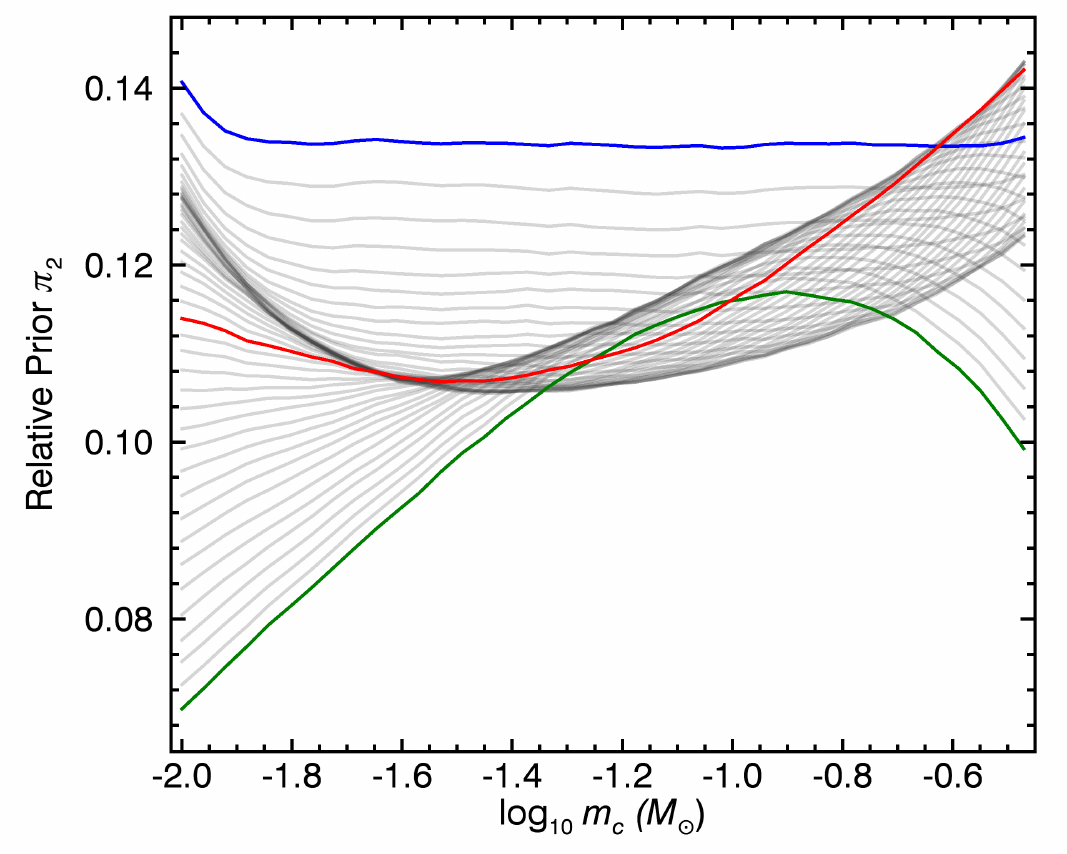}}
\subfigure[$\pi_2(m_c)$ for varying $N$]{\includegraphics[width=0.488\textwidth]{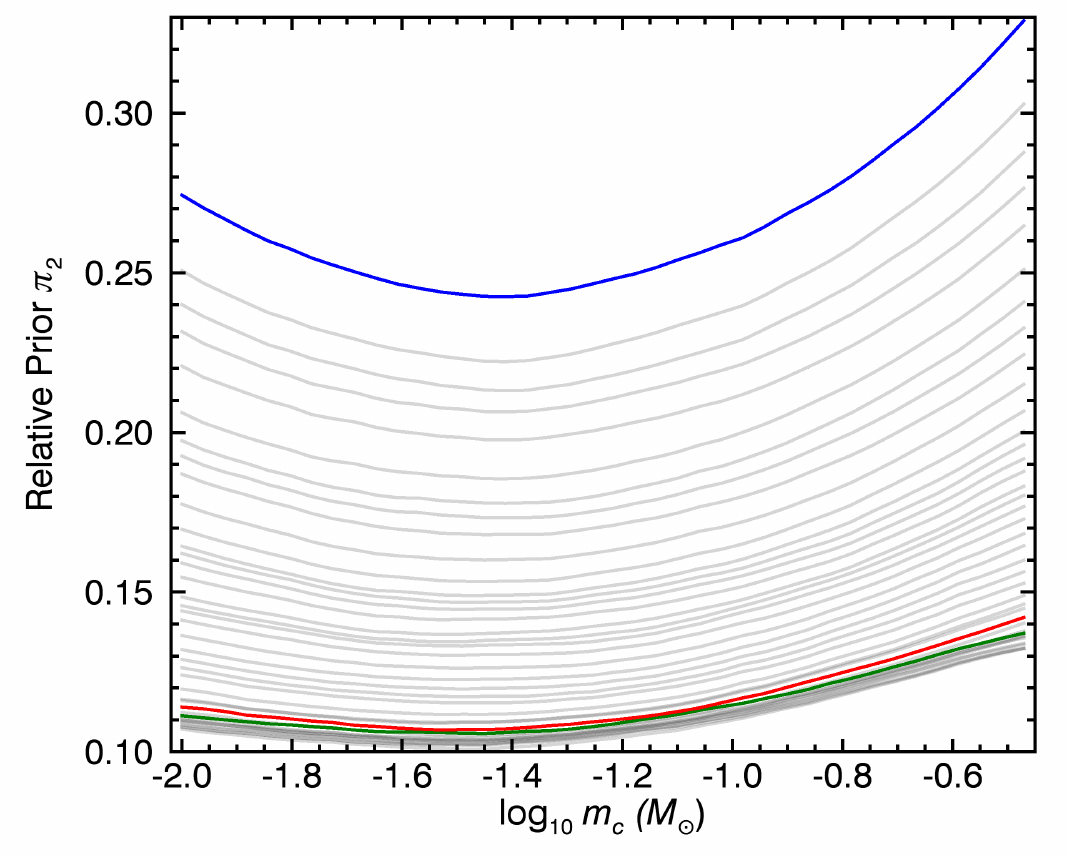}}
	\caption{Reference priors in the log-normal IMF case. The red curves correspond to the cases where 2 out of 3 parameters have been fixed to typical values $\left\{N,\log _{10} m_c,\sigma\right\} = \left\{70,0.27,0.75\right\}$, the blue curves correspond to minimal parameter values, the green curves to maximal parameter values, and the grey curves to other values. Although the priors do not change in a drastic way, lower $N$ tend to be favoured. The regions of $\sigma$ that are favoured are dependent on the value of $\log_{10} m_c$; small central masses favour high $\sigma$ and vice-versa. Low values of $\sigma$ result in a flat prior on the central mass.}
	\label{fig:ref_priors}
\end{figure*}

In the log-normal case, a similar calculation was performed on a 50\,$\times$\,50\,$\times$\,50 grid of parameters $N = \phi_t V_\mathrm{eff}$, $\log _{10} m_c$ and $\sigma$, defined in the ranges $10$--$100$, $-2$--$0$\,dex (in $\log _{10} M_\odot$) and $0.1$--$1.2$\,dex (in $\log _{10} M_\odot$), respectively. Slices of the resulting reference prior cube are displayed in Figure~\ref{fig:ref_priors}.

\bibliographystyle{apj}

\end{document}